\documentclass{emulateapj}
\usepackage{color}
\usepackage{verbatim}

\def\ah{^{\rm h}}
\def\am{^{\rm m}}

\def\pr{^{\prime}}
\def\2pr{^{\prime \prime}}
\def\deg{^{\circ}}
\def\greatsim{\mathrel{\raise.3ex\hbox{$>$\kern-.75em\lower1ex\hbox{$\sim$}}}}
\def\lesssim{\mathrel{\raise.3ex\hbox{$<$\kern-.75em\lower1ex\hbox{$\sim$}}}}
\def\gs{\mathrel{\raise0.27ex\hbox{$>$}\kern-0.70em 
\lower0.71ex\hbox{{$\scriptstyle \sim$}}}}
\def\ls{\mathrel{\raise0.27ex\hbox{$<$}\kern-0.70em 
\lower0.71ex\hbox{{$\scriptstyle \sim$}}}}

\def\sqdeg{{\,{\rm deg}^2}}
\def\vunits{{\,h^3{\rm Mpc}^{-3}}}

\slugcomment{Submitted to {\em Astronomical Journal}}

\shorttitle{BOSS}
\shortauthors{Dawson et al.}

\usepackage{color}
\usepackage{graphicx}
 
\newcommand{\lya}{\mbox{Ly$\alpha$}}

\newcommand{\hmpc}{\mbox{$h^{-1}$Mpc}}

\newif\ifdraftmodep
\draftmodepfalse

\newif\ifapjp
\apjpfalse

\begin{document}

\title{The Baryon Oscillation Spectroscopic Survey of SDSS-III}

\author{
Kyle~S.~Dawson\altaffilmark{1},
David~J.~Schlegel\altaffilmark{2},
Christopher~P.~Ahn\altaffilmark{1},    
Scott~F.~Anderson\altaffilmark{3},
\'Eric~Aubourg\altaffilmark{4},
Stephen~Bailey\altaffilmark{2},
Robert~H.~Barkhouser\altaffilmark{5},
Julian~E.~Bautista\altaffilmark{4},
Alessandra~Beifiori\altaffilmark{6},
Andreas~A.~Berlind\altaffilmark{7}
% ...
}

\altaffiltext{1}{
Department of Physics and Astronomy, 
University of Utah, Salt Lake City, UT 84112, USA.
}

\altaffiltext{2}{
Lawrence Berkeley National Laboratory, One Cyclotron Road,
Berkeley, CA 94720, USA.
}

\altaffiltext{3}{
Department of Astronomy, University of Washington, 
Box 351580, Seattle, WA 98195, USA.
}

\altaffiltext{4}{
APC, University of Paris Diderot, CNRS/IN2P3, CEA/IRFU, Observatoire de Paris, Sorbonne Paris Cite, France.
}

\altaffiltext{5}{
Department of Physics and Astronomy, Johns Hopkins University, Baltimore, MD 21218, USA.
}

\altaffiltext{6}{
Max-Planck-Institut f\"ur Extraterrestrische Physik,
Giessenbachstra{\ss}e,
85748 Garching, Germany.
}

\altaffiltext{7}{
Department of Physics and Astronomy, Vanderbilt University, 
VU Station 1807, Nashville, TN 37235, USA.
}

% ...

\email{kdawson@astro.utah.edu}

\begin{abstract}
The Baryon Oscillation Spectroscopic Survey (BOSS) is designed to measure the scale
of baryon acoustic oscillations (BAO) in the clustering of matter over a larger volume than
the combined efforts of all previous spectroscopic surveys of large scale structure.
BOSS uses 1.5 million luminous galaxies as faint as $i=19.9$ over 10,000 $\sqdeg$ to measure BAO to redshifts $z<0.7$.
Observations of neutral hydrogen in the \lya\ forest
in more than 150,000 quasar spectra ($g<22$) will constrain BAO over the redshift range $2.15<z<3.5$.
Early results from BOSS include the first detection of the large-scale 
three-dimensional clustering of the Ly$\alpha$ forest and a strong detection
from the Data Release 9 data set of the BAO in the clustering of massive
galaxies at an effective redshift $z = 0.57$.
We project that BOSS will yield measurements of
the angular diameter distance $D_A$ to an accuracy of 1.0\% at redshifts $z=0.3$ and
$z=0.57$ and measurements of $H(z)$ to 1.8\% and 1.7\% at the same redshifts.
Forecasts for \lya\ forest constraints predict a measurement of
an overall dilation factor that scales the highly degenerate
$D_A(z)$ and $H^{-1}(z)$ parameters
to an accuracy of 1.9\% at $z \sim 2.5$ when the survey is complete.
Here, we provide an overview of the selection of spectroscopic targets,
planning of observations, and analysis of data and data quality of BOSS.
\end{abstract}

\section{Introduction}\label{sec:intro}
The large scale structure of the universe, as traced by galaxies
in redshift surveys and by intergalactic hydrogen absorption towards
distant quasars, has played a central role in establishing the
modern cosmological model based on inflation, cold dark matter,
and dark energy.  Key steps in this development have included the
Center for Astrophysics redshift surveys \citep{huchra83a,falco99a},
the Las Campanas Redshift Survey \citep{shectman96a}, and the
Two Degree Field Galaxy Redshift Survey \citep[2dFGRS;][]{colless01a}.
The largest and most powerful redshift surveys to date have been those
from the Sloan Digital Sky Survey \citep[SDSS;][]{york00a}, which 
measured redshifts of nearly one million galaxies in spectroscopic
observations between 2000 and 2008 \citep[the phases known as SDSS-I and -II, 
described in the seventh data release, DR7, by][]{abazajian09a}.
The SDSS also obtained the most precise constraints to date on structure
at $2<z<4$ using \lya\ forest absorption towards $\sim 3000$
high redshift quasars \citep{mcdonald05a,mcdonald06a,seljak06a},
building on earlier studies that analyzed much smaller samples
of higher resolution spectra \citep{croft99b,croft02b,mcdonald00a,kim04a}.  

This paper describes the Baryon Oscillation Spectroscopic
Survey (BOSS), the largest of the four surveys that comprise
SDSS-III \citep{eisenstein11a}.  Shorter summaries of BOSS have
appeared previously in an Astro2010 white paper \citep{schlegel09a},
in the SDSS-III overview paper \citep{eisenstein11a}, and in papers
presenting some of the survey's early cosmological results
\citep[e.g.][]{white11a,slosar11a,anderson12a}.

Large-scale structure offers a novel tool to make precise measurements
of cosmic distances via baryon acoustic oscillations (BAO), a feature
imprinted on the clustering of matter by acoustic waves that propagate in the
pre-recombination universe.  A long-standing prediction of cosmological
models \citep{sakharov66a,peebles70a,sunyaev70a},
BAO rose to prominence in recent
years as a tool to measure the expansion history of the universe
\citep{eisenstein98a,blake03a,seo03a}.  The first clear detections of BAO
came in 2005 from analyses of the SDSS \citep{eisenstein05a} and
the 2dFGRS \citep{cole05a} galaxy samples, and even these first discoveries set
few-percent constraints on the cosmic distance scale.  With sufficient
data, the BAO ``standard ruler'' can be used to separately measure
the angular diameter distance $D_A(z)$ from transverse clustering and
the Hubble expansion rate $H(z)$ from line-of-sight clustering.
$D_A(z)$ and $H(z)$ have not yet been measured separately at any redshift.
A particular attraction of the BAO method is its low susceptibility
to systematic errors, a feature highlighted in the report of the
Dark Energy Task Force \citep[DETF;][]{albrecht06a}.
The BAO method is reviewed in detail by
\citet[][see their \S 4]{weinberg12a},
including discussions of the underlying theory, the effects
of non-linear evolution and galaxy bias, survey design and
statistical errors, control of systematics, recent observational
results, and complementarity with other probes of cosmic acceleration.

A number of recent cosmological surveys were designed with BAO
measurement as a major goal.  The recently completed WiggleZ
survey \citep{drinkwater10a}
observed 200,000 emission-line galaxies over $800 \, \sqdeg$,
obtaining the first BAO measurements at $0.5<z<1.0$, with an aggregate
distance precision of 3.8\% at a central redshift $z=0.6$
\citep{blake11b}.
The Six-Degree Field Galaxy Survey \citep[6dFGS;][]{jones09a}
was designed to map structure in the local universe;
it produced a 4.5\% BAO distance measurement at redshift
$z\sim 0.1$ \citep{beutler11a}.
The primary SDSS BAO results came from its redshift survey of
luminous red galaxies \citep[LRGs;][]{eisenstein01a}, with
color selection used to separate high luminosity
targets from nearby galaxies at similar apparent
magnitudes.  This survey, reaching a limiting apparent magnitude
$r=19.5$, provided a sparse sample of 106,000 strongly
clustered galaxies over a large volume, well suited to
measuring structure on BAO scales.  The LRG survey achieved a
roughly constant comoving space density of $10^{-4} \vunits$ from $z=0.16$ to
$z=0.40$, with a declining density out to $z \sim 0.45$.
\citet{percival10a} derived the power spectrum from the
LRG sample, the DR7 galaxy sample, and the 2dFGRS sample to obtain a 2.7\%
measurement of the BAO scale at $z=0.275$ using a total of
893,319 galaxies over 9100 deg$^2$.
\citet{kazin10a} used the full LRG sample from DR7 to measure the
galaxy correlation function and obtain a 3.5\% measurement of the
BAO distance scale at $z=0.35$.

The BAO measurements described above suffer from a degradation in the detection significance
due to large-scale flows and non-linear evolution in the density field.
As matter diffuses due to non-linear evolution and galaxy peculiar velocities,
the BAO peak becomes broader and more difficult to constrain, particularly at low redshift.
These effects can partly be reversed by ``reconstruction'', a
technique by which the observed galaxy field is used to estimate the large-scale
gravitational potential \citep{eisenstein07a}. 
Using the Zel'dovich approximation \citep{zeldovich70a},
the diffusion of galaxies can then be measured
and the density fluctuations can be shifted back to their Lagrangian
positions, thereby restoring the acoustic peak in both real and redshift space.
Redshift-space distortions \citep{kaiser87a} are removed in the same manner. 
The approach was refined, tested on simulated data, and first applied to survey
data using the full SDSS LRG sample \citep{padmanabhan12a}.
The scale of the acoustic peak was measured from
the reconstructed data to an accuracy of $1.9$\% \citep{xu12a},
an improvement of almost a factor of two over the $3.5$\% measurement before reconstruction.
Combining this distance measurement with the WMAP7
cosmic microwave background (CMB) anisotropies \citep{komatsu11a}
leads to an estimate of $H_0 = 69.8 \pm 1.2$ km s$^{-1}$Mpc$^{-1}$ and $\Omega_M = 0.280 \pm 0.014$
for a flat Universe with a cosmological constant ($w=-1$). 
The cosmological constraints are additionally improved when the reconstructed
SDSS data are combined with Type Ia supernovae
(SN~Ia) from the SN Legacy Survey \citep[SNLS;][]{conley11a}, measurements
of $H_0$ from the Hubble Space Telescope \citep[][]{riess11a},
and BAO constraints from the 6dF Galaxy Survey \citep{beutler11a}
and the WiggleZ Dark Energy Survey \citep{blake11a}.
\citet{mehta12a} use these measurements to
constrain $w = -1.03 \pm 0.08$ for a cosmological model in which
the equation of state for dark energy is allowed to vary.

As the name suggests, BOSS is a survey designed to measure the universe using BAO.
BOSS uses a rebuilt spectrograph from the original SDSS survey
with smaller fibers, new improved detectors, higher throughput, and a wider wavelength range,
enabling a spectroscopic survey to higher redshift and roughly one magnitude deeper than SDSS.
We hereafter refer to the upgraded
instruments as the BOSS spectrographs.  The BOSS spectrographs and their SDSS
predecessors are mounted to the telescope and are described in detail by
\citet{smee12a}.  In brief, there are two double spectrographs, each
covering the wavelength range 361 nm -- 1014 nm with
resolving power $\lambda/\Delta\lambda$ ranging from 1300 at the
blue end to 2600 at the red end.
Both spectrographs have a red channel with a 4k $\times$ 4k, 15 $\mu$m pixel CCD \citep{holland06a}
from Lawrence Berkeley National Laboratory (LBNL).
We denote these channels R1 and R2 for the first and second spectrographs, respectively.
Similarly, both spectrographs have blue channel with a 4k $\times$ 4k, 15 $\mu$m pixel CCD from e2v
denoted B1 and B2.
The instrument is fed by 1000
optical fibers (500 per spectrograph), each subtending
$2\2pr$ on the sky.  (The original spectrographs used 640 $3\2pr$ fibers.)
When the survey is complete, fibers will have been plugged into more than 2000
unique spectroscopic plates that each cover a circular field of view with $3\deg$ diameter.
Aluminum cast cartridges support
the optical fibers, spectrograph slithead, and spectroscopic plug plate.

BOSS consists primarily of two interleaved
spectroscopic surveys observed simultaneously: a redshift survey of 1.5 million luminous
galaxies extending to $z=0.7$, and a survey of the \lya\ 
forest towards 150,000 quasars in the redshift range
$2.15 \leq z \leq 3.5$.
BOSS uses the same wide field, dedicated telescope as was employed
by SDSS-I and -II, the 2.5{~}m-aperture Sloan Foundation
Telescope at Apache Point Observatory in New Mexico \citep{gunn06a}.
Those surveys imaged more than $10000 \, \sqdeg$
of high latitude sky in the $ugriz$ bands \citep{fukugita96a},
using a mosaic CCD camera with a field of view spanning $2.5\deg$
\citep{gunn98a}.  As discussed in \S\ref{subsec:imaging},
SDSS-III completed roughly $2500 \, \sqdeg$ of new imaging to enlarge the
footprint available to BOSS.  All of the imaging, including that
from SDSS-I and -II, is available in SDSS Data Release 8 (DR8;
\citealt{aihara11a}), the first data release from SDSS-III.
The forthcoming Data Release 9 \citep{ahn12a}
will present the first public release of BOSS spectroscopic
data, containing observations completed prior to August 2011.

The defining goal of the BOSS galaxy survey is to produce BAO
measurements limited mainly by sample variance (as opposed to
galaxy shot noise) over the volume available to the Sloan 2.5m
telescope, a volume defined by the area of accessible high latitude
sky and by the maximum practical redshift depth.
The BOSS survey observes primarily at $\delta > -3.75\deg$
in the SDSS high-Galactic latitude footprint.
The spectroscopic footprint is approximately 10,000 $\sqdeg$; larger areas
begin to stray into regions of high Galactic extinction.
Given the configuration of the BOSS experiment, we can exceed
the density of objects in the LRG sample and observe a comoving space
density $\bar{n} = 2-3\times 10^{-4}\vunits$
with strongly clustered galaxies (bias factor $b \sim 2$).
In one-hour exposures under good conditions, the BOSS spectrographs
can measure redshifts of these luminous galaxies with high completeness
to $i = 19.9$, which for the desired space density puts
the outer redshift limit at $z=0.7$.
BOSS is designed to observe 1.5 million galaxies to these limits,
including 150,000 galaxies that satisfy the BOSS
selection criteria but were previously observed during SDSS-I and -II.
For BAO and other large scale power spectrum measurements, the
comoving volume of the completed BOSS galaxy survey,
weighted by the redshift-dependent number density and galaxy power spectrum,
will be 6.3 times that of the SDSS-I and -II LRG sample.
The higher density of BOSS galaxies relative to SDSS-I and -II LRGs
also improves the performance of the reconstruction methods
discussed above that correct for non-linear degradation of the BAO signal.

The BOSS quasar survey is pioneering a novel method of measuring
BAO at high redshift ($2.15<z<3.5$) using \lya\ forest absorption
towards a dense grid of background quasars.
The redshifted \lya\ line becomes detectable in
the BOSS spectral range just beyond $z= 2$ and becomes highly opaque
at $z > 4$, motivating the targeted redshift range.
Even at moderate resolution, transmitted flux in the forest provides a
measure of hydrogen along the line of sight that
can be used to infer clustering of the underlying dark matter
distribution \citep{croft98a,croft99b,mcdonald03a}.  Prior to BOSS,
cosmological measurements of the forest treated each
quasar sightline in isolation, since samples were too small
and too sparse to measure correlations across sightlines except
for a few cases of close pairs or other tight configurations.
\citet{mcdonald03a,white03b} suggested that three-dimensional \lya\ forest correlations in
a large quasar survey could be used to measure BAO.
\citet{mcdonald07a} developed this idea in detail, constructing
a Fisher matrix formalism that helped guide the design of the
BOSS quasar survey.  \citet{slosar11a} used the first year of
BOSS data to obtain the first three dimensional measurement of large scale
structure in the forest, detecting transverse flux correlations to comoving
separations of at least $60\hmpc$.  The goal of the BOSS
quasar survey is to obtain spectra of at least 150,000 \lya\ quasars
selected from about 400,000 targets.
Extrapolations based on data taken to date suggest that BOSS will
ultimately observe about 175,000 \lya\ quasars over 10,000$\sqdeg$.
Ultraviolet, near-infrared, and multiple-epoch
optical imaging data increased the selection efficiency beyond the minimum goal.

This paper is one of a series of technical papers describing the BOSS survey.
The BOSS imaging survey data are described in \citet{aihara11a},
while BOSS is described in the context of SDSS-III in \citet{eisenstein11a}.
The selection of galaxy and quasar targets are described in \citet{padmanabhan12b}
and \citet{ross12a}, respectively, the spectrograph design and performance in
\citet{smee12a}, and the spectral data reductions
in \citet{schlegel12a} and \citet{bolton12a}.
Here, we describe the details of the BOSS survey itself,
with an emphasis on survey strategy and operations. 
In \S\ref{sec:TS}, we present the BOSS footprint and the selection of the
galaxies and quasars that will be used to measure the BAO feature.
In \S\ref{sec:tilesandplates}, we review the process
by which we design the spectroscopic plates to observe those targets. 
In \S\ref{sec:observations} we describe the procedures for observation,
the real-time data quality assessment, and the data processing pipeline.
Much of the information found in \S\ref{sec:tilesandplates} and
\S\ref{sec:observations} was never described for SDSS-I or -II
and remains unchanged from those earlier surveys.
We discuss the redshift efficiency
and the strategy to complete observations of the full 10,000 $\sqdeg$
in \S\ref{sec:simulations} and provide examples of the spectral data quality in
\S\ref{sec:data}.
Finally, we compare the BOSS data to SDSS
spectra, highlight some recent BOSS science results, and provide
forecasts for cosmological constraints in \S\ref{sec:conclusion}.

\section{Spectroscopic Targets}\label{sec:TS}
\newcommand{\datasweep}{\url http://data.sdss3.org/datamodel/files/PHOTO\_SWEEP/RERUN/calibObj.html}

The SDSS-I and -II imaging programs provide two large, contiguous regions of sky that
lie away from the Galactic plane.  Additional imaging in the Falls of 2008 and 2009
extended this footprint and increased the sky volume observable with BOSS.
BOSS is designed to survey the full SDSS imaging footprint with dense spectroscopic
coverage over five years.
The first goal of BOSS is a redshift survey of $\sim 1.5$ million
luminous galaxies at $0.15 <z< 0.7$, at a surface density of 150
deg$^{-2}$. This sample is divided into two subsamples.  The first is a
low redshift sample at $0.15 <z< 0.43$, with a median redshift of
$z=0.3$ and a surface density of 30 deg$^{-2}$. This
sample is a simple extension of the SDSS-I and -II LRG sample \citep{eisenstein01a}
to lower luminosities, where the brightest $\sim 10$ deg$^{-2}$ have already been
observed by SDSS. A higher redshift sample expands the
2dF-SDSS LRG and QSO Survey \citep[2SLAQ;][]{cannon06a} galaxy
selections and covers $0.43 <z< 0.7$, with a surface density of
120 deg$^{-2}$ and a median redshift of $z=0.57$.   
The second goal of BOSS is to survey at least 150,000
quasar spectra at $2.15 <z< 3.5$ to probe the intergalactic medium along
each sightline through \lya\ absorption.  
Approximately 5\% of the BOSS fibers are allocated to obtain spectra of
targets that would otherwise not be included in the BOSS target selection.
Dubbed ancillary science targets, these spectra provide the data for smaller
research programs proposed by members of the collaboration.  These programs are described
in Appendix~\ref{appendix:ancillary}.

\subsection{Imaging Data}\label{subsec:imaging}

In its first five years of operation (2000$-$2005), SDSS obtained 5-band imaging over
$7600 \, \sqdeg$ of the high Galactic latitude sky in the Northern
Galactic hemisphere and $600 \, \sqdeg$ along four disjoint stripes
in the Southern hemisphere.
During the falls of 2008 and 2009, the same camera \citep{gunn98a}
was used to grow the Southern imaging to a contiguous $3100 \, \sqdeg$
footprint.
As with the original SDSS imaging survey, we
obtained CCD imaging in scans with an effective exposure time of 55 seconds in
each of five filters \citep[$ugriz$;][]{fukugita96a}.  Imaging was performed when
the moon was below the horizon, under photometric conditions, and seeing was
better than $2 \2pr$ in the $r$ filter.
The prioritization was to image north of the celestial equator in September and
October, and south of the equator in November and December.
This prioritization makes the imaging depth more uniform by
roughly canceling two effects: the southern declinations are looking
through more atmosphere, but we did so in the winter months when the
atmosphere is typically more transparent.
The seeing requirements were somewhat relaxed from SDSS by accepting
conditions to $2\2pr$, although the minimum scan time was increased
from 0.5 to 1 hour to ensure sufficient astrometric and photometric
calibration stars.
The typical 50\% completeness limit for detection of point sources is $r=22.5$.

The final SDSS imaging data set, including the new BOSS imaging,
was released as part of DR8.
These data were reduced with a uniform version of the photometric
pipeline \citep{lupton01a,stoughton02a} with improved sky-subtraction
described in the DR8 release \citep{aihara11a}.
The photometric calibration \citep{smith02a, ivezic04a, tucker06a}
has been improved with a global re-calibration \citep{padmanabhan08a}.
The astrometric \citep{pier03a} calibration was tied to the UCAC2
system for declinations below approximately $41 \deg$ and the USNO-B
system at higher declinations.
This astrometry is less accurate than the earlier DR7 release or the later
re-calibration of the entire sky in the DR9 release.
The BOSS coordinates for all target classes corresponds to the DR8 astrometric
system which is internally consistent and only introduces small relative offsets
on the $3 \deg$ scale of a plate.
These coordinates are offset 240 mas to the North and 50 mas to the West
relative to DR9 coordinates at higher declinations, with additional sources
of error introducing scatter of $\sim 50$ mas \citep{ahn12a}.

\begin{figure}[h]
\vspace*{-1.5in}
\begin{center}
\includegraphics[scale=0.40]{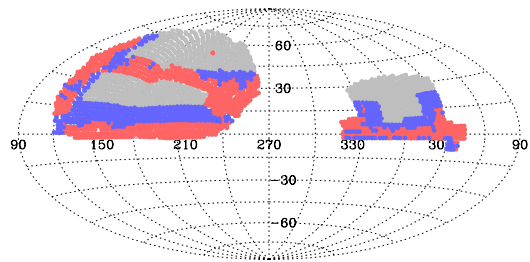}
\end{center}
\caption[Footprint]{\label{fig:footprint}
Location of pointing centers for the 2208 spectroscopic plates in the BOSS
survey footprint in an Aitoff projection in J2000 equatorial coordinates.  Gray
circles represent the location of plates that remained to be drilled after
Summer 2011.  Blue circles represent plates that were drilled and
were ready to observe in the third year of the survey.  Red circles represent
the plates that were completed in either the first or second year of the
survey, whose observations are released in DR9.\\
}
\end{figure}

These imaging data define the baseline goal for the BOSS spectroscopic survey.
We used a trimmed subset called the ``BOSS footprint'', which is two contiguous
regions of low extinction centered in the north Galactic cap (NGC)
and south Galactic cap (SGC).
To avoid regions of high extinction, the footprint lies above $\pm 25\deg$ Galactic
latitude for all regions except around a Right Ascension of $120 \deg$ ($b >15\deg$)
and Right Ascension of $330 \deg$ ($b >20\deg$).
Several $0.2\deg$-wide strips were also excised from this
overall footprint for the quasar targets only to reject imaging
data when $u$-band amplifiers were not functioning,
as explained in DR2 \citep{abazajian04d}.
The BOSS footprint is presented in Figure~\ref{fig:footprint}.

\begin{deluxetable*}{lccccc}[!h]
\centering
\tablewidth{0pt}
\tabletypesize{\footnotesize}
\tablecaption{\label{tab:survey} Survey Parameters}
\tablehead{ \colhead{Region} & \colhead{Contiguous Imaging} & \colhead{Area for Galaxy} & \colhead{Area for Quasar} & \colhead{Area after} & \colhead{Number} \\
\colhead{} & \colhead{Area ($\sqdeg$)} & \colhead{Targets ($\sqdeg$)} & \colhead{Targets ($\sqdeg$)} & \colhead{Tiling ($\sqdeg$)} & \colhead{of Plates} }
\startdata
SGC & 3172 & 2663 & 2663 & 2634 & 633  \\
NGC & 7606 & 7606 & 7578 & 7426 & 1575  
\enddata
\end{deluxetable*}

In total, galaxy spectroscopic targets were chosen from $3172 \, \sqdeg$ in
the SGC and $7606 \, \sqdeg$ in the NGC.  A total of $7578 \, \sqdeg$ in the NGC were used
for the quasar target selection due to the loss of $u-$band imaging
in certain regions.
Because this footprint is too large for the BOSS survey, it was
was further trimmed by removing most of the area south
of the celestial equator, reducing the SGC footprint to $2663 \, \sqdeg$.
The coverage of the SGC is
presented in Figure~\ref{fig:footprint} as the region with $\alpha >300\deg$
or $\alpha <60\deg$.  The coverage of the NGC is included in the region
with $60\deg < \alpha <300\deg$.
The spur in Figure~\ref{fig:footprint} located at approximately $30\deg < \alpha <40\deg$
and $\delta <-2 \deg$ was chosen to overlap with the W1
region of the Canada-France-Hawaii Telescope Legacy Survey
\cite[CFHTLS;][]{goranova09a}.  Using $ugriz$ photometry from these 10,269
$\sqdeg$, we choose galaxy, quasar, and ancillary science targets for spectroscopic follow-up.
A summary of the area covered in imaging and spectroscopy is provided in Table~\ref{tab:survey}.

\subsection{Imaging Pre-selection} \label{subsec:preselect}

We briefly describe those photometric parameters measured for each object
in the imaging data that are relevant to target selection.
More detailed descriptions are available in the online SDSS documentation
and the data release papers,
especially EDR \citep{stoughton02a} and DR2 \citep{abazajian04d}.

\begin{itemize}

  \item {\bf PSF magnitudes} are designed for point sources and are a fit of the
  point spread function (PSF) model evaluated at the location of each object.
  We refer to these magnitudes measured in any filter ``$X$''
  as $X_{\rm PSF}$.

  \item {\bf Fiber2 magnitudes} are designed to simulate the flux
  captured by the BOSS fibers for either point sources or extended sources.
  The images are convolved with Gaussians to simulate $2\2pr$ seeing,
  and the flux is measured within a $2\2pr$ circular aperture representing
  the size of the BOSS fibers.
  Unlike the other magnitudes that are measured on individual objects
  deblended from their neighbors, the fiber magnitudes include flux
  from those neighbors.
  A caveat of these magnitudes is for the small amount of SDSS imaging
  data taken in seeing worse than $2\2pr$ seeing, the fiber magnitudes
  are measured without convolution or deconvolution of those images.
  In those cases, the fiber magnitudes are fainter than they should be.
  Throughout the text we refer to these magnitudes measured in any filter ``$X$''
  as $X_{\rm fib2}$.
  These magnitudes are analogous to the $3\2pr$ fiber magnitudes from SDSS
  which simulated the flux captured by the original SDSS $3\2pr$ fibers.

  \item {\bf Model magnitudes} are designed to measure galaxy colors
  that are unbiased relative to the signal-to-noise ratio (SNR) of the image or unmodeled
  substructure in the galaxies.
  A library of de Vaucouleurs \citep{devaucouleurs48a}
  and exponential \citep{freeman70a} profiles with varying
  radii, ellipticities and position angles are fit to
  each object's image.  The best-fit profile from the $r$ band,
  convolved with the PSF
  in each band, is used as a matched aperture for the model magnitudes.

  \item {\bf Cmodel magnitudes} (composite model)
  are designed to produce a good estimate
  of the ``total'' flux of galaxies.
  These magnitudes are based upon the best-fitting 
  positive linear combination of exponential and de Vaucouleurs profiles, convolved
  with the PSF.  These are measured independently in each band.

  \item {\bf r$_{deV,i}$} is the best fit de Vaucouleurs effective radius in
  the $i$ band (in units of arcseconds).

\end{itemize}

Every object in the SDSS imaging catalogs has a series of flags
indicating the nature and quality of the data, allowing one to define
a single unique measurement of objects in regions observed more than
once, and to identify any problems with processing and measuring the
properties of each object. We used these flags to remove questionable
objects from consideration for spectroscopy.  

Repeat observations and overlapping scans resulted in multiple observations for
some objects; only unique observations identified by the {\bf SURVEY\_PRIMARY}
bit of the {\bf RESOLVE\_STATUS} flag assigned to each object were used for the
main targeting algorithms, as described in \citet{aihara11a}.  Multiple imaging
observations were used for some special programs, such as quasars targeted
based on their variability.  

For the galaxy, quasar, and standard star targets selected for the primary BAO
program (denoted ``main targets'' in which follows), we rejected objects for
which flags set during photometric pipeline
processing\footnote{http://www.sdss3.org/dr8/algorithms/flags\_detail.php}
indicated that their photometry was unreliable.  
All objects with the SATUR flag set
(which indicates that the object includes saturated pixels)
are dropped from consideration for galaxies, quasars, and standard stars.

The quasar target selection algorithm is most sensitive to objects with
unreliable photometry (stars are far more numerous, leading to
a very large contamination rate), and thus it does the most careful checking.  The flags used
are described in detail in \citet{bovy11a} and Appendix A of \citet{ross12a}
(see also \citealt{richards06a}), and include flags designed to indicate
problems due to deblending, interpolation over cosmic rays and bad pixels and
columns, and proper motion between exposures in different filters (due to
asteroids).  

For galaxy target selection, we also use a specialized set of set of flag cuts
to ensure good target selection.  The detailed galaxy flags for galaxies are
given in \citet{padmanabhan12b}. 

Objects flagged in the following ways were not considered for followup as 
standard stars.
These flag cuts are in addition to {\bf SURVEY\_PRIMARY} and {\bf SATUR}
listed above.
\begin{itemize}
\item Objects observed in non-photometric conditions (as
    indicated by the {\bf CALIB\_STATUS} flag) in any band.
\item Objects labeled {\bf INTERP\_CENTER}, {\bf PSF\_FLUX\_INTERP}, 
    indicating that interpolation over cosmic
    rays or bad pixels significantly affected the measured flux of the
    object.
\item Objects flagged in association with deblending of overlapping objects.
    Objects labeled as {\bf BLENDED}; such objects are the parents of objects subsequently
    deblended into children.  Objects flagged {\bf DEBLEND\_TOO\_MANY\_PEAKS}; 
    this flag indicates that there were more objects to deblend in this 
    family than the deblender could handle.  Objects marked
    {\bf PEAKS\_TOO\_CLOSE}, indicating 
    the peak for this object was too close to another peak.
\item Objects flagged {\bf CR}, meaning the object contained a cosmic
    ray that was interpolated over.
\item Objects flagged with {\bf BADSKY}, meaning the estimate of the background sky
    at this position is suspect.
\item Objects labeled {\bf NOTCHECKED\_CENTER}, meaning the 
    center of the object lies in a region not checked for peaks; 
    often due to bad deblending.
\item Objects flagged as {\bf !STATIONARY}, indicating that the object 
    is an asteroid.
\end{itemize}

Further details of these flags are found in \citet{stoughton02a},
\citet{richards06a}, and the SDSS-III
website.

As described in \citet{padmanabhan12b}, we tracked regions of rejected
objects and 
bad photometry to define the exact area where spectroscopy was performed.
It is recommended that users apply these masks when performing statistical studies of
large-scale structure with the galaxy sample or investigations that require complete understanding
of the selection function, such as a measurement of the quasar luminosity function.
Since the \lya\ quasars provide random sightlines, it is unlikely that the masks
will affect studies of large scale structure in the \lya\ forest.

\subsection{Galaxy Target Selection}\label{subsec:galaxyTS}

The dominant cosmological volume in the SDSS-I and -II surveys was mapped by the
Luminous Red Galaxy sample \citep{eisenstein01a} and the BOSS survey uses a
similar philosophy of color-magnitude and color-color plots to select the galaxy
sample. Unlike the LRG sample, the higher redshift BOSS (``CMASS'', see below)
sample is not restricted to a sample of red galaxies, but instead
attempts to select a stellar mass-limited sample of objects of all intrinsic colors,
with a color cut which selects almost exclusively on redshift.
We lower the luminosity cut relative to the LRG sample and allow a wider color range to achieve
a higher density and also provide a less biased sample for studies of massive galaxy evolution.
We therefore do not use the designation ``LRG'' for these
galaxies, even though their selection criteria are similar in spirit to previous LRG
surveys \citep{eisenstein01a, cannon06a}.

The BOSS galaxies are selected to
have approximately uniform comoving number density of
$\bar{n} = 3\times 10^{-4}\vunits$ out to a redshift $z=0.6$, 
before monotonically decreasing to zero density at $z \sim 0.8$.
Galaxy shot noise and sample variance make roughly equal
contributions to BAO errors when $\bar{n} P_{\rm BAO} =1$, where $P_{\rm BAO}$ is the galaxy power spectrum
at the BAO scale, approximately $k=0.2 h {\rm Mpc}^{-1}$.
For the strongly clustered galaxies observed by BOSS, $\bar{n} = 3\times 10^{-4}\vunits$
yields roughly $\bar{n} P_{\rm BAO} =2$, making shot-noise clearly sub-dominant.
Increasing to $\bar{n} P_{\rm BAO}=2$ should also improve the performance of the
reconstruction approach described in the introduction.
Generating a higher space density would require substantially more observing time with only moderate gain
in clustering signal-to-noise.
We therefore expect significant improvements in BAO constraints over the 
SDSS-I and -II LRG sample ($\bar{n} = 1\times 10^{-4}\vunits$),
even in the redshift range where the two samples overlap.

The galaxy target selection algorithms will be documented in detail in
\citet{padmanabhan12b}; we summarize the principal points below. We
select these targets from SDSS imaging according to the photometric parameters from the processing pipeline.
All magnitudes are corrected for Galactic extinction using the \citet{schlegel98a}
models of dust absorption.
Target galaxies are chosen based on color and apparent magnitude cuts motivated by the
stellar population models of \citet{maraston09a}.  These cuts are based on the expected
track of a passively evolving, constant stellar mass galaxy as a function of redshift.

Targets are selected based on the following set of model magnitude colors
\citep{eisenstein01a, cannon06a} :
\begin{eqnarray}
  c_{||} &= &  0.7(g-r) + 1.2(r-i - 0.18) \\
  c_{\perp} &= & (r-i) - (g-r)/4 - 0.18 \\
  d_{\perp} &= & (r-i) - (g-r)/8.0 \,\,.
\end{eqnarray}

These color combinations  are designed to lie parallel or perpendicular to
the locus of a passively evolving population of galaxies,
with $c_{\perp}$ and $c_{||}$ being the appropriate colors to select galaxies below
$z\sim 0.4$ and $d_{\perp}$ at higher redshift. 
To a good approximation, the perpendicular colors track the location of the 4000 \AA\ break,
which is redshifted from $g-$band to $r-$band at $z \sim 0.4$, and provide an initial redshift selection.

Galaxy target selection 
mirrors the split at $z \sim 0.4$ and selects two principal samples: ``LOWZ'' and ``CMASS''.  
The LOWZ cut targets the redshift interval $0.15 <z< 0.43$.
This sample includes color-selected galaxies with
$16<r<19.5$, $r < 13.6  + c_{||}/0.3$, and
$| c_{\perp} | < 0.2$ where $r$ is the cmodel magnitude.
Over most of the BOSS footprint, about one third of the LOWZ galaxies
already have spectra from SDSS-I and -II; these objects are not re-observed.
However, in the SGC, the
SDSS-I and -II spectroscopy was limited primarily to low declination, resulting in a 
larger fraction of LOWZ targets in the South.
Cut II (CMASS for ``constant mass'') includes color-selected galaxies in the
magnitude range $17.5<i<19.9$, $d_{\perp} > 0.55$, and
$i < 19.86 + 1.6 \times (d_{\perp} - 0.8)$ where $i$ is the cmodel magnitude.
The CMASS sample is designed to select galaxies at $0.43<z<0.7$, although it
extends beyond these nominal redshift boundaries. In addition, to better
understand the completeness of the CMASS galaxies and the impact of the
color-magnitude cut, we also target CMASS\_SPARSE galaxies. These galaxies share
the same selection as the CMASS galaxies, except that the color-magnitude cut
has been shifted to $i < 20.14 + 1.6 \times (d_{\perp} - 0.8)$. We target
these galaxies at a rate of 5 deg$^{-2}$.

In a manner similar to that described in \citet{stoughton02a},
we track the source of the selection for galaxy objects and all spectroscopic targets
with flag bits encoded in the new quantities
{\bf BOSS\_TARGET1} and {\bf BOSS\_TARGET2}.
Targets that satisfy the LOWZ criteria are denoted {\bf GAL\_LOZ},
while CMASS targets fall into several categories labeled with
a {\bf GAL\_CMASS\_} prefix.
A description of all flag bits for the main targets is found in Appendix~\ref{appendix:mainflags}.

\citet{maraston12a} present photometric stellar masses for a large sample of BOSS galaxies
by fitting model spectral energy distributions to $ugriz$ photometry. 
They demonstrate that the main BOSS galaxies have a narrow mass distribution
which peaks at log$(M/M_\odot) \sim 11.3$.
The results confirm that the target selection has successfully produced a sample of galaxies
with fairly uniform mass from $0.2<z<0.6$. 

\subsection{Quasar Target Selection}\label{subsec:QSOTS}

The primary goal of the BOSS quasar survey is to 
map the large scale distribution of neutral hydrogen via absorption
in the \lya\ forest.  Measurements of BAO in the three-dimensional
correlation function in this intergalactic neutral hydrogen will provide the first
direct measurements of angular diameter distance
and the Hubble parameter at redshifts $z>2$.
However, for the \lya\ forest to adequately sample the
three-dimensional density field for a BAO measurement, the BOSS quasar density
must be an order of magnitude larger than SDSS
\citep[for the density of SDSS quasars, see][]{schneider10a} over the same redshift
range \citep{mcdonald07a, mcquinn11a}.
The sample must also provide targets with a luminosity distribution
that enables adequate signal-to-noise ratio spectroscopy over the forest.
At a minimum, 15 quasars deg$^{-2}$ at redshifts $2.15<z<3.5$ and $g_{\rm PSF}< 22.0$
are required to make this measurement.
Since the precision of the BAO measurement shows an approximately linear improvement
with the surface density of quasars at
fixed spectroscopic SNR, we attempt to obtain
as many quasar sightlines as possible.
Fortunately, because the quasars are nearly uncorrelated with the
intervening density field, \lya\ measurements are insensitive to the exact details
of quasar target selection, and do not require the uniform sample that
is essential to the galaxy BAO measurement.

Extrapolating from the \citet{jiang06a} quasar luminosity function,
we find that quasar targets must be selected to a magnitude
limit of $g_{\rm PSF} < 22.0$ to obtain a surface density of
15 high-$z$ quasars deg$^{-2}$.
However, identifying quasar
targets from photometric data is complicated by 
photometric errors and the similarity of quasar
colors (particularly at $z\sim$2.7) to colors of A stars and blue
horizontal branch stars \citep[e.g.][]{fan99b, richards02a,mcgreer12a}.
This suggests that a more sophisticated
method for target selection is required than
that used in SDSS-I and -II \citep{richards02a}.

In the first year of BOSS observations, we compared and tested a variety of
methods to optimize the efficiency of quasar target selection.  These
methods included: a ``Kernel Density Estimation'' \citep{richards09a}
which measures the densities of quasars and stars in color-color space
from training sets and uses these to select high probability targets;
a ``Likelihood'' approach which determines the
likelihood that each object is a quasar, given its photometry and
models for the stellar and quasar loci \citep{kirkpatrick11a}; an
``Extreme Deconvolution'' \citep[XDQSO;][]{bovy11a,bovy11b}
selection, which performs a density estimation of stars and quasars by incorporating
photometric uncertainties; and an artificial neural network \citep{yeche10a},
which takes as input the SDSS photometry and errors from a training set in order to run a
classification scheme (star vs. quasar) and generate a photometric redshift estimate.
XDQSO includes data from the UKIRT Infrared Deep Sky Survey
\citep[UKIDSS;][]{lawrence07a} and the Galaxy Evolution Explorer
\citep[GALEX;][]{martin05a} when available,
showing the greatest improvement in selection efficiency with the inclusion of the UKIDSS data.

As with galaxies, quasars targeted by BOSS are tracked with the {\bf BOSS\_TARGET1} flag bits;
details of each selection and flag are found in \citet{ross12a}.
Quasar target selection falls into five distinct categories:  

\begin{itemize}
    \item {\bf QSO\_CORE\_MAIN}, hereafter {\bf CORE},
      includes targets selected by a uniform method that can easily be reproduced
      for studies of the luminosity function, clustering, and other quasar science.
      The XDQSO method was selected as {\bf CORE} at the beginning of the second year of 
      BOSS observations and applies to spectroscopic plates numbered 4191-4511, and
      4530 and above.
    \item {\bf QSO\_BONUS\_MAIN}, hereafter {\bf BONUS},
      includes targets selected in a non-uniform way to utilize the full
      complement of ancillary imaging data and maximize the spectroscopic quasar density. 
      As mentioned above, heterogeneous selection does not bias cosmology constraints because the
      BAO measurement in the \lya\ forest is not dependent on the properties of the
      {\it background} illuminating quasar.
    \item {\bf QSO\_KNOWN\_MIDZ} includes previously known $z>2.15$ quasars, the majority of which
      are from SDSS.  Given the higher throughput of the BOSS spectrographs, these new
      observations from BOSS provide much deeper spectra in the \lya\ forest region,
      and comparison to SDSS allows calibration tests and studies of spectroscopic variability.
    \item {\bf QSO\_FIRST\_BOSS} includes objects from the FIRST radio survey \citep{becker95a}
      with colors consistent with quasars at $z>2.15$.
    \item {\bf QSO\_KNOWN\_SUPPZ} includes a subset of previously known quasars (mostly from SDSS-I and -II)
      with $1.8<z<2.15$.  Previous studies of the \lya\ power spectrum
      \citep[e.g.][]{mcdonald06a} indicate that metal absorption contributes a small
      amount of spurious clustering power in the \lya\ forest.
      This lower redshift sample, re-observed with BOSS, allows a 
      measurement of the spectral structure from metal lines that appear in the
      \lya\ forest at the low redshift range of the quasar sample, allowing the excess power to
      be modeled and removed from \lya\ clustering measurements.
\end{itemize}

In addition, in the limited regions where imaging stripes overlap in the
SDSS imaging survey, quasar targets were selected based on photometric variability,
with color selection to choose objects at $z>2.15$.
These regions cover approximately 30\% of the BOSS footprint
and are identified by the eighth bit in the {\bf ANCILLARY\_TARGET2}
flag developed for BOSS ancillary programs.
This bit is referenced by name as the {\bf QSO\_VAR\_SDSS} target flag
and is found in plates numbered higher than 4953.
The density of targets added by this method varies significantly from one region
to another, and is on average $\sim$ 4 deg$^{-2}$.
These targets lead to an increase in the density of spectroscopically confirmed quasars
at $z>2.15$ of two to three deg$^{-2}$ in the overlap regions.

The {\bf CORE}, {\bf BONUS}, {\bf FIRST} and {\bf KNOWN} samples account for an average
of 20, 18.5, 1, and 1.5 targets deg$^{-2}$, respectively. 
40 objects deg$^{-2}$ are targeted as candidate quasars across the full
survey \citep[see Figure~8 and Figure~9 in][]{ross12a},
although quasars were targeted at a somewhat higher density early in the survey.
The {\bf QSO\_KNOWN\_SUPPZ} and {\bf QSO\_VAR\_SDSS} samples are used to fill unused
fibers at a very low priority, resulting in a greatly varying target
density.  Observations in the second year included a near-final target selection
for quasars and produce average densities that
range from 15 to 18 quasars deg$^{-2}$ at $z>2.15$, depending
on the amount of ancillary imaging data (e.g., UKIDSS) and the density of stars.  The
selection of these targets and the target bits for the various quasar
selection schemes is explained in detail in \citet{ross12a}.

Multi-epoch imaging data in Stripe 82
($-43 \deg < \alpha < 45\deg$,$-1.25\deg< \delta <1.25\deg$)
was used to determine an effective upper limit of quasar detection efficiency.
By using photometric variability to select quasars,
we found 24 quasars deg$^{-2}$ at $z>2.15$ with $g_{\rm PSF}<22$
\citep{palanque-delabrouille11a} across the $220 \sqdeg$ region.
This analysis recovered quasars in the $2.5<z<3.5$
redshift range that were systematically missed by color-selection
techniques because of proximity to the stellar locus and larger photometric
errors in single-epoch imaging.

\subsection{Ancillary Science Targets}

As in the original SDSS spectroscopy survey,
several special observing programs were designed to pursue
science goals not covered by the primary galaxy and quasar targets.
A total of 206 special plates were dedicated to these programs in SDSS
\citep[as described in the DR4 paper;][]{adelman-mccarthy06a}.
These SDSS plates were primarily along Stripe 82
because of the additional science goals enabled by multiple imaging epochs 
and because of the amount of available spectroscopic observing time
in the Fall months when that region was visible.
In BOSS we dedicate roughly 5\% of fibers
on each plate to a new series of ancillary programs.
We track the selection of each object with flag bits encoded in the quantities
{\bf ANCILLARY\_TARGET1} and {\bf ANCILLARY\_TARGET2}.

These ancillary programs are intended to support studies that require sizable samples
over large regions of sky, making them difficult to complete in conventional observations
at shared facilities.
However, because these programs are given a lower priority for target selection than the
primary science drivers of BOSS, the sample selection is often not complete.
These programs fall into two categories:  those using the repeat and deep imaging of
Stripe 82 and those selected from the rest of the BOSS imaging area.
Spectroscopic observations are now complete for targets in Stripe 82 and 
are included in DR9.
Observations of the remaining programs will continue through the end of the survey
and will be included in part in DR9 and in future data releases.
A description of each of the ancillary programs, including their
scientific rationale, approximate density,
and algorithm for target selection is found in Appendix~\ref{appendix:ancillary}.
Additional ancillary programs may appear in future BOSS observations;
these will be documented in the corresponding SDSS Data Release 
papers and in the science papers that exploit them.

\section{Designing the Observations}\label{sec:tilesandplates}
\begin{deluxetable*}{lcc}[!t]
\centering
\tablewidth{0pt}
\tabletypesize{\footnotesize}
\tablecaption{\label{tab:collision_priority} Fiber Assignment Priority
for Resolving Fiber Collisions\tablenotemark{a}}
\tablehead{ \colhead{Chunks} & \colhead{Plates} & \colhead{Target Bit Priority Order} }
\startdata
boss1-boss2 & 3536-3750, 4526-4529 & quasar, galaxy \\
boss3-boss9 & 3751-4097 & galaxy, quasar \\
boss10-boss13 & 4174-4428, 4530-4556, & {\bf QSO\_FIRST\_BOSS}, galaxy, \\
& 4656-4669 & {\bf QSO\_KNOWN\_MIDZ}={\bf CORE}={\bf BONUS} \\
boss14-boss20 & 4440-4525, 4557-4655, & {\bf QSO\_KNOWN\_MIDZ}, {\bf CORE}, \\
& 4670-5140 &  {\bf QSO\_FIRST\_BOSS}, galaxy, {\bf BONUS}
\enddata
\tablenotetext{a}{Target types are listed in order of fiber collision
  priority from highest to lowest.  ``quasar'' and ``galaxy'' refer to all
  target bits within those target classifications; i.e., all quasar or
  galaxy target bits have equal priority for resolving fiber
  collisions. A ``='' symbol between two target bits means those target
  bits have equal collision priority. The priorities in the last row
  are defined for plates designed and drilled through the end of DR9
  and will likely apply through the end of the BOSS survey.}
\end{deluxetable*}

As with the previous SDSS surveys, the BOSS spectra are obtained
through observations of $1.5\deg$ radius spectroscopic plates.
The SDSS-I and -II plates
accommodated 640 fibers ($180\mu$m diameter) that
projected to a $3\2pr$ diameter on the sky to match the profile of the main
SDSS galaxies ($z_{\rm mean} \sim 0.15$).
Each BOSS plate contains 1000 fibers with a smaller ($120 \mu$m) $2\2pr$ diameter
to reduce sky background and match the smaller angular profile of the higher
redshift CMASS galaxies ($z_{\rm mean} \sim 0.57$).
The minimum allowable distance between fibers was $55\2pr$ in SDSS and is
$62\2pr$ in BOSS, set by the cladding around each fiber.
One might worry that pointing errors and seeing losses would overcome the effects of
reduced sky noise for $2\2pr$ fibers, degrading the SNR relative to $3\2pr$ fibers.
We have conducted tests that show that we are in fact gaining on an object-by-object
basis, the spectrograph throughput is significantly improved,
and the increase in fiber number from 640 to 1000 (which fit in the same slithead
design as SDSS) is a large efficiency gain.

Science targets are assigned to these plates in a process referred
as ``tiling'' \citep{blanton03a}.
In SDSS-I and -II, imaging and spectroscopy were interleaved, 
requiring the tiling to progress in pieces as the imaging for each area of
the sky sky was completed.
As explained in \S\ref{subsec:imaging}, the imaging footprint for BOSS was completed
in Fall 2009, making it possible to tile the entire 10,269 $\sqdeg$ footprint
before full spectroscopic observations began.
Changes in the target selection algorithms required occasional retilings of the survey footprint;
we track changes in target selection as described below.
After tiling, the plug plates are designed according
to the estimated time (and thus airmass) of the observation.
Target positions are converted from equatorial coordinates to ($x$,$y$) plate
coordinates for drilling with a Computer Numerically Controlled milling machine
at the University of Washington.
The application of tiling and plate design to spectroscopic observations
is the same as that used during the SDSS-I and -II surveys but was never documented.
The process is fundamental to the reconstruction of the angular selection
function for studies of clustering in galaxies and quasars and is described below.

\subsection{Fiber Assignment for Science Targets}

Fibers are first allocated to science targets using the tiling algorithm
described in \citet{blanton03a}. The tiling process is intended to
maximize the fraction of targets that are assigned fibers (tiling
completeness) while minimizing the number of tiles required to
complete observations (tiling efficiency) without leaving gaps of coverage
in the survey footprint. The process for BOSS is nearly identical to that
for SDSS-I and -II, except for the number of fibers and prioritization of targets.
We describe the process for BOSS here.

We begin with a uniform
distribution of tiles, which are then perturbed to account for angular
variations in the density of targets. Of the 1000 available fibers on
each tile, a maximum of 900 fibers are allocated for science targets;
the rest are reserved for calibration stars and measurements of the
sky background (i.e., fibers placed at locations without detectable objects in the
SDSS imaging data).
Of the science fibers, five on each plate are assigned to targets that are also
assigned fibers on neighboring tiles.  We refer to these as ``repeat observations''
and use them to test the reproducibility of spectroscopic results.
The remaining fibers are reserved for unique quasar, galaxy, or ancillary targets.
Most plates do not use all available fibers because the plates must
overlap to avoid gaps in coverage, thereby increasing the total number of
plates and number of fibers beyond the minimum required by the number of science targets.
The overall tiling efficiency of the survey, defined as the
fraction of these 895 fibers assigned to unique, tiled science targets, is
$0.927$ with a plate-to-plate dispersion of 0.067.
The remaining fibers are used for lower priority targets such as additional repeat
observations, {\bf QSO\_VAR\_SDSS} targets, and additional sky fibers.

Groups of targets that are linked together within the $62\2pr$ fiber collision limit are
denoted collision groups.
The priorities for assigning fibers within a collision group vary with target type,
and have varied throughout the survey as shown in Table~\ref{tab:collision_priority}.
The decollided set of targets is
the subset that does not lie within $62\2pr$ of any other target.
Targets may also be precluded from
fiber assignment because they lie within a $92\2pr$ radius of the center of each
tile, which is the region covered by the centerpost of the cartridge.
Given these conditions, the tiling algorithm attempts to maximize the
number of targets assigned fibers.  
In regions covered by a single tile, fiber collisions
limit the galaxy and quasar tiling completeness to $\sim 90$\%.
Once the galaxy and quasar
target sets have been assigned fibers, ancillary targets are
assigned to the remaining fibers. In this way, galaxy and quasar targets are
never competing for fibers with the lower-priority ancillary set.
Each point of the sky may be covered by up
to four tiles (although overlap by more than two tiles is fairly rare).
In tile overlap regions, targets within collision groups may be assigned to multiple
tiles, bringing the completeness near unity in these regions (Table~\ref{tab:completeness}).
The tiles and the imaging footprint define the geometry of the survey. We use the
software package {\tt mangle} \citep{swanson08a} to create and
manipulate the survey geometry, as described briefly in \citet{aihara11a}.

The fiber assignment in BOSS was not performed as a single process over the
full spectroscopic footprint; instead, small
regions of the footprint were tiled at different times to accommodate changes
in the target selection algorithms.
Each of these regions is called a ``chunk'' and is denoted ``bossN'', where 
N is the chunk number.
These chunks vary in size from
a few dozen to a few hundred plates, depending on the region and the 
status of the target selection testing at the time the plates were needed.
Within each chunk, a sector is defined as a region covered by a unique set of tiles. 

While the fiber assignments changed within tiles, the 
tile centers were set around December, 2010 in boss15 and will not change.
The final tiling of the BOSS footprint results in a total of 2208
plates as shown in Figure~\ref{fig:footprint} and Table~\ref{tab:survey}; 1575 plates lie in the
NGC and 633 plates lie in the SGC. After accounting for spatial patterns in tiling
and incomplete regions, the tiled area of the
10,269 $\sqdeg$ imaging footprint (\S\ref{subsec:imaging})
amounts to a total of 10,060 deg$^2$.
The mean area covered per tile is 4.7 deg$^2$ in
the NGC and 4.2 deg$^2$ in the SGC. The higher tile density in the SGC
is required to include approximately 20 deg$^{-2}$ additional galaxies
from the LOWZ sample that lie in regions not covered in previous
SDSS spectroscopy.
The area of the survey covered by single tiles is 5417 deg$^2$, while the
remaining 4643 deg$^2$ is covered by two or more tiles.
The statistics of completeness for each target type are found in
Table~\ref{tab:completeness}.
Because of the connection between completeness and tile overlap,
and the varying efficiency among the chunks, the incompleteness
has strong spatial patterns that must be included in clustering analyses.
As examples, \citet{anderson12a} describe the method to account
for incompleteness in galaxy clustering due to fiber collisions
and \citet{white12a} describe the process in a measurement of quasar clustering.

\begin{deluxetable}{lcccc}
\centering
\tablewidth{0pt}
\tabletypesize{\footnotesize}
\tablecaption{\label{tab:completeness} Completeness by Target Type and Sector Type}
\tablehead{
\colhead{Target Type} & \multicolumn{2}{c}{Single-tile Sectors} & \multicolumn{2}{c}{Overlapping Sectors} \\
\colhead{} & \colhead{Mean} & \colhead{Dispersion\tablenotemark{a}}
           & \colhead{Mean} & \colhead{Dispersion\tablenotemark{a}}
}
\startdata
Galaxy     & 0.891 & 0.010  & 0.988  & 0.001   \\
Quasar     & 0.914  & 0.030  & 0.998  & 0.002   \\
Ancillary  & 0.680  & 0.143  & 0.850  & 0.158    
\enddata
\tablenotetext{a}{RMS computed from area-weighted chunk-to-chunk
  variation in boss1 through boss20.}
\end{deluxetable}

\subsection{Plate Design}\label{subsec:platedesign}

In the tiling stage described above, approximately 160-200 fibers per
plate are dedicated to the main quasar targets, 560-630 to
galaxy targets, and 20-90 to ancillary science targets.
In plate design, additional fibers are assigned to each plate for the
purpose of sky subtraction and calibration.

We assign each plate at least 80 ``sky'' fibers placed at locations
that contain no detections from the SDSS imaging survey. A random selection
of such locations is output from the photometric pipeline, as described in
\citet{stoughton02a}.  The sky fibers are
used to model the spectroscopic foreground for all science
fibers. Their distribution is constrained to cover the entire focal
plane, by requiring at least one sky fiber per bundle of 20 fibers.
The fibers within a bundle are anchored to a fixed point on a cartridge,
easing the handling of fibers but limiting the reach of a bundle.
The spacing of sky fibers allows sampling of the varying sky background
over the focal plane and over the optical system of the spectrograph,
allowing a model to be constructed for any spatial structure and leading to improved sky subtraction.

We assign each plate 20 ``standard'' fibers to sources with
photometric properties consistent with main sequence F stars,
to serve as spectrophotometric standards. These stars allow
calibration of throughput as a function of wavelength
for each exposure, including atmospheric telluric absorption
corrections and the spectral response of the instrument.  We choose
these stars to have $15 < r_{\mathrm{fib2}} < 19$ and
$m_{\mathrm{dist}} < 0.08$, where $m_{\mathrm{dist}}$ is a scaled
distance in color space from the color of a fiducial F star:

\begin{eqnarray}
m_{\mathrm{dist}} &=& \left[ \left((u-g)-0.82\right)^2 + \left((g-r)-0.30\right)^2  \right. \nonumber \\
                 && \left. + \left((r-i)-0.09\right)^2 + \left((i-z)-0.02\right)^2  \right]^{1/2} 
\end{eqnarray}

We use PSF magnitudes corrected for Galactic extinction to calculate this
quantity.  Typically there are at least 10 deg$^{-2}$ standard stars
(or 70 per plate), although their density on the sky is
a function of Galactic coordinates. These relatively hot stars provide
a well-understood and mostly smooth continuum that allow the spectra
to be calibrated.
A stellar template of appropriate temperature and surface gravity
is derived for each star and used to derive the spectral response as
described in \citet{schlegel12a}.

During plate design, we choose 16 ``guide'' stars for each plate. During
observations, each hole for a guide star is plugged with a coherent
fiber bundle that is constructed with alignment pins to
track orientation. Two guide stars near the center of the plate have large
bundles ($30 \2pr$ diameter) and are used to acquire the field.  The remaining
fourteen guide stars use smaller bundles ($10 \2pr$ diameter)
and are used to guide during exposures.
We choose guide stars from the SDSS imaging with
$13<g_{\rm PSF}<16.5$ and colors $0.3<g-r<1.4$,
$0<r-i<0.7$ and $-0.4 < i-z<1$ determined from PSF magnitudes. 
The positions are corrected for proper motion when
data from the US Naval Observatory catalog \citep[USNO-B;][]{monet03a} are available.
We use the DR8 version of the SDSS astrometry both for the guide stars and the
science targets. Because of errors in that astrometry, there are some
offsets at declinations $\delta>+41\deg$. However, as 
\citet{aihara11b} explains, these offsets are coherent on large
scales, and the expected contribution of these errors on three degree scales
(corresponding to the plate size) is 60 mas per star, well below our tolerances. 
Finally, we drill holes (3.175 mm, corresponding to $52.5\2pr$ diameter) at the locations of bright stars
to minimize light scattered from the surface of the plug plate.
Stars brighter than $m_V=7.5$ are chosen from
the Tycho-2 catalog \citep{hog00a}.
These light trap holes are not plugged with fibers (they are larger size to prevent accidental
plugging) and allow light to pass through the focal plane unobstructed.

The fibers assigned to standard stars, guide stars, and sky fibers are
distributed uniformly over each plate to ensure consistent data
quality for all spectra, regardless of their position in the focal
plane. Because the fibers in the instrument are distributed in 50 bundles of
20 fibers each, we assign each hole to a particular bundle for ease of plugging.
Unlike the cartridges used in SDSS-I and -II, half of the BOSS fibers are colored
red and half blue; the red and blue fibers alternate in position along the slit head.
By tracking the holes associated with each target,
we assign red fibers preferentially to CMASS targets, which are typically very faint at short wavelengths.
By doing so, we minimize cross-talk on neighboring fibers inside the spectrograph between quasars and
other quasars, low-redshift galaxies, or ancillary targets that have substantial
blue light (since they are usually separated by at least one fiber on
the CCD).
Most bundles do not have exactly ten high-redshift galaxies for the ten red fibers,
so the extra high-redshift galaxies are assigned to
blue fibers, or the extra red fibers are assigned to other objects. 
Other than the bundle assignment and the red/blue fiber designation,
there is no requirement that particular fibers be placed in particular
holes. As described in \S\ref{subsec:plateplugging}
the exact hole where each fiber is plugged is determined shortly before observations.

In planning the survey, we estimate the Local Sidereal
Time (LST) at which each plate will be observed.
We use the corresponding hour angle and altitude predictions of the field
during observation and determine each hole position on the plate
accounting for atmospheric differential refraction (ADR). For galaxies,
standard stars, or ancillary targets, we center the hole position at the
optimal focus and ($x$,$y$) position for 5400 \AA\ light.
Since ADR depends on wavelength, redder or bluer
light will be offset in ($x$,$y$) coordinates away from the center of the fiber.
For quasars, we center the hole position on the
4000 \AA\ light to maximize the SNR in the \lya\ forest.
This difference corresponds to about $0.5\2pr$ 
typically, and the overall throughput difference at wavelengths shorter than 4000 \AA\
can be about 50\%.
In the DR9 data model (plateDesign and spAll files), the quantity
{\bf LAMBDA\_EFF} records
the wavelength for which the hole position of each object was optimized.

In addition to the wavelength-dependent ADR offset, we also
account for the wavelength dependence of the focal plane when
observing the quasar targets.  The focal plane for 4000 \AA\
light differs in the $z$-direction from the focal plane for 5400 \AA\ light by 0-300
microns, depending on the distance from the center of the plate.
To account for this difference, small, sticky washers are
adhered to the back of the plate for quasar targets, where the fibers are plugged.
The washer causes the fiber
tip to sit slightly behind the 5400 \AA\ focus. No washers are used
for holes within $1.02\deg$ of the plate center. Between $1.02\deg$ and $1.34\deg$,
$175 \mu$m washers are used, and between $1.34\deg$ and $1.49\deg$,
$300 \mu$m washers are used. These washers only became
available after Modified Julian Date (MJD) 55441 (September 2, 2010), and were not consistently used until
MJD 55474 (October 5, 2010).
In the DR9 data model, the quantity {\bf ZOFFSET} (plateDesign and spAll files)
records the intended usage of washers, but not the actual usage.
The exact washer usage for each observation during this transition
period (including plates observed both before and after) is documented
on the publicly available software
webpage\footnote{www.sdss3.org/svn/repo/idlspec2d/trunk/opfiles/washers.par}.
The discrepancy will be resolved with DR10 in the summer of 2013.
By optimizing the focal plane position,
and thus the SNR, for 4000 \AA\ light, we are also perturbing the
spectrophotometry relative to the standard stars as discussed in
\S\ref{subsec:specphoto}.
Only the main quasar targets are optimized for 4000 \AA\ focal plane and ADR offsets in DR9.
{\bf QSO\_VAR\_SDSS} targets, and a few other programs assigned the ancillary target flags,
will be similarly affected in the future.
Otherwise, the plate design remains the same
as it was in the SDSS-I and -II surveys \citep{stoughton02a}.

During exposures, the guider adjusts the offsets and plate scale according
to changes in ADR, and adjusts rotation according to changes in altitude and azimuth.
In addition, thermal expansion of the plate due to temperature changes and stellar aberration       
create purely radial shifts in the position of objects, these effects are corrected with
changes to plate scale as predicted by the guider 
by adjusting the primary mirror axially and refocusing the secondary mirror.
At the design hour angle, all guide star and science
target images will be centered in each fiber. However, because observations
typically begin before the design hour angle and complete after the design hour
angle, the image of each object will drift across the center of each fiber, and there will
be no adjustment the guider can make to center all of the guide stars.
To compensate, we apply changes to the plate scale to minimize the effect.
The differential change in position of the image centers across the plate
constrains the hour angles over which the plate is observable.
We define the plate observability window such that the
maximum offset of any hole relative to its image
(in perfect guiding at the design wavelength, 5400 \AA\ or 4000 \AA)
is less than $0.3\2pr$.
The typical visibility window lasts more than two hours.
This window is longest for plates designed to be observed at transit, 
and it gets progressively longer at higher declination, where the
rate of change in airmass with time is slower.

\section{Spectroscopic Observations and Data Reduction}\label{sec:observations}
The plates are designed as explained in the previous section and machined at the
University of Washington months in advance of the observations.
The plates are prepared by the staff at APO before observations begin.
Experience from the earlier surveys motivated the procedures below to produce a
survey of uniform coverage and data quality.
The process of plate drilling, observing, and data processing is nearly
identical to SDSS but described for the first time here.

\subsection{Plate Drilling and Preparation}\label{subsec:plateplugging}

Plates are drilled at a machine shop operated by the University
of Washington, where up to eight plates can be drilled in one day.
The plug-plates are an aluminum alloy, 3.2 mm thick, 0.813 m in diameter
and weigh 4.3 kg.
Because the telescope focal plane is not perfectly flat,
the plates are deformed during drilling to align the hole axes with the optical axes.
When the plate is observed, it is similarly deformed by the cartridge to match the best-focus surface.
Typical drilling position errors are $<0.15 \2pr$ RMS, although
during observations the exact angle at which the fiber rests in the hole
can contribute larger errors in the focal plane position.

\begin{figure}[!ht]
\begin{center}
\includegraphics[scale=0.3,angle=90]{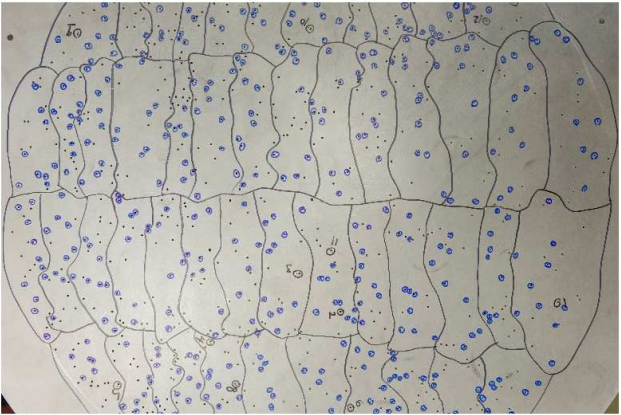}
\end{center}
\caption[Times]{\label{fig:markedplate}
Image of plate 3552 immediately after the marking stage.  Bundles are separated
by black bounded edges, and holes are marked blue to reduce
contamination between nearby emission line galaxies or quasars.
Holes for guide star fibers are marked in black and
denoted by the corresponding number ranging from 1-16.
}
\end{figure}

Plates are shipped to APO where they are received, unpacked, and ``marked''.
In the marking stage, the original plate design is projected
onto the drilled aluminum plate.
Using felt-tipped markers, the staff at APO trace the groupings of 20 fibers 
in a bundle from the plate design onto the aluminum plate
to ensure that all 1000 bundled fibers can reach all 1000 holes.
Following the projected plate design, the APO staff install
washers that are manufactured with an adhesive on one side
around the holes of quasar fibers that need to be offset from the focal plane.
They also mark the locations of holes for guide star fibers.
The entire process takes around 30 minutes; an image of a marked plate
is shown in Figure~\ref{fig:markedplate}.
In SDSS-I and -II, plates were not observed during bright time and the staff
at APO had more time to mark plates.
In SDSS-III, the staff at APO are typically occupied with Marvels and APOGEE observations
in bright time.
We therefore typically try to complete preparation of a BOSS plate at least one month
before it is observed to ease the scheduling of marking and observations.
However, early commissioning plates and changes in target selection often forced
last-minute shipment and plate preparation during the first year.

Once a plate is marked, a database is updated to
indicate that it is available for observation.
When a plate is chosen for observation on a given night based on its visibility window,
a message is sent from the nighttime staff to the daytime staff
requesting that the fibers on this plate be ``plugged''.
In the plugging stage, two staff members install the plate onto the cartridge and
plug fibers from each bundle into the appropriate region of marked holes.
When the plugging is complete, they measure the profile
of the plate surface to confirm that the cartridge has deformed the plate to properly match
the surface of the telescope focal plane.
Finally, a custom-designed machine maps the location of the fibers in the focal plane
by illuminating the output of each fiber with a light and recording the position of
the corresponding hole with a camera that is focused on the surface of the plug plate.
Repeat tests have proven the mapping process to be extremely reliable.
Fibers are occasionally broken, and after plugging occasionally a fiber slips out
of its hole enough to not receive light.  Such cases occur $\sim 2$ times
per plate, are detected during mapping and recorded
in the spectroscopic reductions.
The process of plugging and mapping typically takes about 45 minutes.
A full description of the marking and mapping is found in \citet{smee12a}.
Up to nine BOSS cartridges can be plugged at any time.

\subsection{Procedures During Observations}\label{subsec:procedures}

BOSS observations are performed by a rotating staff of
two or three nighttime staff members known as observers.
Preparations for observing begin in the afternoon;
one observer checks the status of all instruments,
including CCD temperatures and controller connections,
followed by a bias exposure to confirm that all
four detectors (two for each spectrograph) are functioning correctly.
On nights when BOSS observations are planned, the afternoon observer focuses the
spectrographs using a pair of Neon-Mercury-Cadmium arc exposures behind a Hartmann mask.
The Hartmann mask successively obstructs the top half
and the bottom half of the collimated beam.  A simple analysis
of the cross-correlation between the two arc exposures reveals the optical correction
required to obtain optimal spectrograph focus.
The required correction to the total path length can be applied with an adjustment to the
position of each collimator mirror,
thereby changing the focus on the red and blue cameras by an identical amount.
The afternoon adjustment confirms the performance of the instrument;
the collimator mirrors are adjusted throughout the night each time observations begin for a new plate.
In cases where there is a differential in focus between the two cameras,
the correction is applied through adjustment of the focus ring on the blue camera, changing the path length
to the blue detector with no impact on the focus on the red side of the spectrograph.
Because the blue focus rings must be moved manually,
they are typically only adjusted during the afternoon.
Using typical nightly temperature changes and associated focus drift, the blue focus
rings are adjusted every afternoon to compensate for the decrease in temperature and
minimize the effect of differential focus over the course of the night.
Currently, the predictive adjustment is
$-4\deg$ on the B1 focus ring and $+8\deg$ on the B2 focus ring.
Following the focus routine, a five second arc and a 30 second
flat field exposure are obtained to measure the spectral profile and
confirm the focus of the instrument.

When the sun is $\sim 12\deg$ below the horizon in the evening, the observers fill the liquid
nitrogen receptacles in all four cameras and open the telescope enclosure.
The cartridge with the first scheduled plate of the night is mounted
to the telescope, signaling the start of observations.
A script is run to automatically slew the telescope and perform
calibration routines in the following order:

\begin{itemize}
\item Close the eight flat-field petals \citep{gunn06a} that obstruct most of the
opening in the wind baffle and provide a reflective surface for the calibration lamps.
\item Turn on Neon-Mercury-Cadmium arc lamps; they require several minutes to warm up.
\item Slew telescope to field.
\item Take a pair of exposures through the Hartmann masks using a shorter readout
of the detector on a selected sub-region centered on well-known arc emission lines.
\item Adjust collimator positions to account for change in focus since
afternoon checkout and cartridge change.
Because focus rings are not adjusted during the night,
the median between the best blue and red focus
is used to adjust the collimator mirrors accordingly.  
\item Acquire five second calibration arcs and 30 second calibration flats with a quartz lamp.
\item Open petals and take the first guider exposure through coherent fiber bundles. 
\end{itemize}

When the first guider exposure is complete,
the observers center the telescope on the field, using the observed guide stars.
Once the guide stars are roughly centered in their respective fiber bundles,
closed loop guiding begins and
the plate scale of the telescope, the rotation of the telescope relative to
the field, offsets in altitude and azimuth, and focus are adjusted.
At this point, the light from the targeted objects is optimally focused on the fibers
and the observers begin a sequence of 15 minute science exposures.
When observations of the plate are completed according to the
criteria described below, the telescope is moved back to the zenith,
the cartridge replaced with the next scheduled plugged cartridge,
and the series of field acquisition, calibration, and exposures is repeated.
The entire process takes about 15 minutes between cartridges.
In the first year of the survey, approximately 65\% of the available time
for observing was spent with the shutter open on science targets.
In the second year, we automated the scripting of field acquisition
as described above and improved to 75-78\% efficiency, depending on the
amount of exposure time required for each plate.

Ideally, we would acquire signal on each plate until the full data reductions
reveal a data quality that exceeds the threshold required to meet the survey requirements.
However, the full reductions take much longer than the actual observations,
so we instead use a suite of simple reductions performed in real time
to provide quick feedback to the observers.
We refer to these data reductions as the ``Son of Spectro'' (SOS) reductions, 
referring to the full spectroscopic reduction pipeline.
At all times, a daemon automatically identifies new images as they appear
on disk and submits them for flat field calibration, wavelength solution,
and the extraction of the one-dimensional spectra for each fiber.
The SOS reductions perform the same functions as the spectral
extraction in the full data reduction pipeline \citep{schlegel12a},
but they use simpler sky subtraction, use box-car extraction to create
one dimensional spectra,
and perform no redshift or object classification.
Comparing these quick reductions to the full reductions on the data,
we find a reduction of approximately 20\% in the SOS SNR for R1 and R2
and 10\% in SNR for B1 and B2.

\begin{figure*}[!b]
\begin{center}
\includegraphics[scale=0.75]{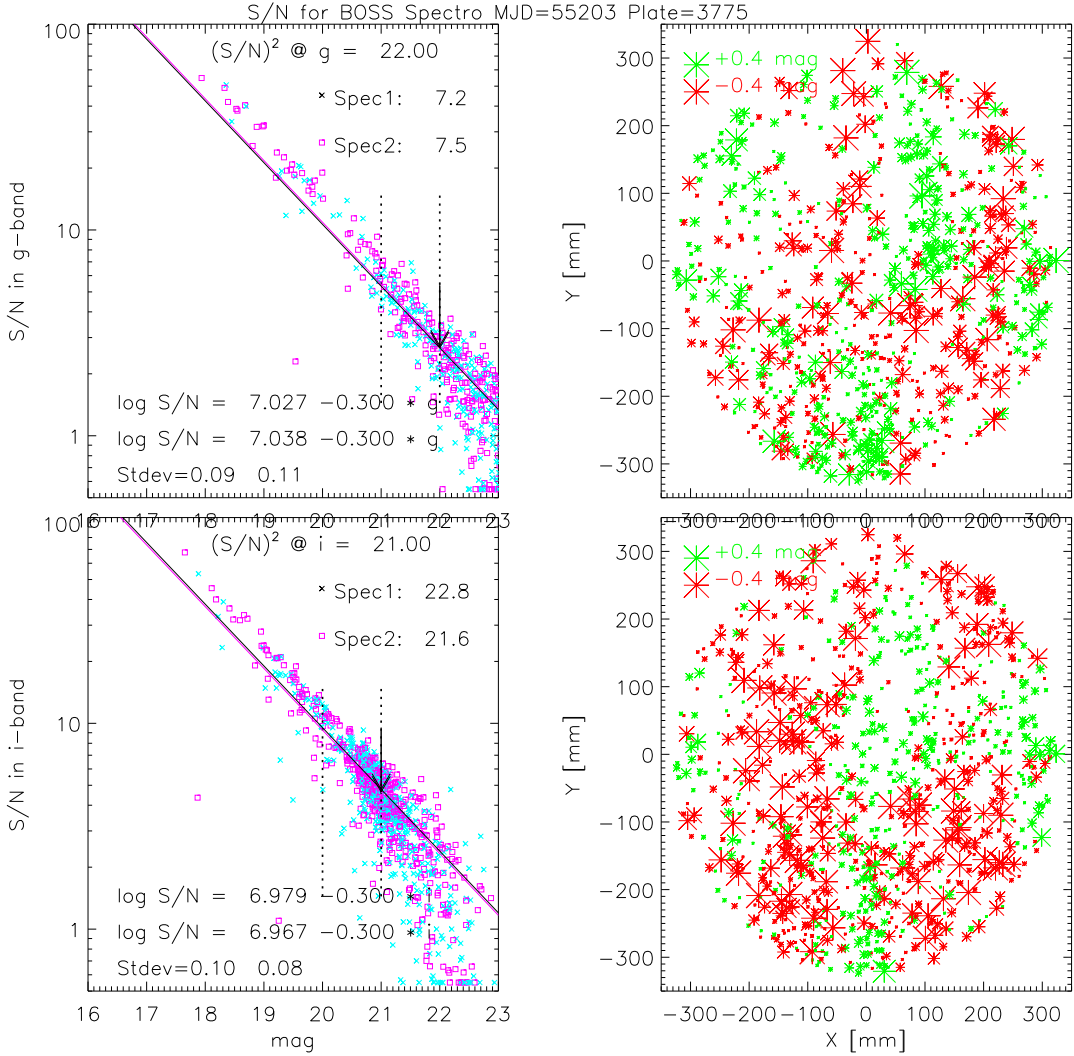}
\end{center}
\caption[Times]{\label{fig:sos}
An example of the diagnostic from a series of four science exposures of plate 3775
that are produced in the quick reductions of data at APO.
Similar plots become available to the observers less than five minutes
after the end of an exposure.
{\bf Left:  }Cumulative signal-to-noise ratio (S/N in the figure captions) as a
function of $g_{\rm fib2}$ (top) and $i_{\rm fib2}$ (bottom).
Fibers from the two spectrographs are indicated by
the blue symbol ``x'' (spectrograph 1) and a magenta square (spectrograph 2).
The intercept of the linear fit of log(S/N) as a function of magnitude and RMS of that fit
are computed for each spectrograph separately.  Only fibers with fiber2 magnitudes in
the range 21-22 (20-21) for the blue (red) cameras are used in the fit as
indicated by the vertical dotted lines.
The slope is held fixed at $-0.3$ as empirically determined from fits to the larger sample.
This magnitude range is near the region of sky-limited noise and near the faint end of the main
galaxy and quasar samples for the red and blue cameras respectively.
{\bf Right:  }The spatial profile of SNR over the plate.  Red symbols represent
fibers that fall below the best fit of the SNR linear solution that is represented
by the solid line in the left hand panels.  Green symbols
represent fibers that fall above the best linear fit of the SNR solution.  The size
of the symbol relates to the amount by which the fiber deviates, growing
larger for fibers with larger deviation from the best fit.
Guiding problems or other systematics can appear as coherent structure in this diagram.
The distribution of objects that fall below the best-fit line is fairly uniform,
indicating a lack of such effects for this particular plate.
}
\end{figure*}

Approximately five minutes are required to process a single flat field exposure
and one minute is required for arc exposures or science exposures.
The results are conveyed to the observers on a webpage that automatically updates every minute.
The webpage provides simple metrics for wavelength
coverage, spectral resolution, profile of each fiber on the CCD,
and, most importantly, the median SNR per pixel
over a synthetic $i-$band bandpass filter for the red cameras and over a synthetic
$g-$band bandpass filter for the blue cameras.
Statistics of this SNR as a function of magnitude allow diagnostics of problems
in the data such as focus or guiding errors.
The magnitudes are estimated from the SDSS imaging data and corrected for
Galactic extinction using the maps of dust infrared emission from \citet{schlegel98a},
scaled to optical wavelengths using the model of \citet{cardelli89a}.
An example of the SNR portion of the report generated for a composite of four
15-minute exposures on a single plate is shown in Figure~\ref{fig:sos}.

The observers use the SOS reports to diagnose performance such as the quality
of focus, the fraction of fibers that dropped from the plug plates
(typically fewer than two or three on a plate), and the rate at which signal is acquired.
They use the accumulated SNR evaluated at a fiducial magnitude in the
sky-dominated regime to determine when a plate is complete.
As shown in the left hand panels of Figure~\ref{fig:sos}, a power law representing the SNR as a function
of magnitude is fit to determine SNR at the fiducial values of $g_{\rm fib2}=22$ and $i_{\rm fib2}=21$.
The SNR$^2$ scales linearly with exposure time and is therefore easier to
use when estimating remaining exposure times.
As explained in \S\ref{sec:simulations}, the SNR$^2$ of the exposures at these magnitudes is
chosen to balance high redshift completeness for the galaxy sample with
the goal of observing 10,000 $\sqdeg$ in five years.
The observers remove the cartridge when the plate is complete by the SNR criteria and request a new plate
to be plugged in that cartridge the following day.
Because of limited visibility windows, it is not uncommon that
observations must cease before a plate is complete.  In these cases, the plate remains
in the cartridge and is observed the next clear night with a new set of calibration frames.
The exposures taken on all nights are considered for determining when a plate is complete.

\subsection{BOSS Data Reduction Pipeline}\label{subsec:pipeline}

At the end of each night of observing, the data are sent to LBNL to be processed.
An automated software routine
checks the data transfer status every 15 minutes and begins the processing
when all of the data have arrived.  The data processing jobs are organized by
plate such that each job extracts, calibrates, coadds, classifies, and fits
the redshift of all 1000 spectra using all exposures of a single plate,
including exposures taken on different nights with the same plugging of fibers.
The data are reduced first by collapsing
them from the two-dimensional image into one-dimensional spectra \citep{schlegel12a}.
In the second step, the one-dimensional spectra are
classified into object types and redshift \citep{bolton12a}.
The software is written primarily in Interactive Data Language
(IDL)\footnote{www.exelisvis.com/language/en-us/productsservices/idl.aspx}
and is collectively referred to as ``idlspec2d''.
This processing typically begins mid-afternoon, and the results are available to
the collaboration by the following morning.

Raw CCD frames are pre-processed by subtracting a bias model, their bias overscan,
subtracting a dark current model, and dividing by a pixel flat-field model
for each CCD.  The models for bias and dark current
are derived from calibration images taken periodically throughout the survey.
The pixel flat-field images are computed about once every six months using a
specially designed slithead that illuminates the entire detector without
fiber-to-fiber spatial structure.
The read noise is measured from the bias overscan region for each CCD
amplifier for each exposure.  Per-pixel variance is estimated using
the measured read noise and the observed photon counts in each pixel.
The inverse variance is multiplied by a known CCD defect mask, and
cosmic rays are identified to mask affected pixels.

The spectra are extracted from the two-dimensional CCD frames into a set of
one-dimensional spectra.  The quartz lamp flat-field spectral images
establish the cross-dispersion profile and initial location of the spectral
traces as projected onto the CCDs.  The locations are then shifted
to match the science frame spectra to account for instrument flexure from the
time of the calibration exposures to the science exposures.
These offsets are typically 0.1 \AA, but can be as large as 1.5 \AA.

The spectra are extracted in groups of 20 according to their respective fiber bundles.
We use a noise-weighted optimal extraction algorithm \citep{horne86a}
with the Gaussian profile cross dispersion widths measured
in the flat field spectra.  A linear cross-dispersion background
term is included for each bundle to account for scattered light not described by the
two-dimensional Gaussian profile.  The centroids and widths of the Gaussians are fixed,
and the extraction solves only for the amplitudes at each wavelength
and the linear background coefficients.

The wavelength solution is initially estimated from the extracted arc-lamp
spectra, then shifted to match the observed
sky lines in each science exposure.
Since the extraction is performed in the native pixel spacing of the CCD,
this results in individual spectral bins that are statistically independent
but not perfectly aligned in wavelength between spectra
or between exposures.
The extracted science spectra from individual exposures are then divided by
the extracted flat-field spectra to correct for fiber-to-fiber throughput variations.
Sky subtraction is performed using a model for background derived from the sky fibers
that were assigned during plate design.  The background varies with fiber position
to account for smoothly
varying differences across the focal plane.
Similarly, the spectral response (i.e., the flux calibration)
is determined over the focal plane using models fit to the
spectra from the standard stars that were assigned in plate design.

Finally, the spectra from individual exposures are combined into a coadded
frame for each fiber on a resampled grid that is linear in log($\lambda$).
Data from both red and blue cameras is used in the coadded frame, generating
spectra that cover the full 361 nm -- 1014 nm wavelength range of the instrument.
The estimated pixel variance is propagated into variance estimates
of the extracted and co-added spectra, but the
covariance terms between different spectra are discarded.
The current pipeline performs this extraction as a single pass, resulting
in a known bias due to estimating the variance from the data rather than
iteratively updating the noise model with the statistics of the extracted spectra.
In the limit of zero flux, the technique systematically assigns a slightly larger variance to
pixels that fluctuate toward higher flux values, leading to a weighted mean that
is suppressed below its true value.  This bias is not corrected in the DR9 sample,
but may be addressed in a future data release.
The process is performed independently for objects observed on different fibers
or different pluggings of a plate, and coadded spectra are recorded for all observations.
Spectra from objects that are observed multiple times are evaluated, and
the observation that produces the best best available unique set of spectra for the object
is identified with the {\bf SPECPRIMARY} flag, as described in the
online documentation\footnote{http://www.sdss3.org/dr8/spectro/catalogs.php}.
Further details about this extraction
pipeline are published in \citet{schlegel12a}.

As described in \citet{aihara11a}, star, galaxy, and quasar templates
are fit to the combined one-dimensional spectra
to determine the classification and redshift of each object.
Redshift classification errors for all objects are reported by the {\bf ZWARNING}
bitmask keyword.
We found that removing the quasar templates from the fits to the galaxy sample
reduces the number of cases of catastrophic failures and classification
confusion.  Allowing only galaxy and stellar templates, we determine a second redshift
for objects targeted as galaxies and report the redshift and 
classification errors with the {\bf Z\_NOQSO} and {\bf ZWARNING\_NOQSO}
keywords, respectively.
Further details about the redshift and classification
of BOSS spectra are published in \citet{bolton12a}.

Throughout the survey, continuous improvements have been made to the spectroscopic data
processing pipeline.  Every few months a new version is tagged and all
of the data are re-processed starting from the raw data.  These tagged
versions are released to the collaboration for internal use and the daily
processing continues with the latest internally released tag.
The majority of processing time is spent on redshift and object classification
due to the large range of redshift over which templates must be fit.
The total processing takes 8 -- 12 hours per plate,
although batch jobs allow an entire data set to be processed in a few days.
Version v5\_4\_45 of the idlspec2d pipeline code is
used to report the results from DR9.

Data reduction results are mirrored to New York University (NYU) on a daily basis
to serve as an offsite backup.
In addition, the daily reductions are processed in parallel on a Linux cluster
at the University of Utah (UU), thus ensuring continuously tested alternate
processing capability.
The standard data flow is from APO to LBNL to NYU and UU; approximately
once a year alternate paths are tested ({\it e.g.}, APO to NYU to LBNL and UU).
Disaster recovery plans with a 24 -- 48 hour turnaround have been
developed and tested, with scenarios ranging from the outage of a single
disk to permanent loss of the LBNL computing center.

DR9 includes all of the spectral products available to the
BOSS collaboration: coadded and individual exposure flux calibrated spectra,
inverse variances per pixel, masks, subtracted sky, calibration vectors, and
model fits.  Intermediate data products such as uncalibrated spectra
and extracted arc/flat lamp spectra are also available.  Catalog data
such as redshifts, classifications, astrometry, and quality flags
are available via
the SDSS-III Catalog Archive Server database and in a fits file binary
table (spAll-v5\_4\_45.fits).  The spectra themselves are available
in a format of one plate-mjd-fiber per file, or in a bundled format of
all spectra of all objects for each plate-mjd.  Details for accessing these
data are at \url{http://www.sdss3.org/dr9/} and \citet{ahn12a}.

\section{Projections for Completing the Survey}\label{sec:simulations}
\begin{figure*}
\begin{center}
\includegraphics[scale=0.45]{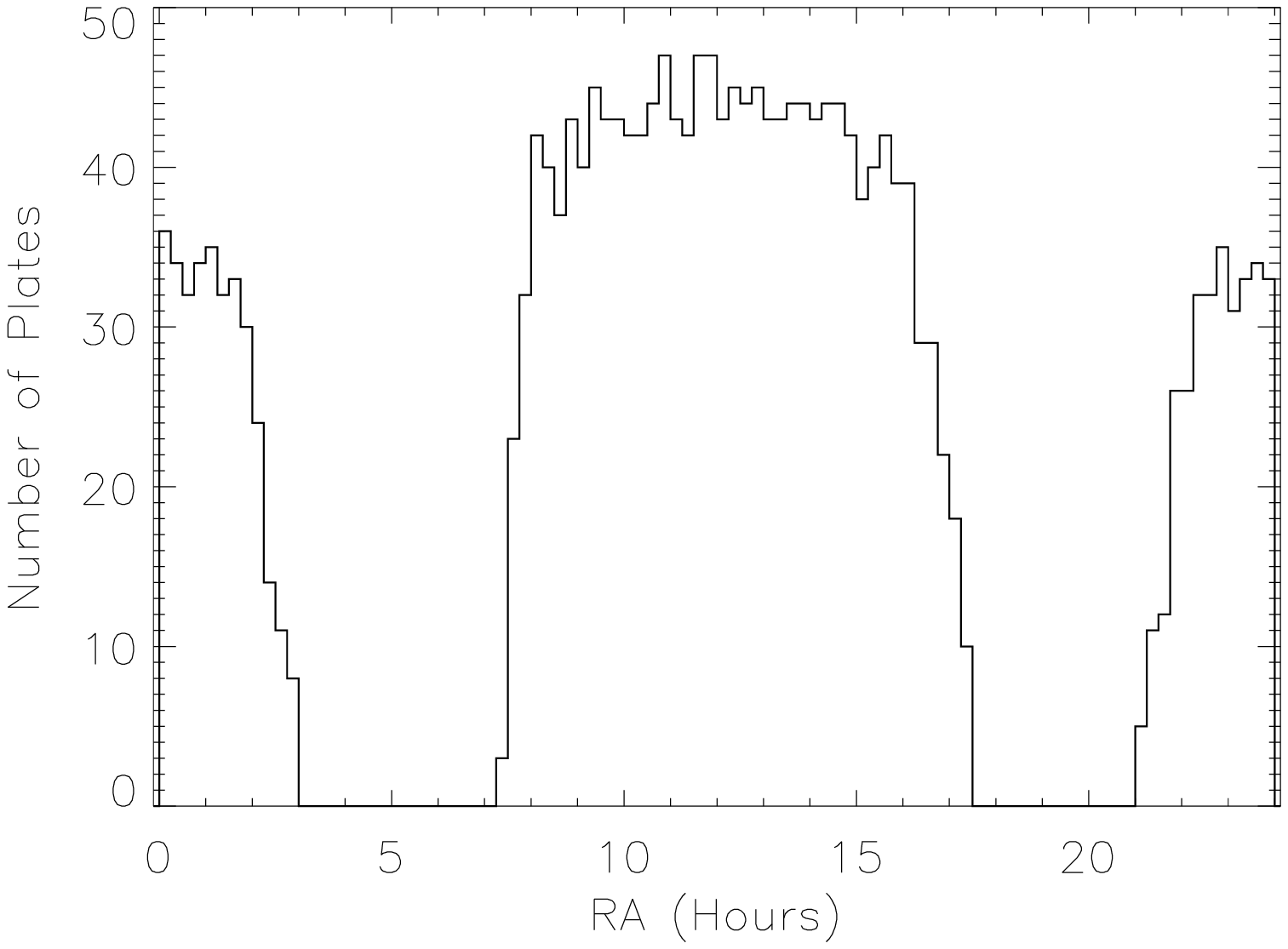}
\includegraphics[scale=0.45]{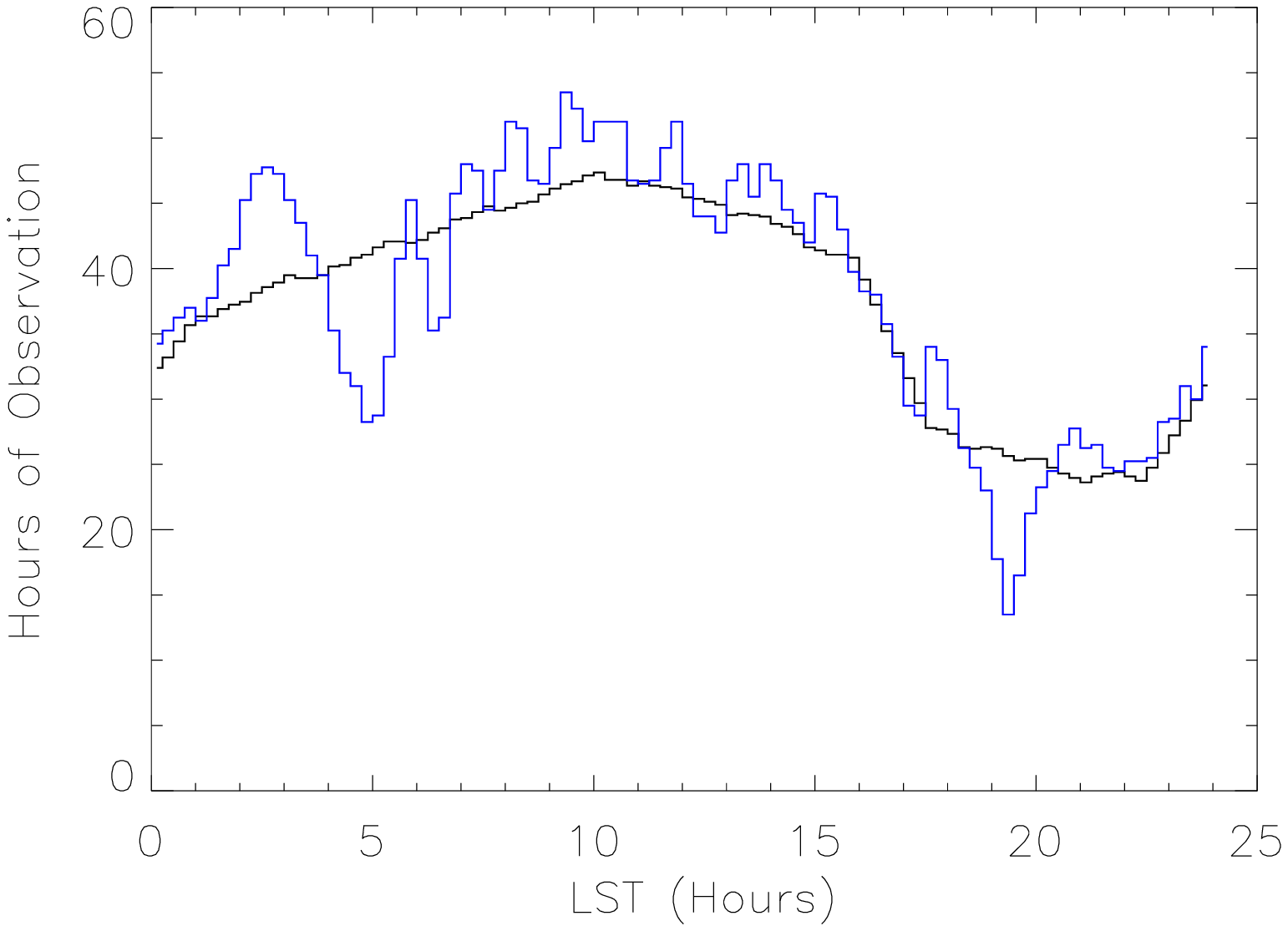}
\end{center}
\caption[Times]{\label{fig:LSTcoverage}
{\bf Left: } Distribution of plates in the BOSS footprint binned in 15 minute
increments in Right Ascension.
{\bf Right: }  Number of hours of observing time available as a function
of LST (black line) and the simulated time required at each LST to observe the full survey (blue line).
The observing time is sampled in 15 minute increments and assumes
a uniformly distributed 45\% efficiency after weather loss.  The 
discrete sum of the entries under the black line is equal to 3585 hours.
The final simulated LST distribution shown in blue
is discussed in \S\ref{subsec:thresholds}.
}
\end{figure*}

BOSS has been allocated the majority of the dark and gray time in the SDSS-III survey
since Fall 2009.
Bright time and a fraction of gray
time are used for high resolution spectroscopy of Milky Way stars to search
for extrasolar planets \citep[MARVELS;][]{ge09a} and to measure abundances in 
evolved, late-type stars from $H-$band spectroscopy to study evolution of the Milky Way
galaxy (APOGEE; Majewski et al., in prep).
These programs are less sensitive to the increased sky background.
On nights when the moon is illuminated at
less than 39\%, all of the observing time is allocated to BOSS.
On nights when the moon is illuminated at more than 56\%,
time is given to BOSS when the moon is below the horizon.
For nights when the moon is between 39\% and 56\% illumination, the allocation
of time depends on the time of year.
In these phases, BOSS is only given time when the moon is below the horizon when
the NGC cannot be observed; BOSS 
is given the full nights when the NGC can be observed.
Time is only split between projects on a given night
if there are at least 1.4 hours allocated to each survey.

Based on historical records, we assume that
55\% of the time will be lost to bad weather and that most of the months of July and August
will be lost to summer monsoon.
Taking the total amount of time allocated to BOSS and the expected weather loss,
we anticipate approximately 3600 hours of observation over the course of the survey
from December 2009 through July 2014.
The distribution of this time as a function of LST is shown in Figure~\ref{fig:LSTcoverage}.
To maximize the survey volume to probe BAO with galaxies and \lya\ quasars,
our goal is to observe the entire 10,060 $\sqdeg$ footprint
tiled with targets from the SDSS imaging program.
Given the total number of spectroscopic plates over this area, and
assuming perfect scheduling, the average target can be observed for 1.62 hours
including overhead due to field acquisition and calibration exposures.
It is essential to characterize the on-sky performance of BOSS
to maximize the efficiency of these integrations over the full five-year survey.

A comparison of the left- and right-hand panels of Figure~\ref{fig:LSTcoverage}
quickly reveals that the distribution of plates in Right Ascension
is not perfectly matched to the
time available to BOSS over the course of the survey.
The differences are particularly large
in the intervals 3 hr $<$ LST $<$ 7 hr and
17.5 hr $<$ LST $<$ 21 hr when the Galactic plane passes directly overhead.
We must then observe the high Galactic latitude BOSS plates at these times
at non-zero hour angles.  
Hour angles are chosen to optimize LST coverage while minimizing
the additional observation time required to account for the higher airmass.

Observations are planned at
a high target density to measure the BAO signal, obtain high redshift completeness,
and finish the survey (Figure~\ref{fig:footprint}) in the allotted time. 
We quantify redshift completeness in the context of 
the value of SNR$^2$ reported by the SOS reductions and find that exposures
must be at least as deep as SNR$^2 > 20$ on the red cameras and SNR$^2 > 10$ on the blue
cameras to obtain the desired completeness in galaxy redshifts.
We estimate the maximum exposure time that allows completion of the full BOSS footprint
and find that we have time to allow for slightly deeper exposures than that minimum requirement.
In this section, we describe the tension between these two constraints and our analysis to
determine the optimal hour angles and SNR$^2$ thresholds.

\subsection{Galaxy Redshift Completeness}\label{subsec:LRG}

\begin{deluxetable*}{lcc}
\centering
\tablewidth{0pt}
\tabletypesize{\footnotesize}
\tablecaption{\label{tab:thresholds} Spectroscopic Classification in Subsets of First Year Data}
\tablehead{\colhead{Data Subsample} & \colhead{SNR$_i^2 \ge 20$ AND SNR$_g^2 \ge 10$} & \colhead{SNR$_i^2 < 20$ OR SNR$_g^2 < 10$}
}
\startdata
LOWZ    &   0.995  &   0.992  \\
CMASS   &   0.942  &   0.920  \\
CMASS ($i_{\rm fib2} < 21.5$)  &   0.966  &   0.955  \\
CMASS ($21.5 < i_{\rm fib2} < 21.7$)     &   0.851  &   0.741  \\
CMASS ($i_{\rm fib2} > 21.7$)   &   0.634  &   0.512
\enddata
\end{deluxetable*}

Meeting the projections for BAO constraints from the galaxy sample described in
\S\ref{subsec:galaxyTS} requires that we measure galaxy redshifts for
$>$94\% of the targets, where the remainder
are either not galaxies or are recognized redshift-fitting failures.
For comparison, the SDSS main and LRG spectroscopic samples
had a redshift success rate of $>$99\% for galaxies at $0<z<0.45$ \citep{strauss02a}.
We first identify a SNR$^2$ threshold at the fiducial magnitude for determining the completion
of a plate that realizes this 94\% redshift success rate for galaxies.

In the first year of BOSS operations, we lacked the information to empirically determine the
quality of a typical exposure or the depth of the data required to obtain redshifts
for faint targets.
We thus intentionally chose a threshold that produced deeper data
than would allow us to finish the survey in the five year window,
with the idea of adjusting the thresholds once redshift success rates could be quantified.
Specifically, we chose a threshold of SNR$^2 > 16$ in the blue
cameras evaluated at $g_{\rm fib2} = 22$, and
SNR$^2 > 26$ for the red cameras evaluated at $i_{\rm fib2} = 21$.
For nights when the plate visibility window expired before meeting these thresholds,
a plate was considered complete if it reached SNR$^2 > 13$ and SNR$^2 > 22$ on the blue
and red cameras, respectively.  Plates that did not reach these thresholds
were kept in their cartridges and observed on the next clear night.

To quantify the minimum SNR$^2$ to reach the survey goals,
observations from the first year were artificially degraded by removing 33\%
of the 15 minute exposures on each plate.
Exposure depths for each plate were determined from the remaining subset of exposures
using the SOS reductions.
The exposures were combined as explained in \S\ref{subsec:pipeline}
and evaluated for redshift completeness.
A successful classification of a LOWZ or CMASS
galaxy target is one that produces {\bf ZWARNING\_NOQSO} $=0$.
A reduction in redshift completeness becomes evident at SNR$^2 \le 20$
at the fiducial magnitude on the red cameras
and SNR$^2 \le 10$ on the blue cameras.
The redshift completeness of the LOWZ sample and the brightest objects in the CMASS sample are
only marginally impacted by the reduced SNR$^2$,
but the faintest objects in the CMASS sample ($i_{\rm fib2} > 21.5$) are
quite sensitive to the SNR$^2$.

The results imply that each exposure must satisfy SNR$^2 \ge 20$ for the red cameras
and SNR$^2 \ge 10$ for the blue cameras to obtain a reliable
reliable classification of the CMASS objects.
These SNR$^2$ values define the absolute minimum threshold that
can be used, beginning in the second year when the observing procedures were updated.
Table~\ref{tab:thresholds} shows the impacts of applying these
thresholds on the redshift success rate for various subsets of the data.
When applying this minimum SNR threshold, we find that 94\%
of the CMASS sample is classified with {\bf ZWARNING\_NOQSO}$=0$.
However, less than 65\% of the subset of
CMASS targets with $i_{\rm fib2} > 21.7$ are classified successfully and only
85\% of the subset of CMASS targets with $21.5 < i_{\rm fib2} < 21.7$ are classified successfully.
The lower efficiency for targets with $i_{\rm fib2} > 21.5$ led
to the decision to remove these targets from the galaxy target selection after the first year,
effectively reducing the size of the CMASS sample by 5.2\%.
The remaining CMASS targets should be classified at an efficiency greater than 96\% as
long as all plates are observed to a depth SNR$_i^2 \ge 20$ and SNR$_g^2 \ge 10$.

\begin{figure}[!ht]
\begin{center}
\includegraphics[scale=0.45]{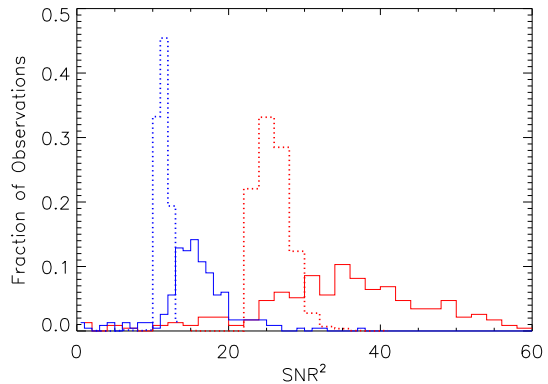}
\end{center}
\caption[Times]{\label{fig:projections}
The fraction of completed plates as a function of SNR$^2$ as reported by SOS.
The SNR$^2$ for the first year of data is shown as a solid line while the SNR$^2$
for the simulated survey,
with somewhat lower thresholds as described in \S\ref{subsec:thresholds}, is shown as the broken line.
In both cases, the SNR$^2$ for the red cameras is shown in red while the SNR$^2$ for blue cameras
is shown in blue.
}
\end{figure}

These completeness estimates are based on the current spectroscopic data reduction,
which does a good but not perfect job of spectral extraction and classification.
We are continuing an effort to implement the ``spectro-perfectionism'' spectral extraction
scheme \citep{bolton10a} and to develop new templates for classifying galaxies and quasars.
We anticipate that the redshift completeness will ultimately exceed the performance in DR9.
Under good observing conditions at low airmass and low Galactic extinction, the SNR$_i^2 \ge 20$
condition is more demanding than the SNR$_g^2 \ge 10$ condition.
However, if the blue criterion is more stringent, then some plates at high airmass or
low Galactic latitude take a very long time to complete.
Figure~\ref{fig:projections} shows the distribution of SNR$_i^2$ and SNR$_g^2$ in
BOSS observations taken in the first year.
Note that the SNR$_i^2$ typically exceeds the threshold (SNR$_i^2 > 26$ for year one)
by a small margin because exposures are taken in 15 minute intervals,
even when the plate is close to threshold.
Some plates are well above threshold because we occasionally had too few
plates available for a given LST and therefore had to ``overcook'' the plates that we had.

\subsection{Quasar Identification and \lya\ Forest SNR}\label{subsec:qso}

While the performance of the galaxy component of BOSS is easily captured by the
redshift efficiency, the performance of the quasar component is complicated
by the fact that the entire \lya\ forest region of each quasar spectrum is important.
While an increase in exposure times leads to deeper spectra in the \lya\ forest,
the extra time also reduces the total sky coverage.
When we considered changing the exposure depths after the first year of BOSS,
scaling arguments based on \citet{mcdonald07a} indicated that for BAO studies,
the gains from additional area would exceed the losses from lower SNR in the \lya\ forest.
In a more thorough recent investigation, 
\citet{font-ribera12a} address the question using detailed simulated \lya\ forest catalogs
constructed from Gaussian random density field realizations.
They simulate quasars with a density 15-17 deg$^{-2}$, $g<22$, exposure
depths comparable to those in BOSS, and a redshift $2.15<z<3.5$.
Noise is added to each quasar spectrum assuming typical sky and readout noise from BOSS.
They find that the errors on the \lya\ forest correlation function increase by 10-15\% if
the exposure times are halved, and conversely, that those errors
would be reduced by 30\% in the limit of zero noise.
Because the error bars on the correlation function scale as the inverse
square root of survey area, the simulations confirm a decision
to favor area over depth in the \lya\ spectra.
We report the quality of the BOSS \lya\ spectra in \S\ref{subsec:lya} in comparison to expectations
from theoretical predictions for an optimal quasar survey.

Investigating the large scale structure of \lya\ absorption relies critically
on our ability to classify quasars and determine their redshifts in the first place.
Quasars at $z>2.15$ are easily identified
due to strong emission lines such as Mg II $\lambda 2798$, C III $\lambda 1908$,
C IV $\lambda 1549$, and \lya\ $\lambda 1215$.
As discussed in \citet{paris12a}, the typical central pixel for a $g_{\rm PSF} \sim 22$ DR9 quasar
in these emission line regions has a SNR$=0.62$, SNR$=1.4$, SNR$=3.0$, and SNR$=4.3$,
respectively.
For comparison, the median SNR per pixel over the whole spectrum at $g_{\rm PSF} \sim 22$ is 0.90.
In many spectra of quasars at $g_{\rm PSF} \sim 22$,
only \lya\ and CIV emission lines are used in identification;
about half of quasar targets are confirmed to be at $z>2.15$.
While incompleteness is substantial, it is important to emphasize that
the vast majority of objects that are not confirmed \lya\ quasars are likely to be stars,
which lack the strong emission features of typical quasars.  Some of the failures
may also be BL Lac objects or other weak-lined active galactic nuclei (AGN).
In any event, the level of spectroscopic incompleteness does not seriously impact
our ability to measure structure in the \lya\ forest.

As an additional test of the spectroscopic completeness of the quasar sample,
we performed spectroscopy on seven plates with quasars selected to $g_{\rm PSF} < 22.5$,
half a magnitude fainter than the typical limit for the main quasar sample.
These targets fall on plates 5141-5147 and were chosen based on variability
without the color selection that was imposed on the other variability-selected
targets in Stripe 82.
Comparing shallow exposures of plate 5141 (the only one of these plates included in DR9)
to deeper exposures, we found
a redshift success rate $\sim 100$\% for quasar targets with $g_{\rm PSF}<22$, and $\sim 80$\% for
$22 < g < 22.5$.  A manual classification revealed that no objects were falsely classified
as quasars in the shorter exposures.
While the quality of the quasar spectra is clearly an essential feature of the BOSS survey,
we conclude that even shallow BOSS exposures are adequate to reliably classify quasars and
that survey area is more important than depth for signal in the \lya\ forest region.
We therefore base our metric for exposure depths entirely on the redshift efficiency for the galaxy sample.

\subsection{Atmospheric and Galactic Extinction}\label{subsec:averagesnr}
Designing the survey, scheduling plates for plugging and observation, and tracking
survey progress against completion of the full footprint require that we have accurate predictions
for the total exposure time that will be needed for each plate to reach at least the SNR thresholds
given in \S\ref{subsec:LRG}.
In order to determine how airmass effects SNR$^2$,
we first predict the impact of observing conditions on the depth of the exposures.  
While the depth of each exposure is affected by seeing conditions, changes in atmospheric
extinction, sky brightness, and cloud cover, these processes are stochastic and cannot
be predicted for a given plate at the time of observation.
On the other hand, airmass and
Galactic extinction only depend on the plate coordinates and design hour angle.
We used the data from the first year of BOSS observations to understand
their effect on the mean SNR$^2$.

Using the reported SNR$^2$ for each exposure in the first year from the SOS
data reductions,
we found that the scaling of the mean SNR$^2$ with Galactic extinction
is well described by the expected relationship:
\begin{equation}
SNR^2 = constant \times 10^{-2 \, A_X/2.5}
\end{equation}
where $A_X$ is the predicted extinction from \citet{schlegel98a} in magnitudes in the synthetic bandpass filter ``X''.
We then quantify the effect of airmass on the SNR$^2$ from the SOS reductions.
Higher airmass degrades the SNR by introducing additional sky background,
increasing the seeing, and reducing atmospheric transparency.
The exposures are binned by airmass and averaged
in each bin to account for all other weather effects.
We find a power law dependence on airmass (Y); the mean SNR$^2$ for a single exposure
at the fiducial magnitudes scales for the red and blue cameras as:
\begin{equation}
SNR^2_i = 7.5 \times Y^{-1.25}
\end{equation}
\begin{equation}
SNR^2_g = 3.6 \times Y^{-1.0}
\end{equation}
These relations accurately describe the data up to an airmass of 1.3.
Beyond that limit, the data become roughly independent of airmass with substantial scatter.

\subsection{Determining Plate Hour Angles}\label{subsec:hadesign}

We attempt to assign the range of LST that each spectroscopic plug plate can be observed
in a manner that covers the available time allocation for BOSS with the highest SNR for the whole survey.
We calculate the Galactic extinction at the center of each tile.
We then determine the hour angle and corresponding airmass at each
available timeslot shown in the right hand panel of Figure~\ref{fig:LSTcoverage} for that tile.
At this stage, we impose several constraints on the observations.
Tiles within $\delta \pm 5\deg$ of the APO latitude of $+32\deg 47\pr$ 
must be designed at $|$HA$| \ge 1$ hour to prevent zenith crossing,
where telescope tracking in altitude-azimuth becomes uncertain.
We then define a maximum hour angle for each tile to ensure that plates
are not observed at an airmass that would lead to visibility windows that
are significantly less than two hours.
Finally, to simplify plate design, we assign hour angles in increments of 20 minutes.
All plates are limited to a design hour angle of less than $\pm 3\ah 20\am$.

Using the models for dependence on airmass and Galactic extinction
in \S\ref{subsec:averagesnr}, we estimate
the effective exposure time required to complete each plate at each timeslot.
We define the weight of each plate as the fractional increase in exposure
time required to achieve a SNR$^2$ identical to a plate at zero Galactic
extinction, airmass unity, mean seeing conditions and mean atmospheric conditions.
We then rank the plates in order of increasing weight at each LST
and determine the most expensive timeslots to perform observations.
The regions that are the most expensive are those that have the highest weight
for the top-ranked plates.
These times occur during
the Galactic plane crossing at $\sim 4.6$ hours and $\sim 20$ hours LST,
where the fields tend to be observed at high airmass.

Given the celestial boundaries in the survey, we assign hour angles to plates in
the NGC and SGC independently.
We start with plates in the NGC with the timeslot
at 4.6 hours LST and the $N$ highest ranked plates for the
$N$ days that are predicted to have observations at that LST.
We simulate observations of each plate using an integral number of 15 minute exposures,
a $75\%$ observing efficiency, and a SNR that evolves according to the change in airmass
over the course of the observation.
The typical simulated observation of a plate with $\alpha \sim 7.5$ hours
takes between 1.5 and 2 hours.  It is these plates that are observed at high hour angle
around 4.6-6 hours LST.
We subtract the integrated exposure time for each plate over the range
of LST in the same manner as actual observations would take place.
We similarly simulate observations for the plates with $\alpha \sim 17$ hours
that are needed for observation at the 20 hour LST timeslot.
We alternate between the eastern and western edges of the NGC, incrementally using plates closer 
to the center of the NGC as the LST approaches 12-13 hours, until all of the available time is used.
We then assign hour angles to plates in the SGC following the same technique.

\subsection{Determining Exposure Depths and Updating Plate Designs}\label{subsec:thresholds}

We next evaluate the amount of time required to complete the plates to determine
how much additional signal beyond the minimum thresholds (SNR$_g^2 > 10$,
SNR$_i^2 > 20$) can be acquired while still completing the survey. 
The process of hour angle assignment is performed iteratively by varying the minimum
depth of each exposure and varying the LST range covered by the NGC and SGC plates.
Simulated fifteen minute exposures are accumulated until the SNR$^2$
exceeds some minimum threshold as described in \S\ref{subsec:procedures}.
We manually vary the thresholds of both the blue and red cameras until the
simulations produce a completed survey in the amount of time allocated to BOSS.
We adjust the LST times that divide plates between the NGC and SGC until both regions
are observed in the amount of time allocated;
plates from the NGC should be observed between
an LST of 4.9 hours and an LST of 19.1 hours
while plates from the SGC should be observed at other times.

We find that we can complete the survey if we set a SNR$^2$ threshold of
22 per pixel for the synthetic $i-$band filter and
10 per pixel for the synthetic $g-$band filter at their respective fiducial fiber2 magnitudes.
As argued in \S\ref{subsec:LRG}, this depth
is sufficient to accurately obtain redshifts for the vast majority of CMASS galaxy targets.
The mean exposure time for the survey will be 1.61 hours per plate,
amounting to a total of 3551 hours of exposure time.
Because the plates in the NGC are located at a higher Declination on average, and in
regions that have smaller amounts of Galactic dust, the typical plate in the
NGC will be completed in 1.49 hours, while the average plate
in the SGC will be completed in 1.92 hours.
The projected time spent observing at each LST is shown
in Figure~\ref{fig:LSTcoverage}
and a histogram of predicted plate SNR$^2$ for the full
survey are shown in Figure~\ref{fig:projections}.

Figure~\ref{fig:LSTcoverage} reveals two additional features of the BOSS survey
projections.
The first feature is a minor shortage of plates in the LST regions covered
by the Galactic plane, appearing as dips around 5 and 20 hours.
It is during these times
that the observations are affected by the limitation on hour angles mentioned above.
This shortage of plates was actually exacerbated during the first two years of observations.
The second feature appears as an excess of plates between 21 and 3 hours LST.
This is the part of the SGC that is visible in the summer months and early Fall
when nights are short and when the telescope is closed for six weeks due
to yearly maintenance.
It is likely that BOSS will have unobserved plates in this region when the survey is complete.
To minimize similar gaps at other ranges of LST,
we update the survey projections every few months to account for the
amount of time remaining in the survey and the plates completed.

\subsection{Maintaining a High Observing Efficiency}\label{subsec:safety}

We require a high observing efficiency to accomplish the goal of completing the
survey footprint within a fixed time window that is set largely by funding constraints.
To do so, we balance nightly observing efficiency against the higher priority
of protecting the telescope and instruments to avoid down-time or catastrophic damage.
Over the years since the beginning of SDSS, we have adopted a number of strategies
to achieve this balance.

We assign two night-time observers at
all times as discussed in \S\ref{subsec:procedures}.
Two observers are scheduled mainly for observing efficiency and safety.
The ``warm'' observer is responsible for setting up the software that controls
the telescope and instruments, starting and monitoring data collection,
checking SOS feedback, and other tasks that can be performed indoors.
The ``cold'' observer's
duties include swapping cartridges, checking data quality and writing the night log.
To minimize the amount of time between cartridge changes,
the cold observer prepares for the change before the final exposure is completed.
As soon as the final exposure is complete, the warm observer moves the telescope to zenith
and the cold observer physically locks the system so that the telescope cannot move.
As soon as the cold observer mounts the new cartridge to the telescope, the
warm observer updates the software control system to reflect the cartridge change.
The cold observer, still outdoors, releases the telescope so that slewing can begin.
With two observers on duty, we shorten each cartridge change
by between five and six minutes (saving 2-4 exposures per night), ensure better data quality,
and guarantee that all parts of the system are functioning correctly.
Operations safety is another important reason for scheduling two observers. While the cold
observer is performing duties outside, the warm observer is monitoring the process from
a camera that feeds to video in the control room.  Therefore, the second observer
can respond quickly in the unlikely case of an accident.
All observers are trained to perform both warm and cold roles to ensure
continuity between observing shift changes.

Both observers monitor the weather conditions and determine when to close the telescope enclosure.
In a normal situation, it takes about seven minutes to close (including
slewing the telescope to stow position and moving the enclosure).
However, our criteria for closing due to external conditions are somewhat
conservative in case we must manually move the enclosure back over the telescope,
which takes about 30 minutes.
The observers will close the telescope during the night if any of the following conditions occur:

\begin{itemize}
\item Any lightning is detected within 15 miles.
\item Precipitation is detected, or a radar return from precipitation is approaching that is
thirty minutes away or less.
\item Smoke or ash is detected.
\item Wind speed is over 40 mph.
\item Humidity is high, with a dewpoint $2.5\deg$ C below ambient temperature.  Observers
also visually check for condensation any time the temperature drops within 4 C of dewpoint.
\item  Integrated dust counts (likely from the White Sands gypsum dunefield or
agriculture in the typically dry valleys below the telescope)
for the night exceeds 20,000-40,000 count-hours.
The exact number depends on humidity, closing at lower integrated dust-hours during high humidity.
\item Ambient temperature is below -12 C.
\end{itemize}

At the end of each night, logs are distributed to a broad list of APO day staff, observers,
software developers, and project leaders (including those for MARVELS, APOGEE, and SEGUE).
These logs are partly generated automatically,
covering most of the telescope, weather and instrument status.
Any problems or unusual situations are recorded manually by the
observers, including relevant error messages and as much relevant
information as is available.

\newpage

\section{Data Quality}\label{sec:data}

\begin{figure*}[h!]
\begin{center}
\centerline{
  \includegraphics[]{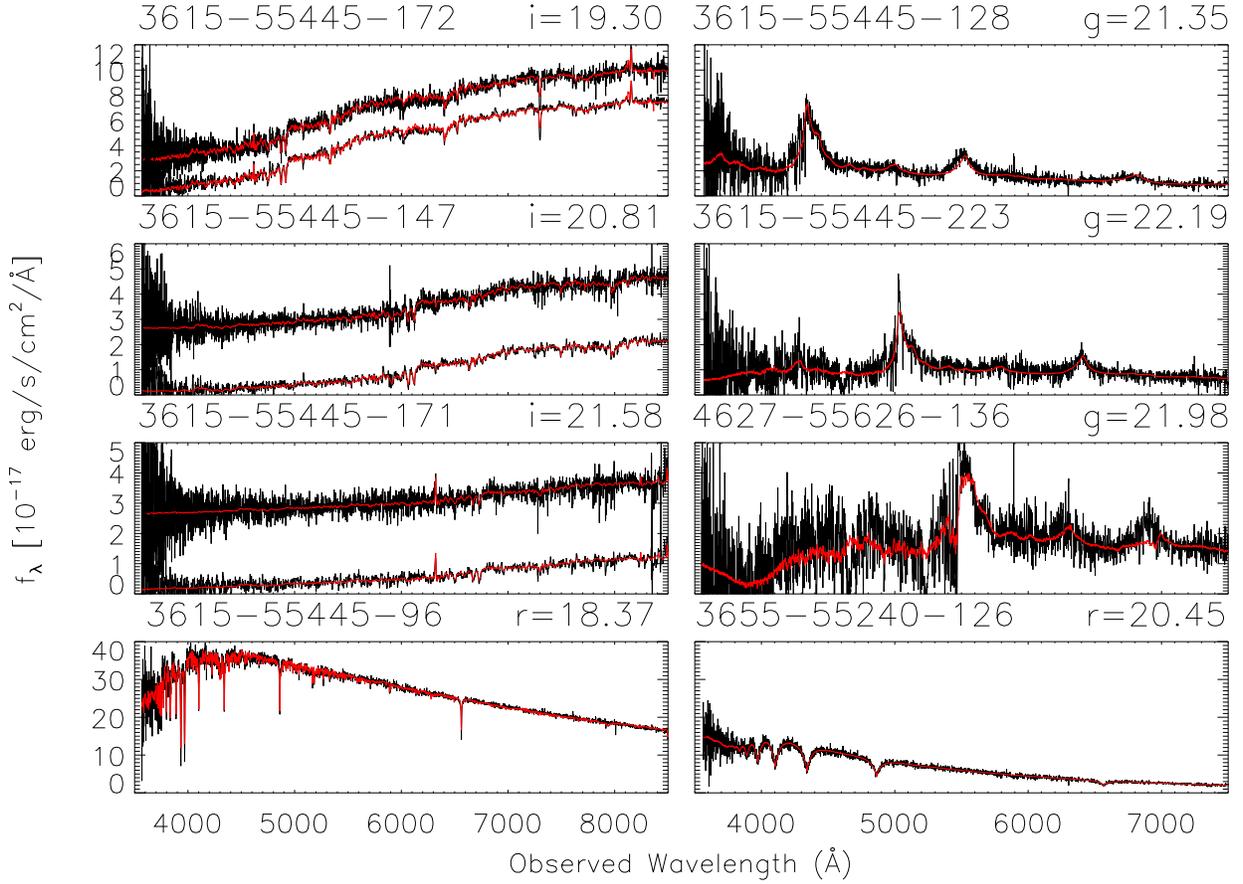}
}
\end{center}
\caption[BOSS Mosaic]{\label{fig:mosaic}
A mosaic of BOSS spectra (black) and best-fit idlspec2d templates (red) of varying luminosity.
The plate, MJD, fiber, and fiber2 magnitude are listed at the top of each panel.
{\bf Left (Top Three Panels):  }Galaxy spectra from the LOWZ and CMASS samples in
order of decreasing luminosity and increasing redshift:
a $z=0.238$ LOWZ galaxy,
a $z=0.541$ CMASS galaxy, and
a $z=0.695$ CMASS galaxy that is slightly fainter than the $i_{\rm fib2}<21.5$ threshold that was imposed
after year one.  In each case, the top spectrum is offset by $2.5 \times 10^{-17}$ erg s$^{-1}$ cm$^{-2}$\AA$^{-1}$
and the bottom spectrum is smoothed with a five-pixel median boxcar filter.
{\bf Right (Top Three Panels):  }Unsmoothed quasar spectra from the CORE and BONUS samples in
order of increasing redshift:
a $z=2.57$ quasar,
a $z=3.14$ quasar, and
a $z=3.53$ quasar.
{\bf Bottom:  }Unsmoothed spectrum of a standard star (Left) and of a white dwarf star (Right).
\\
}
\end{figure*}

The final imaging observations of the BOSS footprint were completed in the Fall of 2009
and are found in DR8 \citep{aihara11a}.
Following the spectrograph rebuild in Summer of 2009,
BOSS commissioning took place in Fall 2009.
During commissioning, spectroscopic observations were performed on nights assigned to
BOSS when conditions were not photometric.
Full spectroscopic survey operations began on December 5, 2009 (MJD 55170) after
the commissioning phase was complete, marking the first spectroscopic
data that are included in DR9.
Several additional improvements were made to the spectrograph during the
survey.  A few bugs in the software for guiding were resolved, residual tilt in the
CCD focal plane was corrected, new triplet lenses were installed for all four cameras,
the red CCDs were replaced, and the collimator mirrors were recoated.
These changes led to improved spectral resolution and throughput, with a combined
improvement to survey efficiency of roughly 25\%.
The dates of these changes are documented in \citet{ahn12a,schlegel12a}.

\begin{figure*}[h]
\begin{center}
\centerline{
  \includegraphics[scale=0.5]{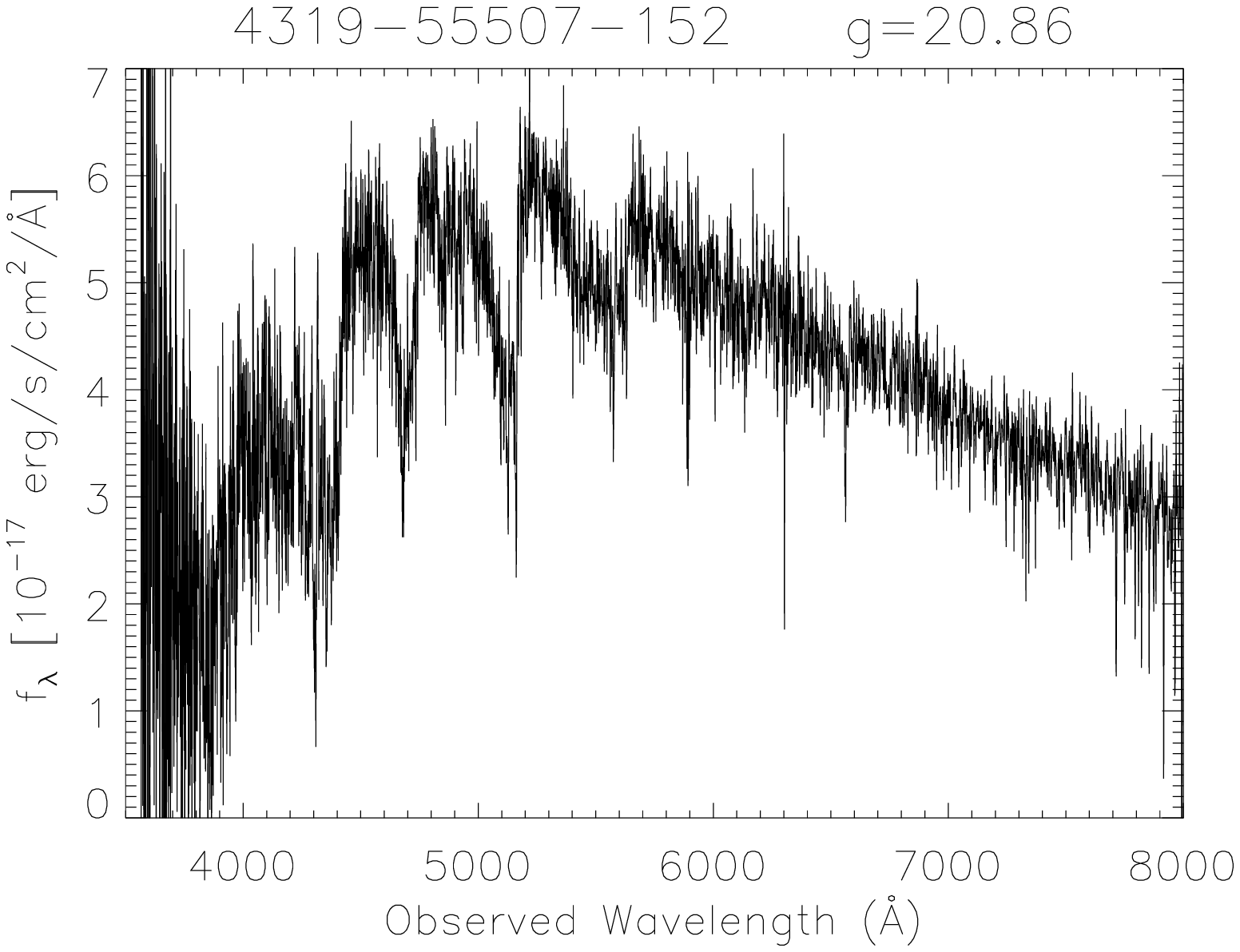}
  \includegraphics[scale=0.5]{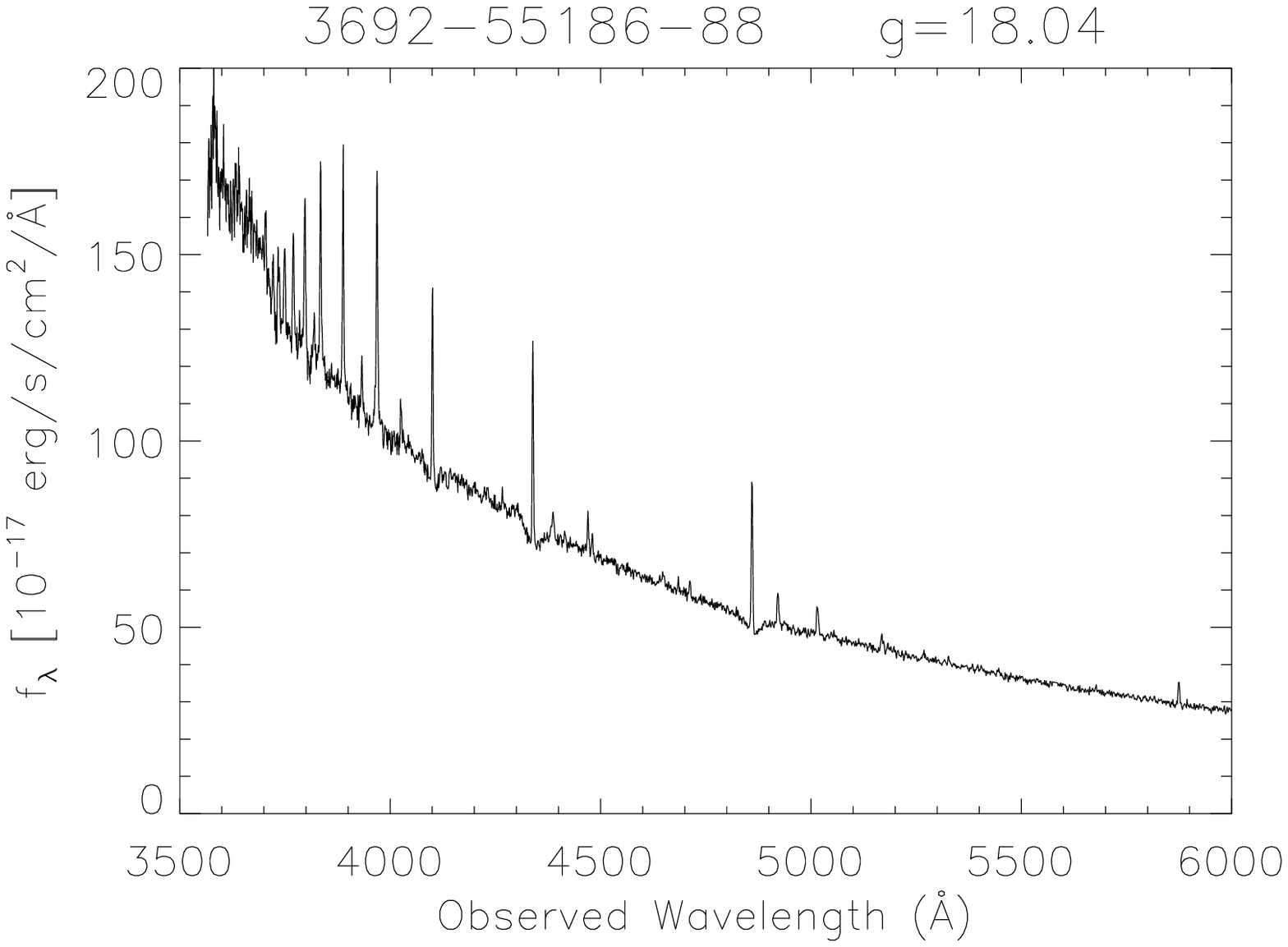}
}
\centerline{
  \includegraphics[scale=0.5]{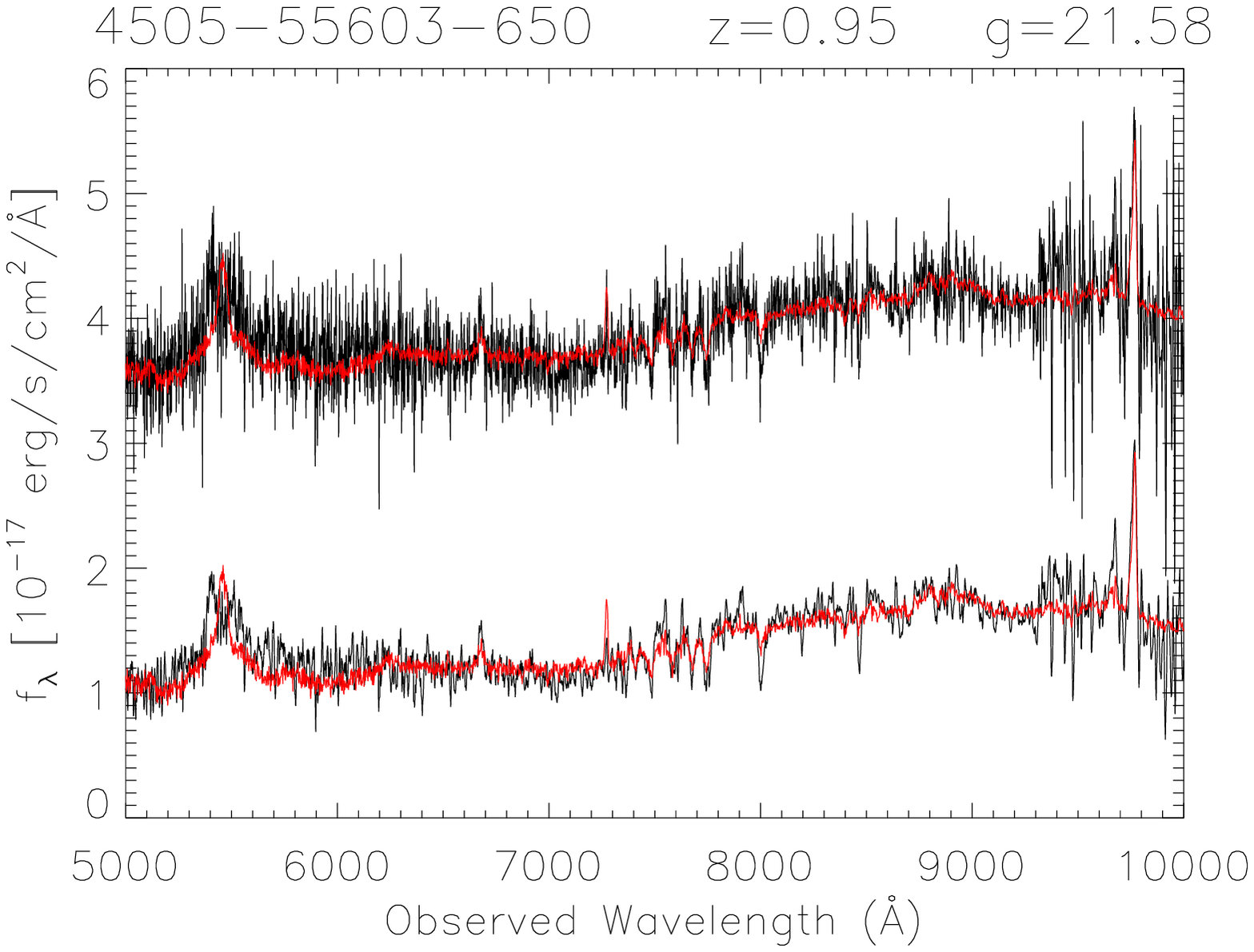}
  \includegraphics[scale=0.5]{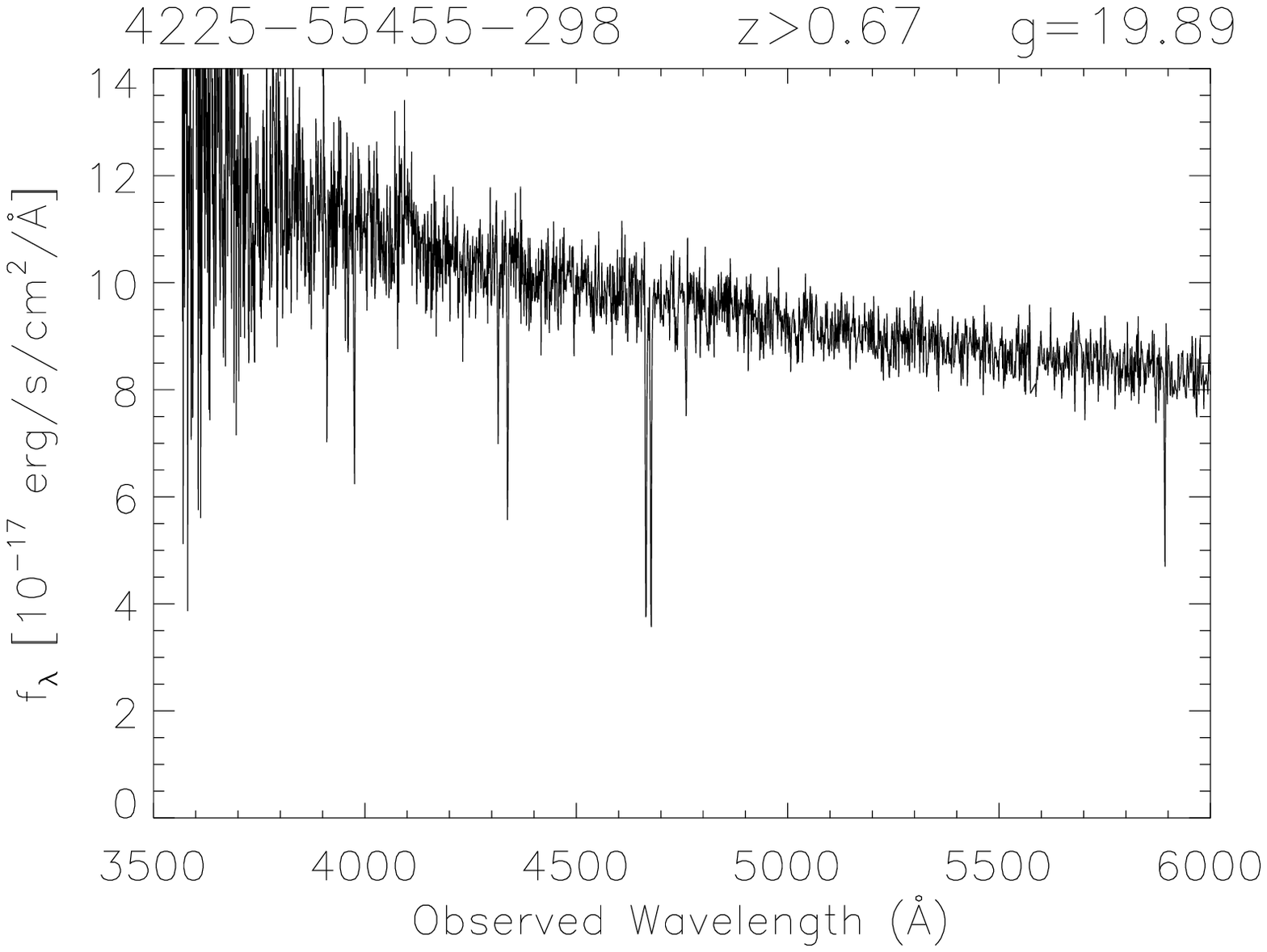}
}
\end{center}
\caption[BOSS Rare Objects]{\label{fig:rarespectra}
A mosaic of example spectra of rare objects discovered in the BOSS spectroscopic sample.
{\bf Top:  }Spectrum of a carbon star (left) and a cataclysmic variable with strong emission lines (right).
{\bf Bottom:  }Spectrum of a post-starburst galaxy at $z=0.95$ with AGN activity (left) and
a featureless BL Lac imprinted with absorption lines from Fe II $\lambda$2382, Fe II $\lambda$2600,
and the Mg II $\lambda$2798,2803 doublet at $z=0.67$ (right).  The idlspec2d template (red) for the post-starburst
galaxy is a good fit to the data.  For this galaxy, the raw data is displayed with an offset of
$2.5 \times 10^{-17}$ erg s$^{-1}$ cm$^{-2}$\AA$^{-1}$ and the spectrum smoothed with a five-pixel median
boxcar filter is shown at true flux density.
}
\end{figure*}

To give the reader a qualitative impression of the BOSS data quality from these first two years
of data, we present a few examples of galaxy, quasar, and stellar spectra in Figure~\ref{fig:mosaic}.
As with SDSS, the enormous size of the BOSS spectroscopic sample also includes many
classes of rare objects.  A few examples of unusual spectra are shown in Figure~\ref{fig:rarespectra}.
In the following section, we describe some global characteristics of the BOSS spectra.

\clearpage

\subsection{BOSS Spectrophotometry}\label{subsec:specphoto}

As described briefly in \S\ref{subsec:pipeline}, the flux calibration for each exposure
is corrected as a function of position in the focal plane using PSF magnitudes of standard stars as a baseline.
Centroiding errors, extended objects, guiding errors, transparency variations and other uncertainties
in the corrections can lead to spectrophotometric errors above the measurement
uncertainty over the $3\deg$ field.
As explained in \citet{tremonti04a}, an analysis of magnitudes synthesized from SDSS galaxy spectra
showed an RMS dispersion of 5\% in ($g - r$)
and 3\% in ($r - i$) relative to the colors measured from $3\2pr$ fiber magnitudes from SDSS photometry.
At the bluest wavelengths (3800 \AA), the error was closer to 12\%.
Comparing stars with a PSF magnitude brighter than 19,
the SDSS spectrophotometry was biased 0.02 magnitudes brighter
with an RMS dispersion of 0.05 magnitudes in $r$ \citep[DR6;][]{adelman-mccarthy08a}.
Similarly, the ($g-r$) colors showed a bias of 0.02 magnitudes with 0.05 magnitude dispersion
while the ($r-i$) colors showed a bias of -0.01 magnitudes with 0.03 magnitude dispersion.

\begin{figure*}[h]
\begin{center}
\centerline{
  \includegraphics[scale=0.45]{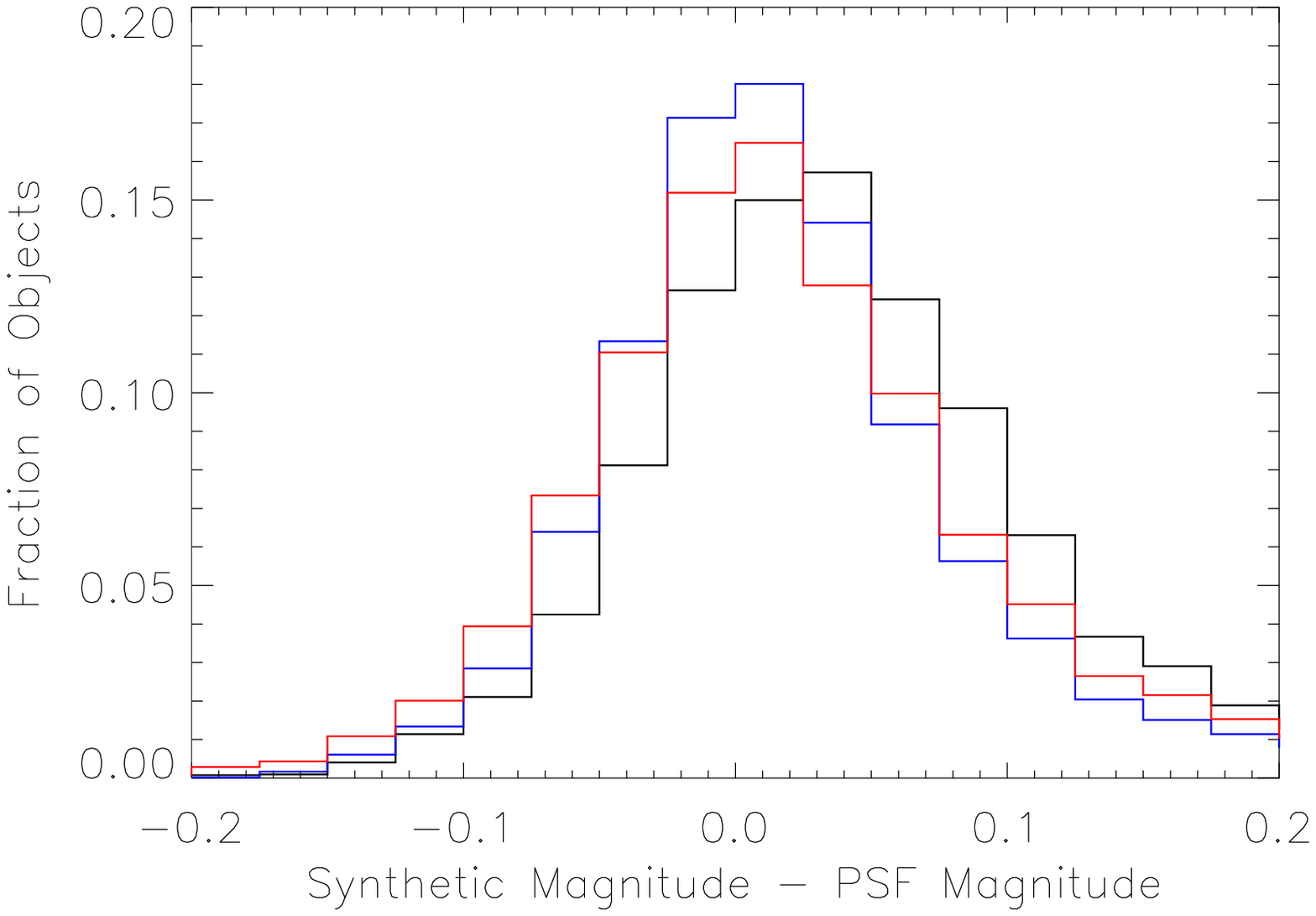}
  \includegraphics[scale=0.45]{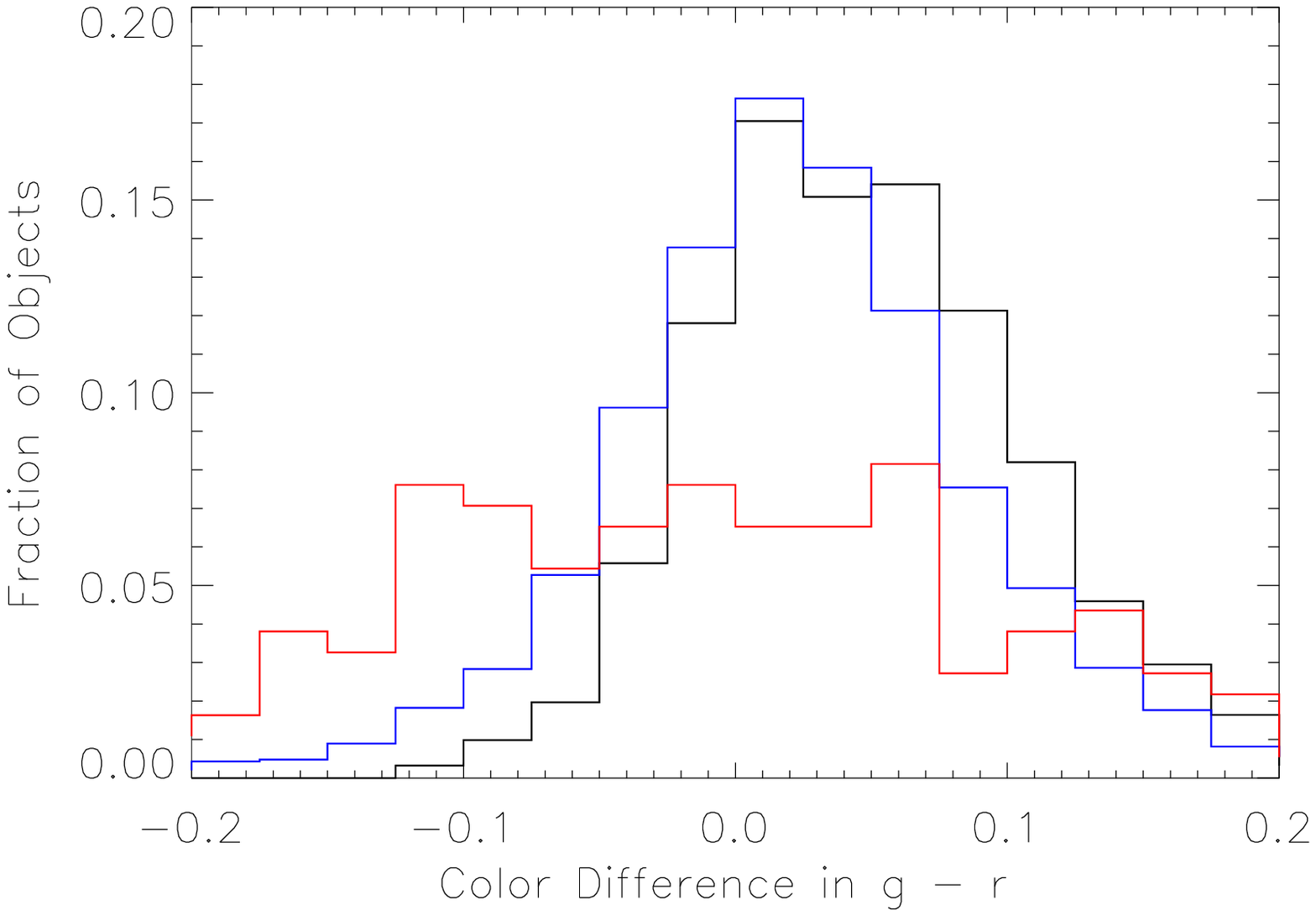}
}
\centerline{
  \includegraphics[scale=0.45]{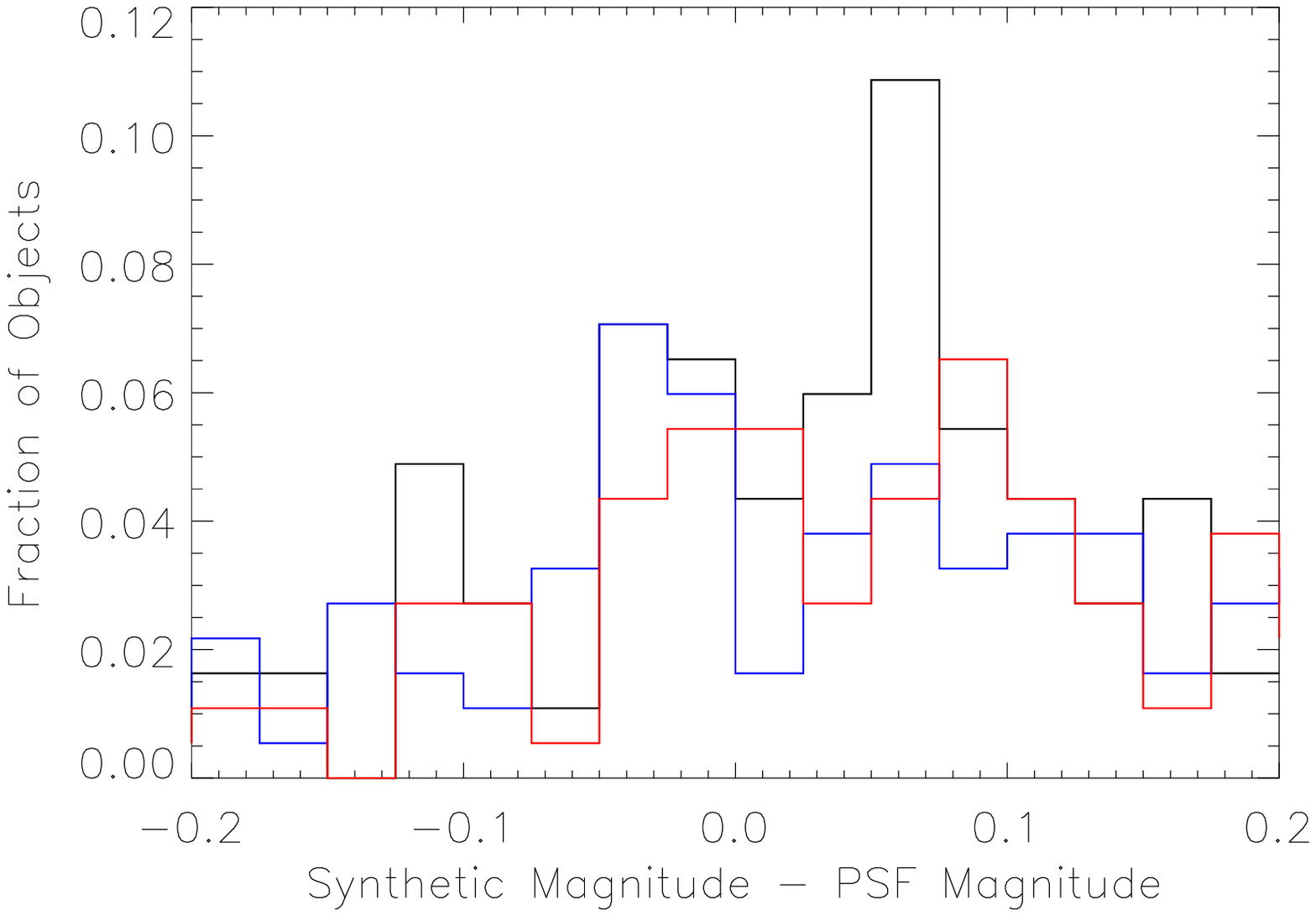}
  \includegraphics[scale=0.45]{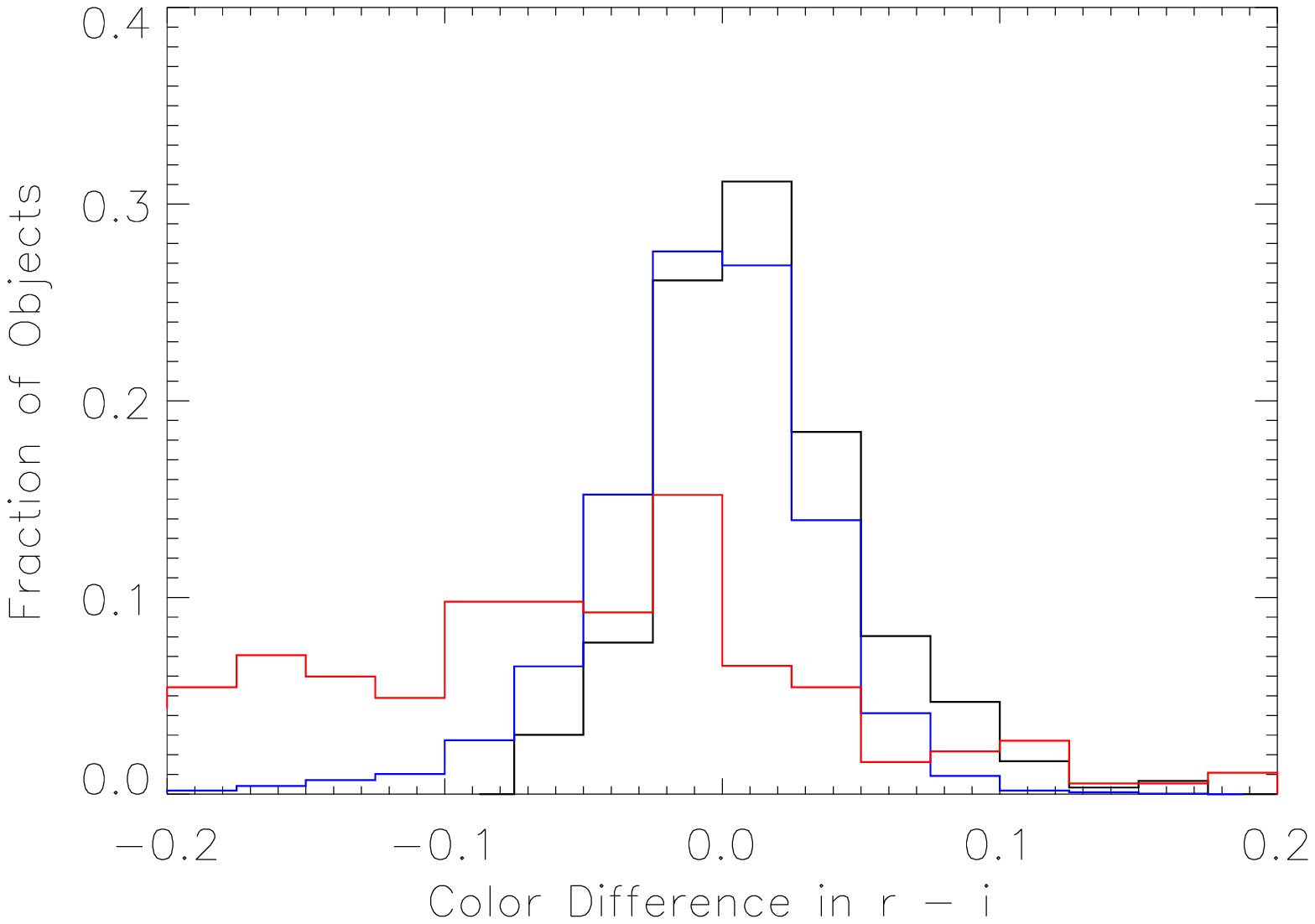}
}
\end{center}
\caption[Fluxing Errors]{\label{fig:fluxerror}
The difference between the synthetic photometry computed from each
spectrum and the measured photometry from the SDSS imaging data.
{\bf Left:  }Histogram of the spectrophotometric offsets
for stars. 
The black line shows $g_{\rm synthetic} - g_{\rm PSF}$ for the standard stars (top)
and for a sample of stars that appeared as contaminants in the {\bf CORE} and
{\bf BONUS} quasar samples (bottom). Similarly, the blue lines show the offsets for the $r$
filter while the red lines show the offsets for the $i$ filter.
{\bf Right:  }Histogram of the color differences
$(g-r)_{\rm synthetic} - (g-r)_{\rm PSF}$ shown in the top panel
and $r-i$ in the bottom panel.
Standard stars are presented as the blue line using PSF magnitudes,
galaxies as the black line (fiber2 magnitudes),
and stellar contaminants in the quasar sample as the red line (PSF magnitudes).
}
\end{figure*}

We performed a similar analysis on the flux calibration of BOSS spectra
using objects with $15<g_{\rm fib2}<19$.
As shown in Figure~\ref{fig:fluxerror}, we find slightly larger bias and RMS
dispersion in fluxing errors for stars and galaxies than was reported in SDSS.
The larger fluxing errors are not surprising because the BOSS fibers are smaller
in diameter and therefore more susceptible to guiding offsets.
For the standard stars
we find BOSS spectrophotometry to be on average $0.014$ magnitudes fainter
than the PSF photometry with an RMS dispersion of $0.058$ magnitudes in $r$.
BOSS spectrophotometry is 0.038 magnitudes fainter (0.068 magnitude dispersion)
than the PSF photometry in $g$.
The ($g-r$) colors are 0.022 magnitudes redder (0.063 magnitude dispersion)
while the ($r-i$) colors are 0.004 magnitudes bluer (0.035 magnitude dispersion).
For galaxies, we compare only colors, using the $2\2pr$ fiber magnitudes from SDSS imaging,
and find that ($g-r$) colors are 0.048 (0.058 magnitude dispersion)
and ($r-i$) colors are 0.013 magnitudes (0.035 magnitude dispersion) redder for the spectra.
As shown in the same figure, the offsets and dispersion of spectra
are much larger for objects targeted as 
quasars but confirmed to be stars.
On average, stellar contaminants in the quasar sample are $0.16$ magnitudes fainter in $r$,
with an RMS dispersion of $0.29$ magnitudes;
these objects are 0.038 (0.158 magnitude dispersion) and
0.070 (0.099 magnitude dispersion) magnitudes more blue in $g-r$ and $r-i$ colors respectively.

The application of washers and the offsets in plate position for the quasar fibers accounts for
the warping of the quasar target spectrophotometry relative to the standard stars.
The quasars have higher throughput in the blue relative to the standard stars at
a cost of reduced throughput in the red, and this difference is not included in the flux calibration.
The effect is exacerbated by the smaller fibers
(now comparable in diameter to the typical $1.5\2pr$ seeing at Apache Point)
relative to those used in SDSS.  

\newpage

\begin{figure*}[h]
\begin{center}
\centerline{
  \includegraphics[scale=0.45]{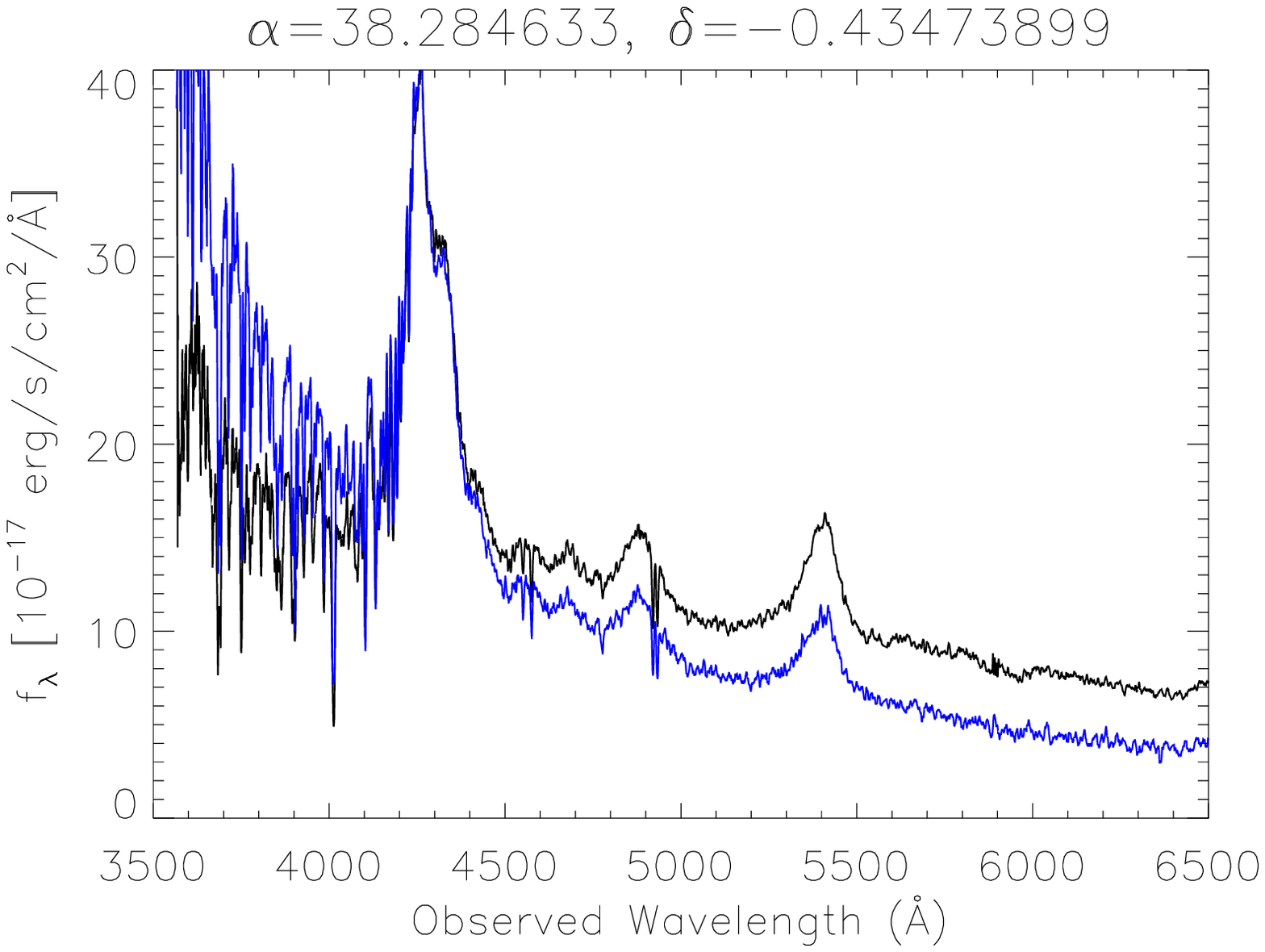}
  \includegraphics[scale=0.45]{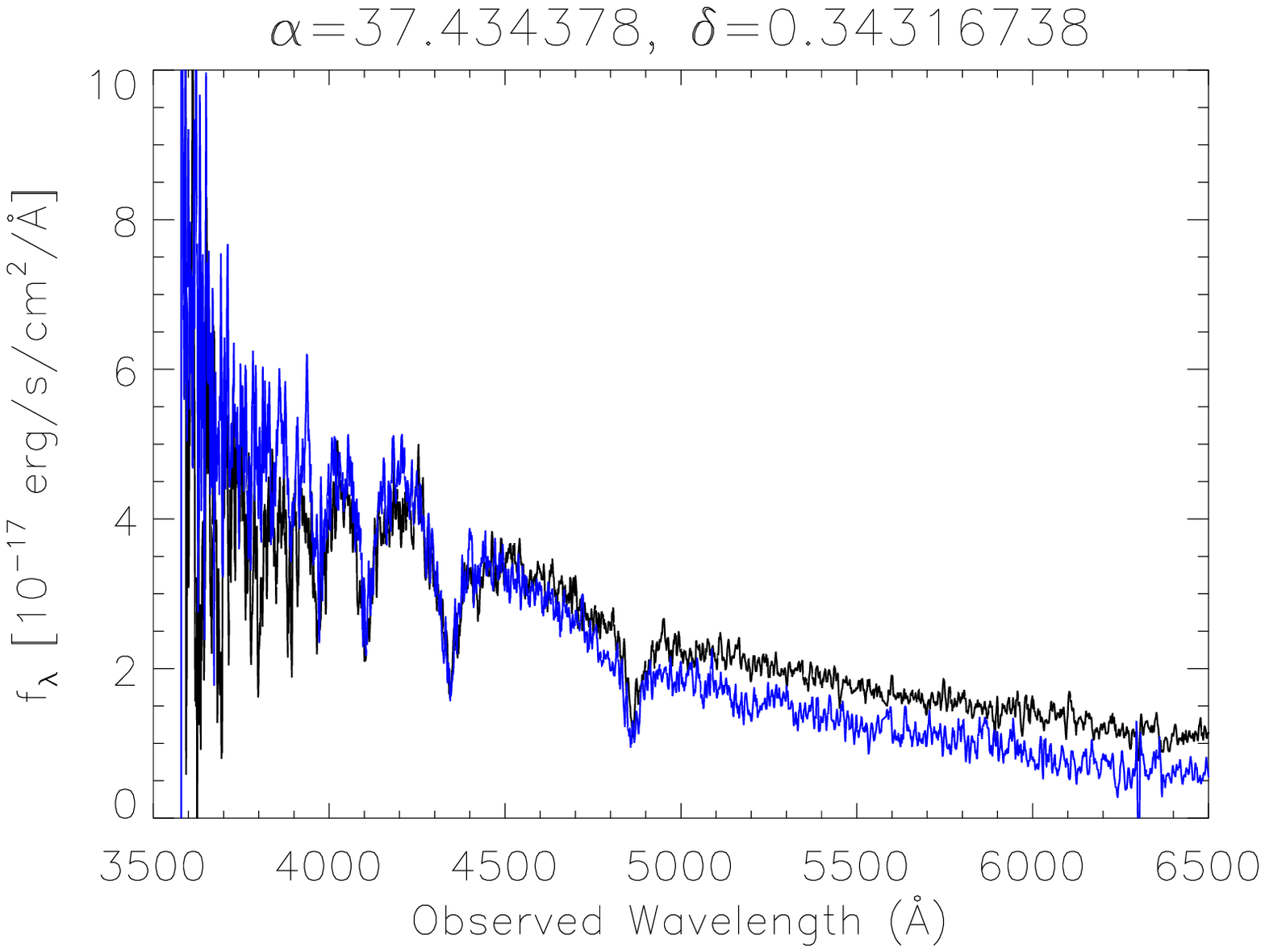}
}
\end{center}
\caption[spectrophotometry]{\label{fig:cal}
Quasar target spectra smoothed with a five-pixel median boxcar filter,
covering the wavelength range 3500 \AA\ through 6500 \AA\
taken at different airmass, demonstrating the effects of atmospheric
differential refraction and other guiding errors on spectrophotometry.
{\bf Left:  }Spectrum of a $g_{\rm fib2}=19.48$ quasar at $z=2.49$ observed at an airmass of 1.2 (black)
and again at an airmass of 1.4 (blue).
{\bf Right:  }Spectrum of a white dwarf with $g_{\rm fib2}=21.53$ on the same observations
with the same color pattern.
}
\end{figure*}

The net effects of these sources of error can be demonstrated in the
spectrophotometry of plates 3615
and 3647, both located on the celestial equator at $\alpha=37\deg$.
The two plates contain identical targets but were drilled for different airmass
and were observed a total of nine times in the first two years to monitor the system.
An example of the spectrophotometric calibration accuracy
from these two plates is shown in Figure~\ref{fig:cal}.
The left panel displays a quasar observed twice at different airmass.
Note the clear trend toward bluer synthetic photometry in
the spectrum that was obtained at larger airmass;
ADR causes the red light to be offset from the center of the
fiber for the quasar target in an opposite sense to the ADR effect
on the spectrophotometric star fibers (optimized for throughput at 5400 \AA).
Also shown in the figure is the spectrum
of a star that was targeted as a likely quasar and therefore positioned in the
offset quasar focal plane.  The stellar profile is known to
be constant over the period of 29 days between the two observations, yet
the spectra show the same trend toward bluer colors.

An understanding of spectrophotometric accuracy is critical for studies of cosmology,
galaxy evolution, and quasar physics.
\citet{yan11a} evaluates small wavelength-scale residuals in the flux calibration that can contaminate
weak emission and absorption features in the spectra. Using SDSS spectra, he estimates that the
wavelength-dependent relative flux calibration is accurate at the 1-2\% level.
A similar analysis has not been published with BOSS spectra, but the similarities in the
data reduction pipeline make it likely that the BOSS spectrophotometry has similar accuracy on small
wavelength scales.
The broad band spectrophotometry accuracy described here is biased at less than 5\% for the
standard stars and galaxy targets, with a comparable amount of scatter.
This precision does not seem to significantly reduce the galaxy redshift efficiency;
as was shown in \S\ref{subsec:LRG},
the $i_{\rm fib2}<21.5$ CMASS spectra are successfully classified for more than 94\% of the objects.
The estimates of color scatter and bias described in this section should be taken into consideration
for science applications that require precise spectrophotometry;
repeat spectra can be used to verify the reproducibility of measurements.
In general, galaxy evolution studies should use the $2\2pr$ fiber magnitudes
for comparison of spectrophotometric color to photometry, as the central region of extended galaxies
contributes the majority of light to the BOSS spectra.
The intrinsic quasar continuum must be estimated before \lya\ forest analysis
and is highly susceptible to spectrophotometric errors because of the fiber offsets.
The \lya\ analysis of \citet{slosar11a} effectively treats the broadband
power introduced by spectrophotometric errors as a nuisance parameter
in continuum fitting.
The continuum fitting procedure outlined in \citet{lee12a} has been applied
to the BOSS data, and will be described in a public release of model quasar continuum \citep{lee12b}.

\clearpage

\subsection{Galaxy Spectroscopic Data Quality}\label{subsec:galaxydata}

\citet{thomas12a} perform a spectroscopic analysis of the 492,450 galaxy spectra
that are part of DR9.
They show that the typical signal-to-noise ratio of BOSS spectra
is sufficient to measure simple dynamical quantities such as stellar velocity dispersion
for individual objects.
The typical error in the velocity dispersion measurement is 14\%, and 93\%
of BOSS galaxies have velocity dispersions with an accuracy better than 30\%.
\citet{thomas12a} further show that emission lines can be measured on BOSS spectra,
but the majority of BOSS galaxies lack detectable emission lines, as is to be expected
because of the target selection design toward massive galaxies.

In order to independently assess 
the performance of BOSS idlspec2d in determining galaxy redshifts, the code
``RUNZ''\footnote{Maintained by Scott Croom} was run on all extracted galaxy spectra.
Comparisons between the RUNZ and idlspec2d redshifts gave important
indications of common failure modes in the idlspec2d redshifts at the beginning of the survey.
The results from RUNZ will be made available with DR9.

RUNZ was originally developed for 2dFGRS \citep{colless01a} and has since been
optimized for several LRG redshift surveys such as the
2SLAQ Survey \citep{cannon06a} and the AAOmega UKIDSS SDSS (AUS) LRG Survey (Wake et al. in prep).
The software has also been used to determine redshifts for emission line galaxies
in the WiggleZ survey \citep{drinkwater10a} and for a broad range of galaxy types
in the Galaxy and Mass Assembly survey \citep[GAMA;][]{driver09a, driver11a}.
RUNZ provides two redshift estimates for each input spectrum; the first
by cross-correlating each spectrum with a series of galaxy and stellar
templates (with the emission lines masked) and the second by fitting
a Gaussian to detected emission lines and identifying
multiple matches at a single redshift.  An integer value ($q$) between one and five
is assigned to reflect the quality of the redshift.
RUNZ was modified for BOSS and tuned to assign quality values
and input templates based on the visual inspection
of one of the commissioning plates.

Selecting all repeat spectra, we identify the primary spectrum (also requiring $q>3$)
and test the reliability of the RUNZ quality flags and redshifts assigned to
the other spectra of the same object.
We attempted to tune the quality value $q=3$ to correspond to a 95\% reliability
and $q=4$ to correspond to a $>99\%$ reliability.
Table~\ref{tab:runzrepeats} shows the fraction of
repeat spectra pairs that yield the same redshifts, where agreement is
defined as a velocity difference $\delta v = (z1 - z2)/(1+z2) * c  < 500$ km s$^{-1}$.
Analogous statistics are reported in Table~\ref{tab:runzrepeats}
for the idlspec2d redshifts, where a reliable redshift is designated as
having {\bf ZWARNING} $= 0$.
The results agree with the findings of \citet{ross12b}, who report
that LOWZ and CMASS galaxies that were observed multiple times
produced idlspec2d redshifts with a deviation $\delta z/(1+z) < 0.001$ in 99.7\% of the
cases when the spectroscopic classification produced {\bf ZWARNING} $=0$.
Visual inspection of a subsample of those data confirmed
this result, indicating that the rate of catastrophic failures among the
galaxy targets is less than 1\%, as required to meet the BAO requirements.

\begin{deluxetable}{llrr}
\centering
\tablewidth{0pt}
\tabletypesize{\footnotesize}
\tablecaption{\label{tab:runzrepeats} Redshift Consistency of Repeat Spectra}
\tablehead{
\colhead{Condition} & \colhead{Condition} & \colhead{Number} & \colhead{Consistency}\\
\colhead{of Object 1} & \colhead{of Object 2} & \colhead{of Pairs} & \colhead{Fraction}
}
\startdata
$q \le 2$ & $q > 3$ &  448 & 0.58\\
$q = 3$ & $q > 3$ & 1298 & 0.93\\
$q > 3$ & $q > 3$ & 24075 & 0.996\\
$q \ge 3$ & $q > 3$ & 25373 & 0.992\\
ZWARN $= 0$ & ZWARN $= 0$ & 27304 & 0.996\\
ZWARN $> 0$ & ZWARN $= 0$  & 771 & 0.68 
\enddata
\end{deluxetable}

\begin{figure}[!h]
\begin{center}
\centerline{
  \includegraphics[scale=0.45]{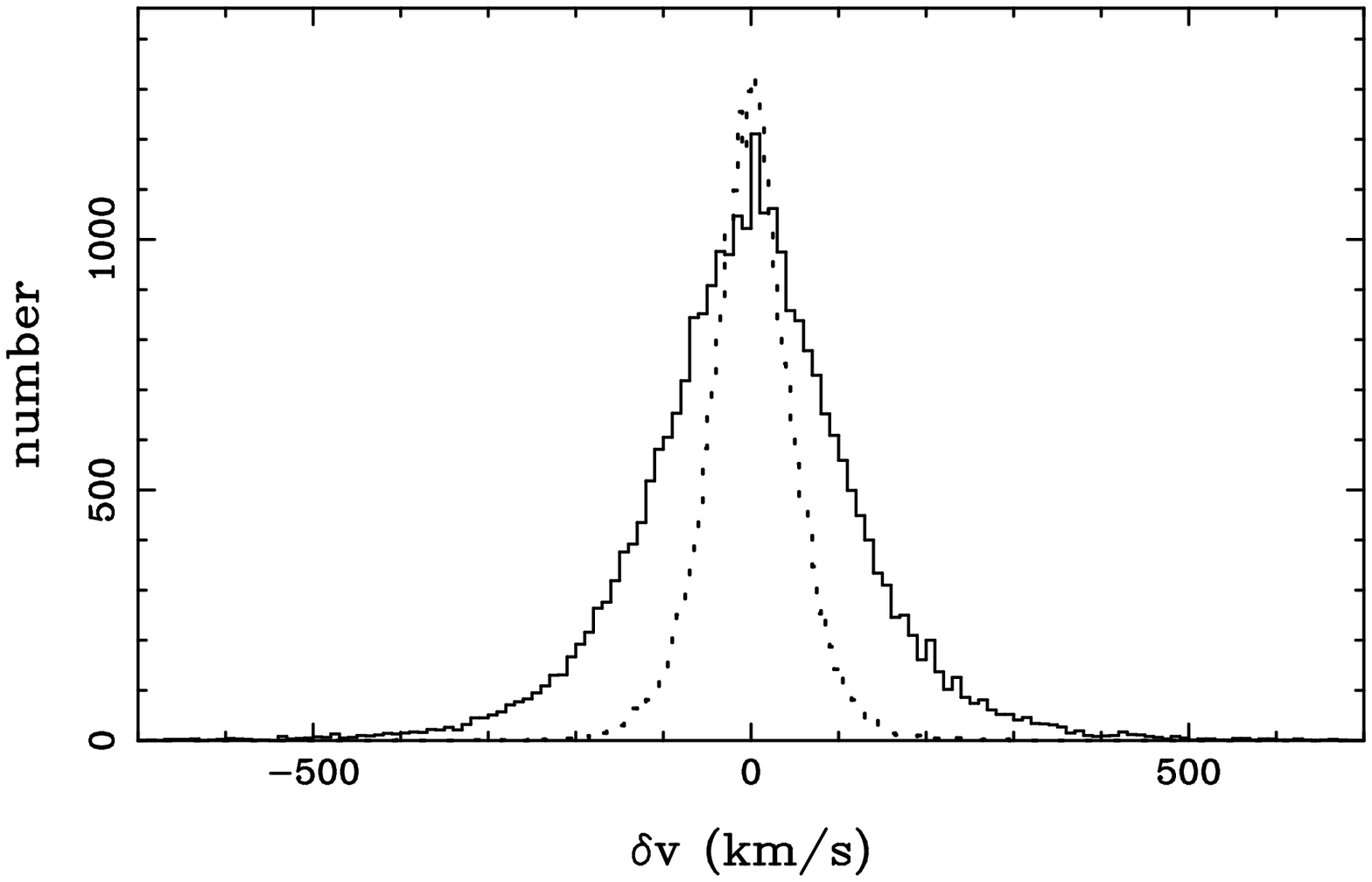}
}  
\centerline{
  \includegraphics[scale=0.45]{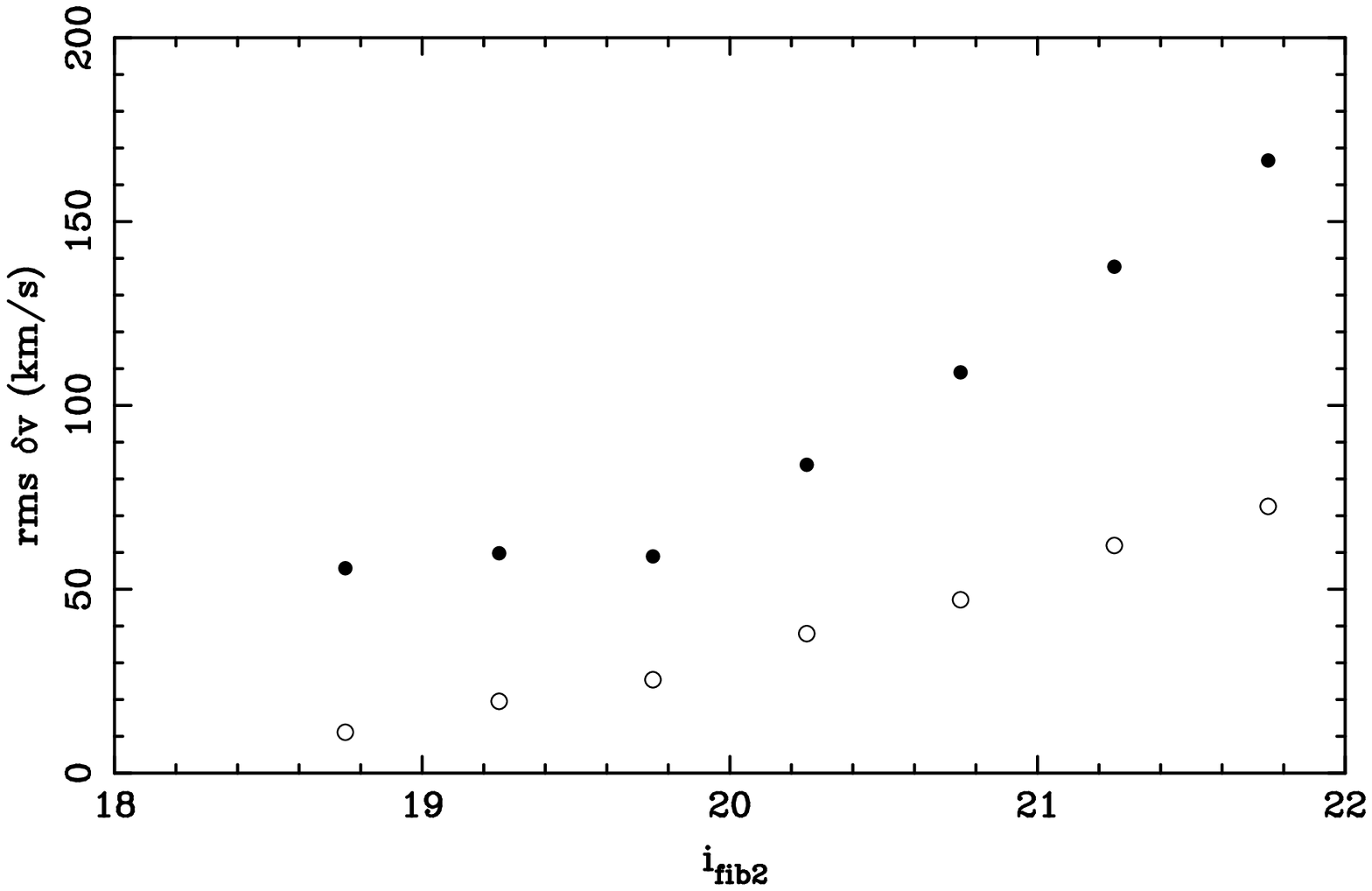}
}
\end{center}
\caption[delta v]{\label{fig:deltav}
Statistics of $\delta v$ in pairs of spectra for the same galaxy.
{\bf Top:}  Histogram of $\delta v$ for RUNZ (solid black) and for
idlspec2d (dashed).
{\bf Bottom:}  RMS width of distribution binned at half magnitude intervals
in $i_{\rm fib2}$.  RUNZ is presented as solid circles while idlspec2d is
presented as open circles.
}
\end{figure}

We estimated the redshift accuracy for RUNZ and for idlspec2d
by using the same repeat spectra and requiring all pairs to have $q>2$ (or {\bf ZWARNING} $= 0$).
The distribution of $\delta v$ for RUNZ and BOSS are shown in 
the bottom panel of Figure~\ref{fig:deltav}.
The idlspec2d redshifts are significantly more repeatable,
indicating a single-epoch RMS uncertainty of 38 km s$^{-1}$ compared to 81 km s$^{-1}$ for the
RUNZ redshifts (after removing pairs with $\delta v > 2000$ km s$^{-1}$).
The bottom panel of Figure~\ref{fig:deltav} shows the dependence of the RMS in
$\delta v$ on $i_{\rm fib2}$. The idlspec2d pipeline now performs better
at all magnitudes, continues to improve at bright magnitudes, and grows more
slowly with decreasing flux when compared to the RUNZ results.

\begin{figure*}[h]
\begin{center}
\centerline{
  \includegraphics[scale=0.45]{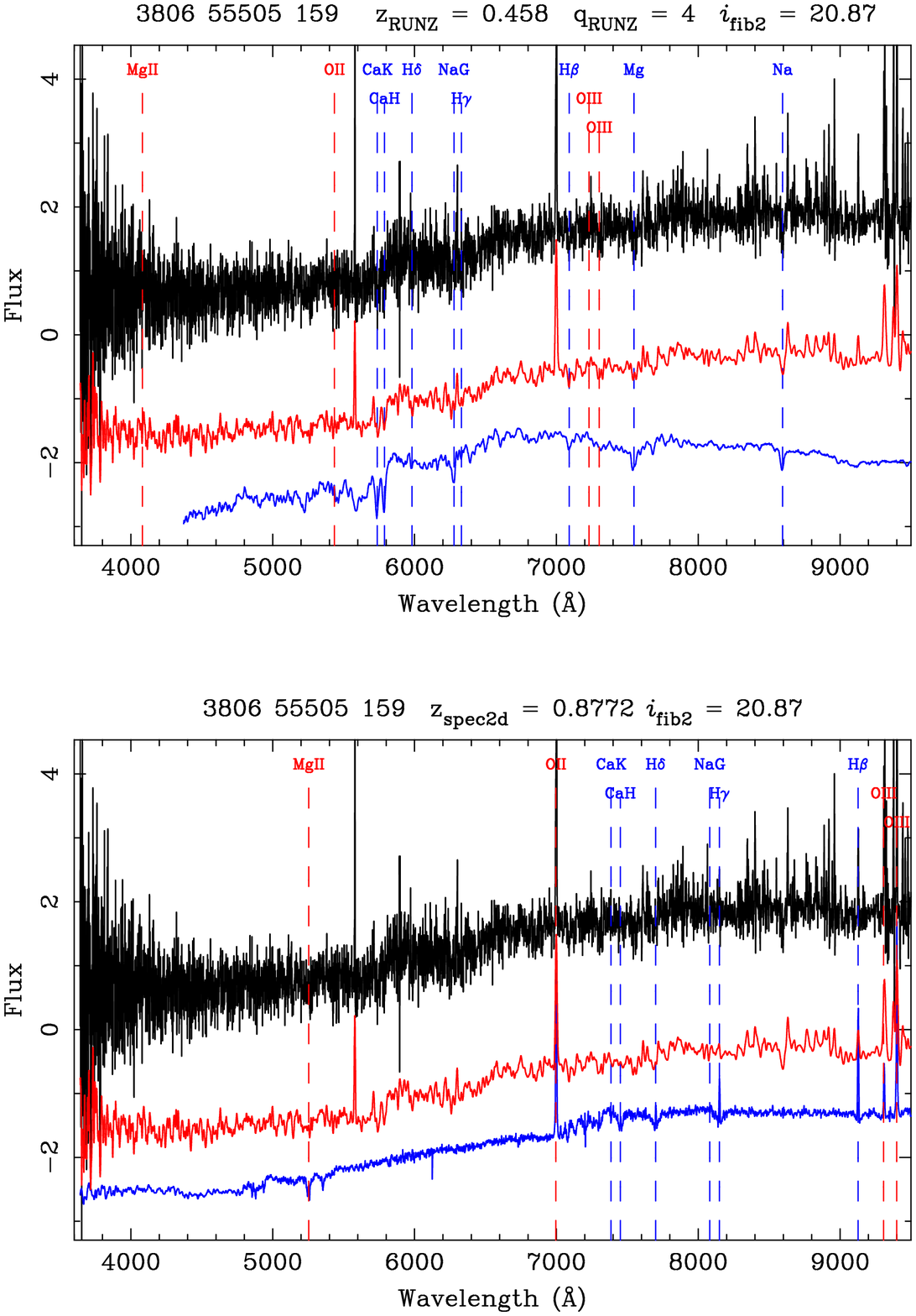}
  \includegraphics[scale=0.45]{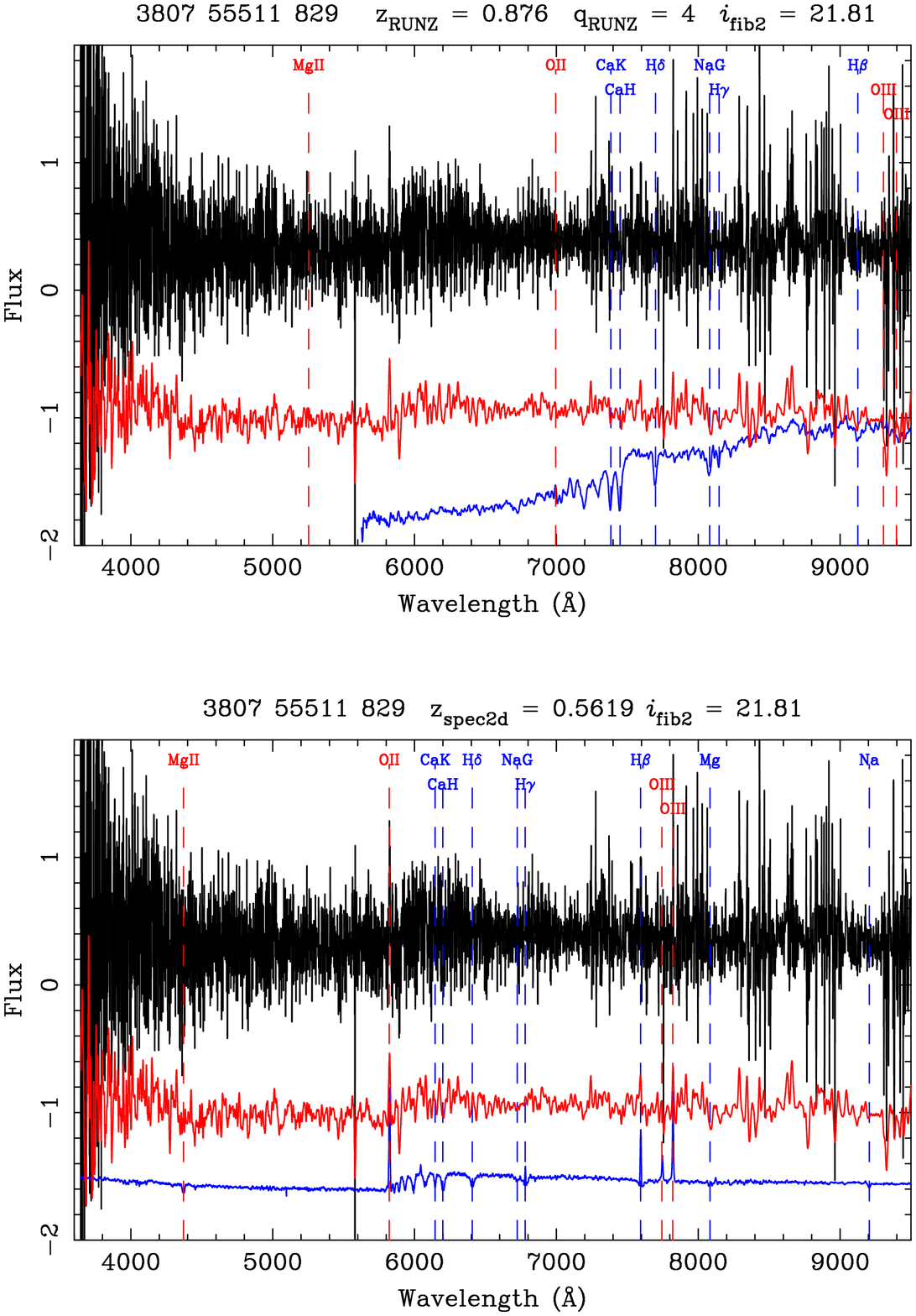}
}
\end{center}
\caption[delta v]{\label{fig:superpose}
Examples of spectra that produced discrepant redshifts between the
RUNZ analysis and the idlspec2d analysis due to a superposition of multiple objects.
The RUNZ fit is shown in the top panels and the idlspec2d fit in the bottom panels.
Three spectra are plotted in each panel, from top to bottom:  raw BOSS spectrum,
BOSS spectrum smoothed with a 7-pixel boxcar filter, and best-fit spectral template.
Typical line features are identified at the top of each panel by name,
flux density is presented in arbitrary units.  As before, the plate, MJD, fiber
combination is presented at the top of each panel.
{\bf Left:  }Example spectrum where the RUNZ redshift corresponds to a foreground
passive galaxy and the idlspec2d redshift corresponds to a background emission-line galaxy.
{\bf Right:  }Example spectrum where the RUNZ redshift corresponds to a background
passive galaxy and the idlspec2d redshift corresponds to a foreground emission-line galaxy.
}
\end{figure*}

\begin{figure*}[h]
\begin{center}
\centerline{
  \includegraphics[scale=0.45]{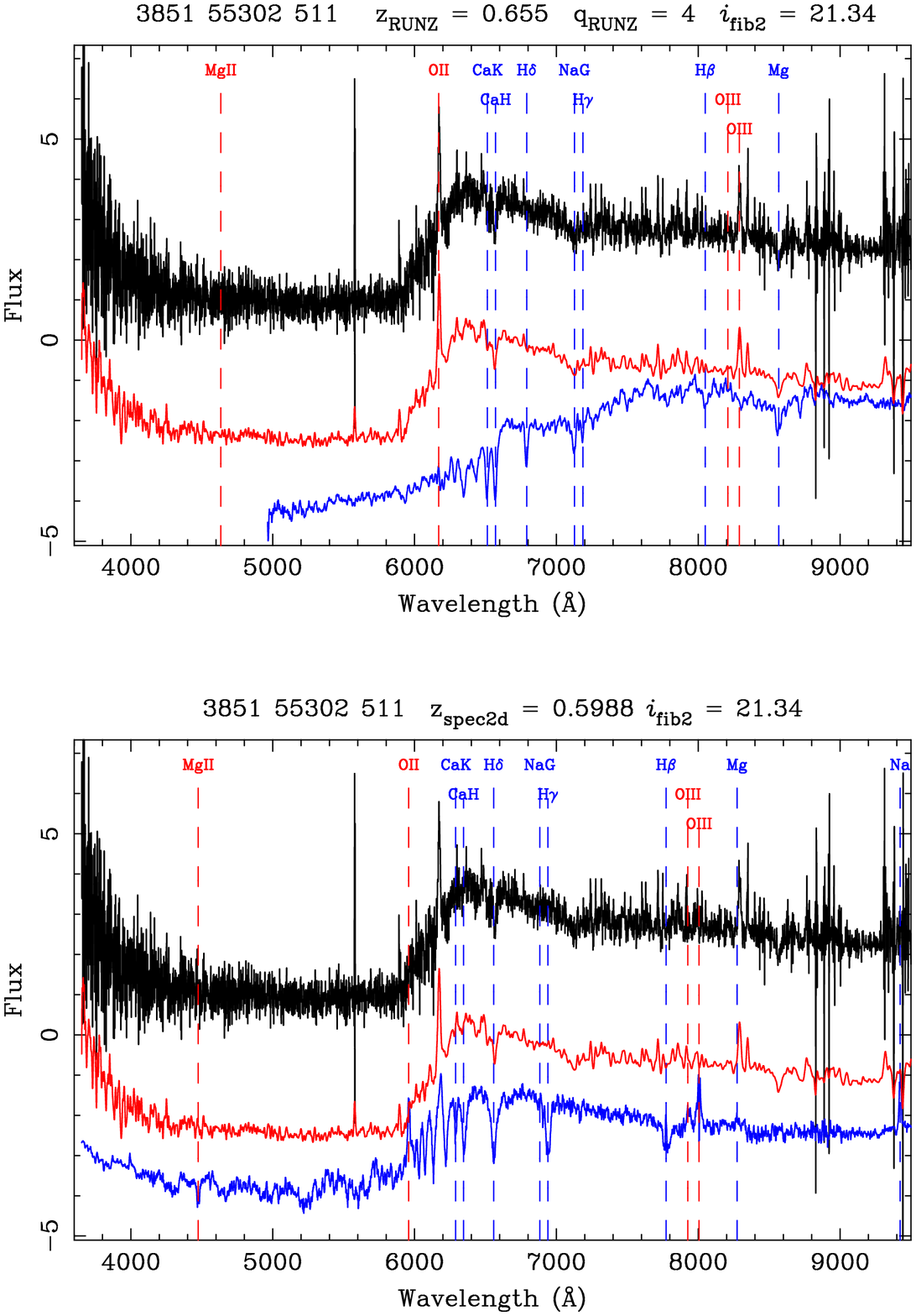}
  \includegraphics[scale=0.45]{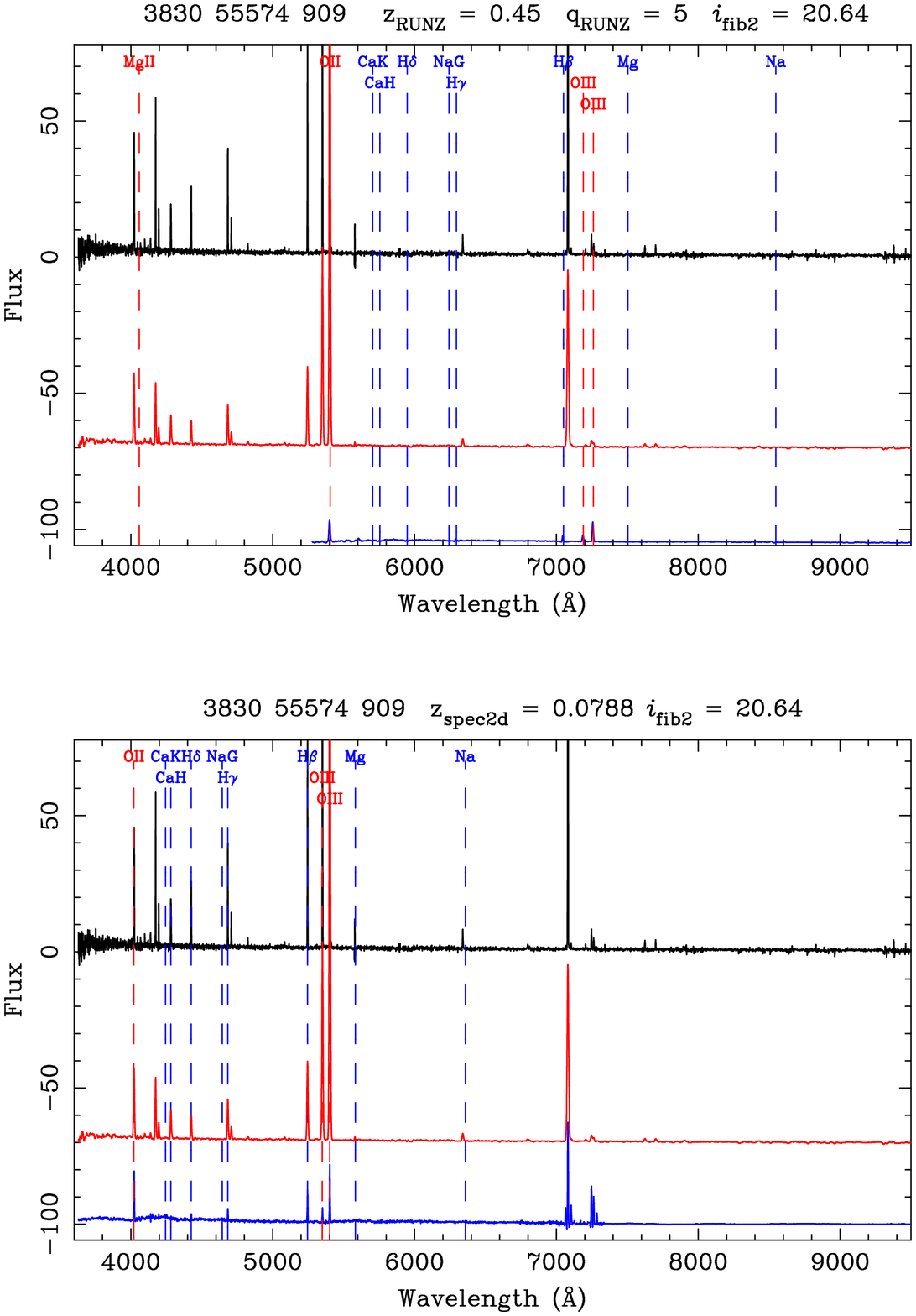}
}
\end{center}
\caption[delta v]{\label{fig:failures}
Examples of spectra that produced discrepant redshifts between the
RUNZ analysis and the idlspec2d analysis where one is incorrect.  The labels
are the same as those in the previous figure.
{\bf Left:  }Example spectrum where manual classification reveals the RUNZ redshift
to be correct and the idlspec2d redshift to be incorrect.  The spectral distortion around 6000 \AA\
is likely due to poor calibration in the dichroic region, a rare occurrence in BOSS spectra.
{\bf Right:  }Example spectrum where manual classification reveals the idlspec2d redshift
to be correct and the RUNZ redshift to be incorrect.
}
\end{figure*}

\clearpage

When comparing the results of RUNZ directly to the idlspec2d pipeline for
identical spectra, we find the RUNZ redshifts are lower by a median of 8 km s$^{-1}$.
The RMS scatter between the two algorithms is 72 km s$^{-1}$, consistent
with the larger RMS scatter observed in the repeat spectra of RUNZ redshifts.
Possible redshift failure modes are identified by comparing instances
where both idlspec2d and RUNZ indicate a correct redshift for a
galaxy but the redshifts differ significantly.
There are 311 such galaxies in DR9 where {\bf ZWARNING}= 0, $q > 3$ and 
$(z_{\rm RUNZ} - z_{\rm idlspec2d})/(1+z_{\rm idlspec2d}) * c   > 1000$ km s$^{-1}$.
Visual inspection reveals a number of causes for the failures.
In many cases, both redshifts are valid because a superposition of two different
objects at different redshifts leads to two correct results;
two such examples are presented in Figure~\ref{fig:superpose}.
In the case of a star-galaxy superposition,
the idlspec2d pipeline mildly favors a redshift assigned to the stars compared to RUNZ.
However, in the case that background emission-line objects are projected on
foreground galaxy light, idlspec2d tends to select
the background emission-line redshift whereas RUNZ selects the foreground redshift.
A genuine failure mode for idlspec2d appears in
instances of bad data, either where the relative normalization of the
red and blue spectra has failed or some broad band shape has been
introduced into the blue spectra.
An example of such a spectrum is shown in the left hand panel of Figure~\ref{fig:failures}.
RUNZ appears to be less affected by such problems, as it doesn't
fit the broad band shape of the spectra.
The most common failure for RUNZ occurs in strong emission-line galaxies.
These failures may result from inadequate templates combined with an
insufficient weight given to the emission-line redshift relative to the cross-correlation redshift
as a result of the LRG optimization of this version of RUNZ.

While RUNZ was used to help diagnose
BOSS galaxy redshifts early in the survey, we concentrated our subsequent software development 
entirely on the idlspec2d data reduction pipeline and redshift classification.
The algorithms employed for redshift determination of BOSS targets
are described in \citet{bolton12a}, including a more thorough description of templates
and redshift efficiency than what is found here.
Even though idlspec2d performs slightly better than RUNZ in terms of
completeness, reliability, and redshift repeatability,
RUNZ would probably perform better if it were fully optimized for the BOSS data.

\subsection{Quasar Classification}\label{subsec:qsoclassify}
Due to the variety of quasar spectral features and the complexity 
of automatic classification, members of the French Participation Group
supplemented the idlspec2d results with a visual inspection of the spectra.
The visual inspection provides a secure identification and
a reliable estimate of the redshift for each object in addition
to a characterization of other fundamental quasar properties.
In total, 189,018 spectra were inspected in the DR9 sample.
We provide a brief summary of the inspections; full details
are found in \citet{paris12a}.

All objects that were targeted as quasars or classified as quasars with
$z \geq 2$ by idlspec2d were visually inspected and added to the DR9 quasar catalog
\citep[DR9Q;][]{paris12a}, which is included with DR9.
The inspections include not only {\bf CORE} and {\bf BONUS} quasars, but also
objects from ancillary programs focused on quasars.
A fraction of objects in the galaxy sample were identified as quasars by idlspec2d
and visually inspected as well.
The catalog provides human classifications of the objects, refined redshift estimates and
emission line characteristics.
Peculiar spectral features that could affect any \lya\ forest analysis, such as damped
\lya\ (DLA) systems and broad absorption line (BAL) quasars, are first flagged in visual inspection.
DLA systems are then evaluated with an automated characterization to determine
column densities \citep{noterdaeme09a}.
Those results will be reported in a separate catalog \citep{noterdaeme12a}.
BAL quasars are processed for automatic estimates of the balnicity index
\citep[BI;][]{weymann91a} and absorption index \citep[AI;][]{hall02a}
of C IV troughs to quantify their strength.
Problems identified in visual inspection such as the presence of artificial breaks in the spectrum,
poor flux calibration or bad sky subtraction are flagged as well.

The visual inspection starts from the output of idlspec2d for both classification and redshift.
We find that fewer than 0.3\% of objects classified as quasars with {\bf ZWARNING} $=0$
by idlspec2d have misidentified the observed emission lines (typically at $z< 2$, where
\lya\ emission is not observable), leading to redshift errors $\Delta z > 0.1$.
About half of the objects classified as quasars by idlspec2d
have redshifts adjusted by less than $\Delta z > 0.1$
($\bar{\Delta z} <0.005$) after their visual inspection,
when the best fit idlspec2d template misses the position of the Mg II
emission line peak or when the maximum of the C IV emission line determines the redshift.
The precision of quasar redshifts are further refined using
a linear combination of carefully crafted principal components
fit to each spectrum \citep[see, e.g.,][]{paris11a}.
Indeed, known shifts between emission lines are intrinsically imprinted in the eigenvectors
and this method also takes into account the quasar-to-quasar variation.
This redshift estimate is also provided in addition to the idlspec2d results as part of the DR9Q.

About 12\% of the quasar targets in the {\bf CORE} and {\bf BONUS} samples are assigned a non-zero
{\bf ZWARNING} flag in the idlspec2d pipeline.
Visual inspection reveals that 13\% of these objects are truly quasars and about 7\% have $z > 2.15$.
Of those objects that are classified with {\bf ZWARNING} $=0$ as quasars in idlspec2d,
approximately 1\% are revealed to be stars.
Similarly, only 23 objects classified as stars with {\bf ZWARNING} $=0$
are shown to be quasars in visual inspection, only five of which are $z > 2.15$ quasars.
A summary of the classifications in DR9Q is provided in Table~\ref{tab:DR9Q}.

\begin{deluxetable}{lcc}
\centering
\tablewidth{0pt}
\tabletypesize{\footnotesize}
\tablecaption{\label{tab:DR9Q} Key Statistics from DR9Q}
\tablehead{\colhead{Classification} & \colhead{\# Visually Inspected} & \colhead{Fraction}}
\startdata
Quasar & 87,822 & 1.00  \\
Quasar with $z \geq 2.15$ & 61,933 & 0.70 \\
BAL Quasar & 7,532 & 0.086 \\
DLA Quasar & 7,492 & 0.085 
\enddata
\end{deluxetable}

\subsection{Lyman $\alpha$ Forest Measurements}\label{subsec:lya}

\begin{figure*}[!t]
\begin{center}
\centerline{
  \includegraphics[scale=0.45]{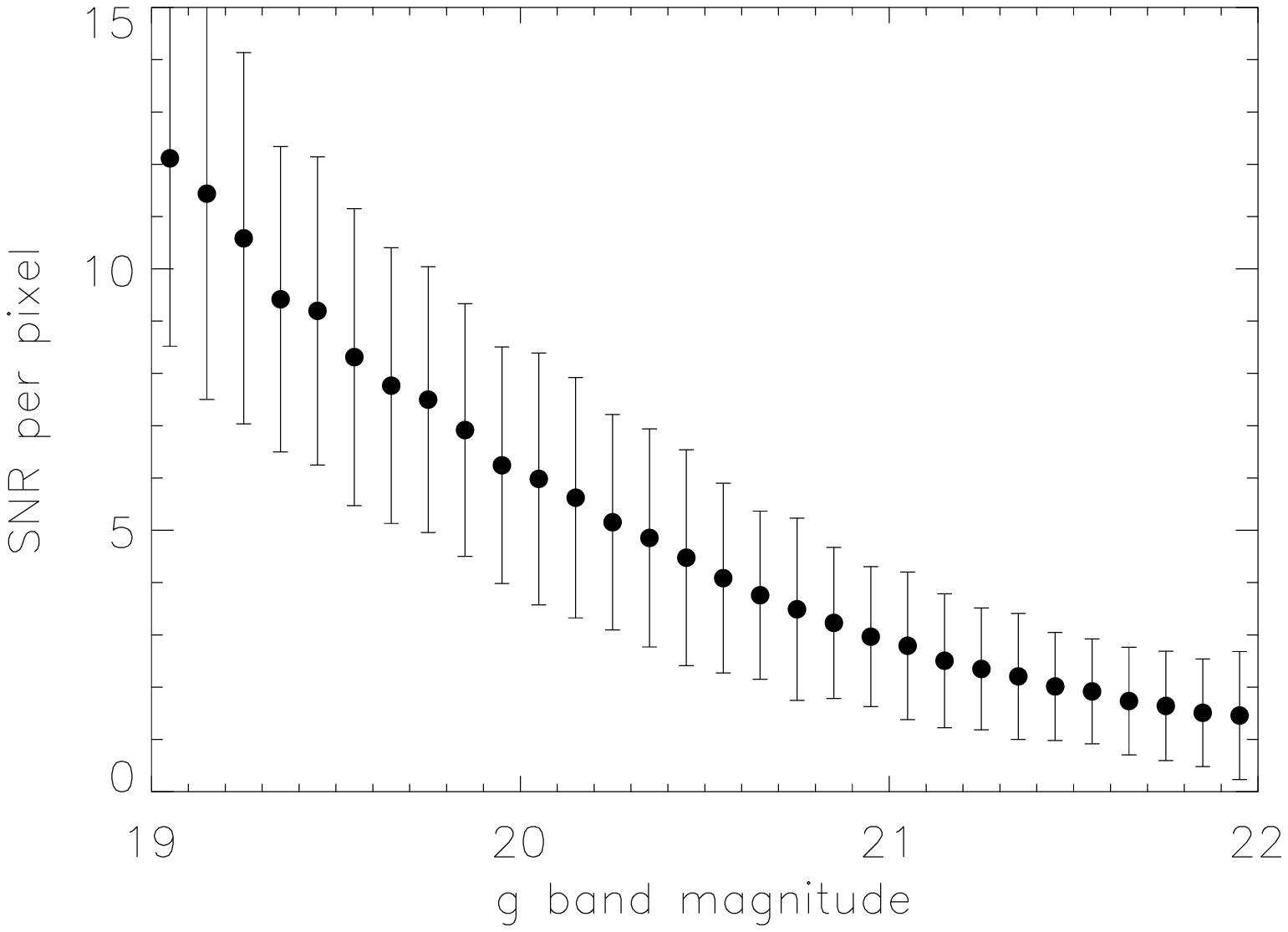}
  \includegraphics[scale=0.45]{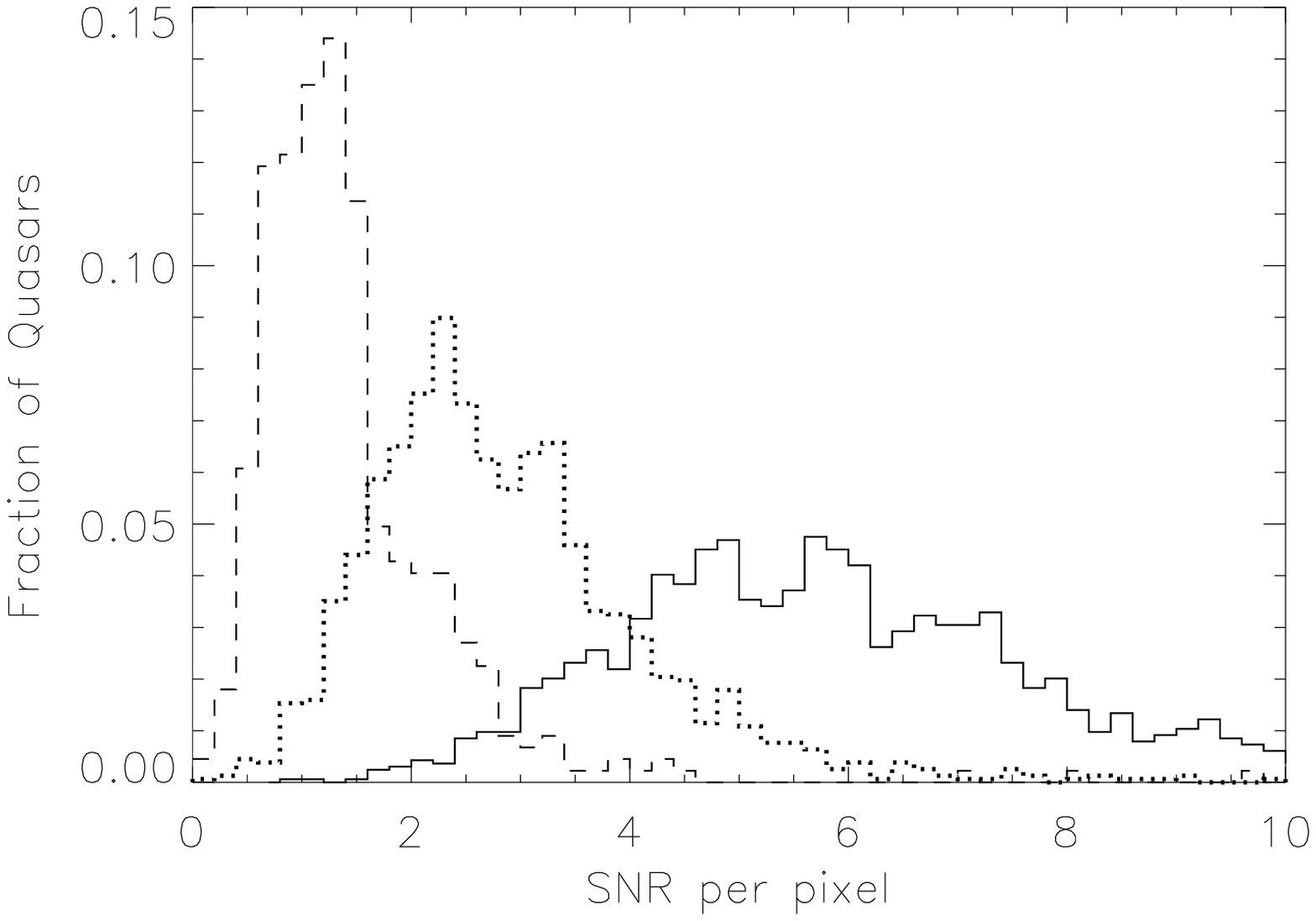}
}
\end{center}
\caption[Depth of lyman alpha spectra]{\label{fig:lyastats}
Statistics of the depth of the spectra in the \lya\ forest region
1041 \AA\ to 1185 \AA\ (rest frame).  Only quasars at $z>2.5$ are included,
so that the entire \lya\ forest lies within BOSS wavelength coverage.
On average, the sampling is bit more than 1 \AA\ pixel$^{-1}$ in this subsample.
{\bf Left:  }Mean SNR per pixel as a function of extinction-corrected $g_{\rm PSF}$ magnitude.
The error bars represent the standard deviation of all quasars in each
bin of width 0.1 magnitudes.
{\bf Right:  }Distribution of the SNR per pixel.  The solid line
is taken from a sample of quasars near the bright end ($g_{\rm PSF} \sim 20$) of the sample.
The dotted line represents a typical quasar ($g_{\rm PSF} \sim 21$).
The dashed line shows quasars near the faint magnitude limit $g_{\rm PSF} \sim 22.0$
of the {\bf CORE} and {\bf BONUS} target selection algorithms.
}
\end{figure*}

As described in \S\ref{subsec:qso}, the exposure depths over the full survey
are sufficient to ensure a nearly 100\% redshift completeness among true quasars.
\citet{mcquinn11a} also address the question of a survey's sensitivity
in the \lya\ forest using a metric $\nu_n$.
The parameter $\nu_n$ ranges from zero to one and represents the effective value
of each quasar to constraining the 3D power spectrum
depending on the redshift and SNR in the \lya\ forest region.
Assuming the quasar redshift and luminosity distribution of \citet{hopkins06a},
they find that $\nu_n$ roughly doubles as SNR per \AA\
increases from one to two for a quasar at $z \sim 3$.
Similar to \citet{font-ribera12a}, they argue for a survey that
favors area over depth, citing specifically that the improvement in sensitivity
is marginal once the noise exceeds SNR$= 2$ per \AA\ for the faintest quasars.
The specific \lya\ data quality modeled in these projections lends themselves to a direct
comparison to the BOSS data obtained in the first two years of the survey.

Using the DR9 quasar sample, we evaluate the depth of the spectra
in the \lya\ forest for comparison to the \citet{mcquinn11a} projections.
Using only objects that were classified as quasars by idlspec2d
with {\bf ZWARNING}$=0$, we compute the SNR per coadded BOSS pixel between 
1041 \AA\ and 1185 \AA\ in the rest frame of each quasar.
In the co-added spectra, the sampling at 4300 \AA\ corresponds to roughly
1 \AA\ per pixel, leading to a sampling that is easily compared to the
projections of \citet{mcquinn11a}.  The sampling of the \lya\ forest is somewhat higher for
quasars at $z=2.15$ and slightly lower for quasars at $z=3.5$.
The SNR statistics of the \lya\ forest pixels are shown in Figure~\ref{fig:lyastats}.
The typical quasar with $g_{\rm PSF} < 21.5$ exceeds SNR$= 2$ per pixel while
the fainter quasars exceed SNR$= 1$ on average, leading to 
values $\nu_n=0.5$ and $\nu_n=0.2$ at $z=2.6$ for the two samples respectively.
The target density and survey depth results in a SNR
that is sufficient for \lya\ constraints on the BAO peak.

\section{Conclusion}\label{sec:conclusion}

\begin{figure*}[!t]
\begin{center}
\centerline{
  \includegraphics[scale=0.33]{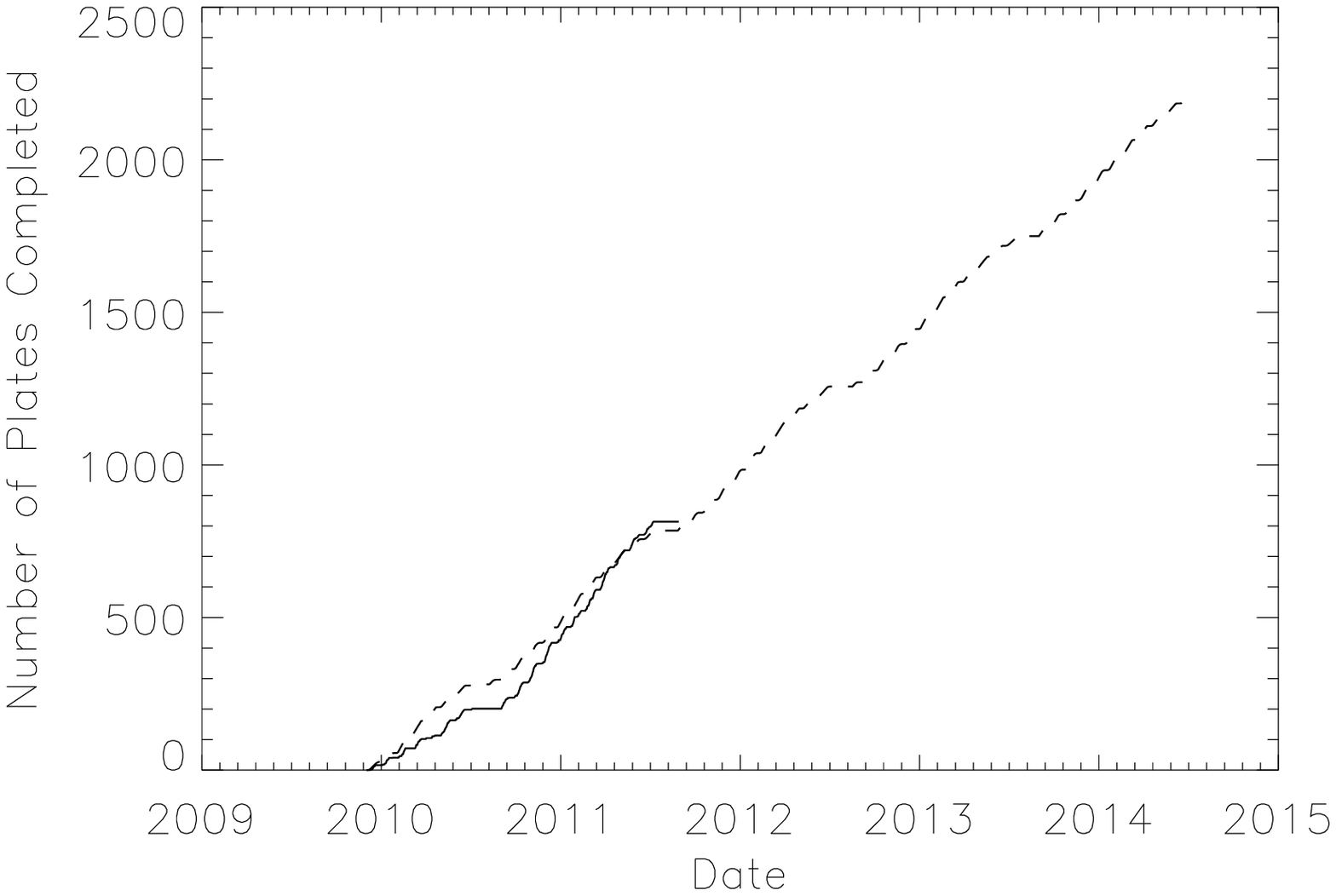}
  \includegraphics[scale=0.33]{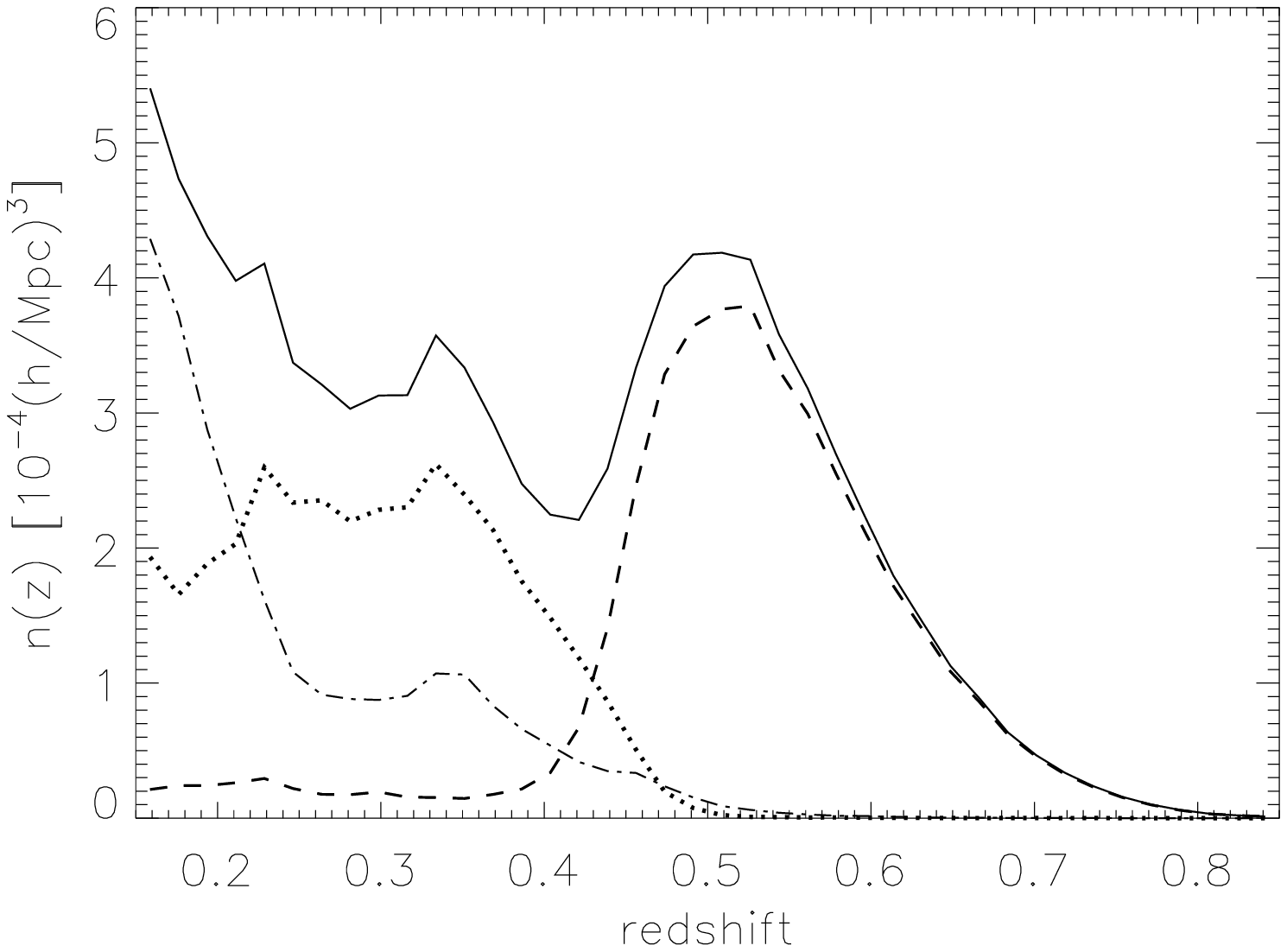}
  \includegraphics[scale=0.33]{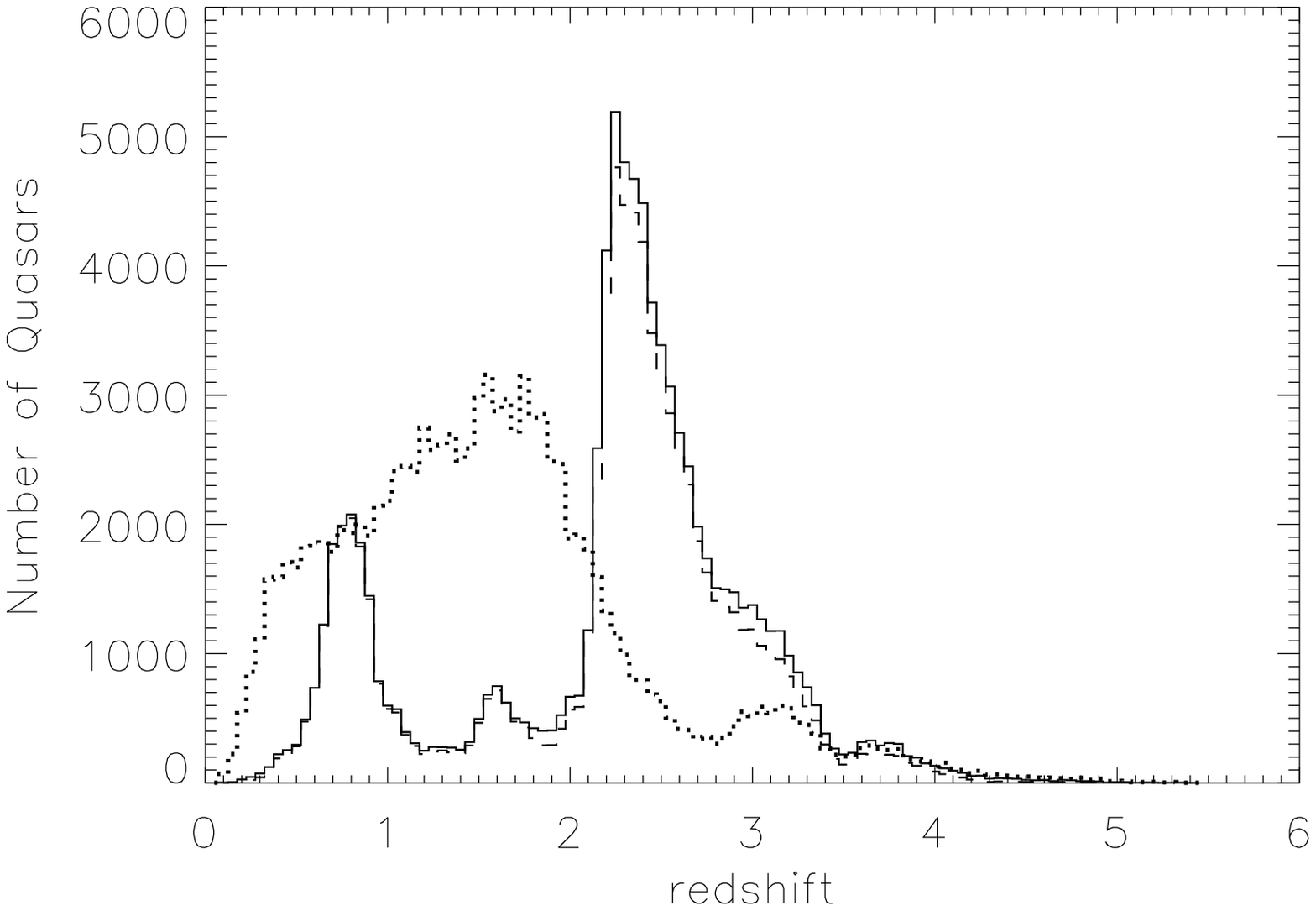}
}
\end{center}
\caption[Histogram]{\label{fig:progress}
{\bf Left:  }Progress of the survey at the time of DR9.  The solid curve represents
the number of unique plates completed as a function of time in the first two years of BOSS.
The dashed curve represents the projected rate required to complete the baseline survey of 2208 plates.
The yearly ``pauses'' are the result of the summer shutdown for telescope maintenance during the
New Mexico monsoon season.
The plate completion thresholds were changed in the summer of 2010 from SNR$_R^2 = 26$
to SNR$_R^2 = 22$ and from SNR$_B^2 = 16$ to SNR$_B^2 = 10$.
{\bf Center:  }Average number density as a function of redshift for the SDSS (dash-dotted),
LOWZ (dotted), and CMASS DR9 galaxies with {\bf ZWARNING\_NOQSO}$=0$.
{\bf Right:  }Redshift distribution of DR7 (dotted), DR9 (solid), and unique DR9 (dashed)
quasars.
}
\end{figure*}

\begin{deluxetable*}{lccccc}[!t]
\centering
\tablewidth{0pt}
\tabletypesize{\footnotesize}
\tablecaption{\label{tab:dr9stats} DR9 Redshift Determination for Main Targets}
\tablehead{\colhead{Sample} & \colhead{Unique Targets} & \colhead{Repeat Observations} & \colhead{Failed Redshifts} & \colhead{Purity} & \colhead{Contaminating Objects}  }\\
\startdata
LOWZ & 103,729 & 7,646 & 119 & 102,890 & 720 \\
CMASS  & 324,198 & 29,493 & 4167 & 309,307 &  10724  \\
Quasar & 154,433 & 11,601 & 31945 & 79,570  &  42918   
\enddata
\end{deluxetable*}

The data from the 2009-2010 and the 2010-2011
BOSS observing seasons are included in DR9.
Observations followed a pace that was close to the survey projections;
236 unique plates were completed in first year compared to an expectation of
331.  While this rate was about 30\% behind the required
rate to complete the survey, the slow start can largely be explained by the deeper
exposures that were acquired.
The SNR$^2$ thresholds were reduced by 37.5\% on the blue cameras and 15\% on the red cameras
after that first year;
577 unique plates were completed in the second year, compared
to an expectation of 454.  The rate of plate completion benefited from better than average
weather in the spring of 2011.
In total, 813 unique plates have been completed, 3.5\% ahead of the projected progress
to complete the survey by the summer of 2014.
As reported in \citet{anderson12a}, these plates account for $3275\, \sqdeg$ of unique sky coverage,
approximately 35\% of the full footprint.
The rate of plate completion as a function of time during the
first two years of BOSS is shown in Figure~\ref{fig:progress}.

The plates completed as a function of equatorial coordinates are presented in
Figure~\ref{fig:footprint}.
This figure reveals a high rate of completion in the
tGC and a lower rate in the SGC.
The difference between the NGC and SGC is due to the unusually good weather in
the Spring of 2011 and the fact that commissioning occurred during most
of the time that the SGC was visible in 2009.
The difference is more acute when comparing the $110\deg < \alpha <130\deg$ 
region of the NGC to the $\alpha>330\deg$ region of the SGC.
As explained in \S\ref{subsec:hadesign}, the Galactic plane crosses at
LST $\sim 4.6$ hours, and many of those NGC fields must be observed near that time.
As shown in Figure~\ref{fig:projections}, the shorter nights and New Mexico monsoon season
result in a smaller amount of observing time available to observe those SGC plates.
Looking beyond the second year and assuming average weather, we 
predict that we will complete the NGC region of the survey in less than
the allotted time, but will likely need better than average weather to
finish the $\alpha>330\deg$ region of the SGC.

The DR9 sample is the first sample of BOSS spectra released to the public
and is the data that define the initial sample for the BOSS galaxy and \lya\ BAO analyses
described below.
DR9 contains 324,198 unique CMASS targets and 29,493 repeat observations
from the CMASS sample.
Exceeding survey requirements, the spectra from 98.7\% of the CMASS targets 
produced {\bf ZWARNING\_NOQSO} $=0$ in at least one of the observations,
and 95.4\% were confirmed as galaxies.
Contamination from stars accounts for the 3.3\% of successfully classified objects
that were not galaxies.
As expected, the statistics from the LOWZ sample are even better;
103,729 unique targets (7,646 repeat observations)
produced a 99.9\% successful rate of classification;
99.2\% of LOWZ targets were successfully classified as galaxies.

There were 154,433 unique quasar targets from the main sample, with 11,601
repeat observations.  
The idlspec2d pipeline classified 79.2\% of these objects successfully and 
determined 51.5\% of the 154,433 objects to be quasars.  The numbers are similar to the
numbers found in manual classification described in \S\ref{subsec:qsoclassify},
with minor differences explained by the inclusion of ancillary programs in the
manual inspections and occasional disagreement between the two techniques.
A summary of the DR9 statistics is found in Table~\ref{tab:dr9stats}
and the redshift distribution for all successfully classified main galaxies and quasars is shown
in Figure~\ref{fig:progress}.

\subsection{BOSS in the Context of Previous SDSS Spectroscopy}

As discussed in \S\ref{sec:intro}, the improvements to the BOSS spectrograph
and increased cosmological volume of BOSS produce a new
sample that is both complementary and a significant expansion beyond SDSS.
The SDSS-I and II surveys obtained roughly 1.8 million spectra of galaxies, stars, and quasars
\citep[DR8: ][]{aihara11a}.
There are now about as many galaxies in the BOSS LOWZ sample as in
the LRG sample used to derive the BAO constraints from SDSS,
while the sample of CMASS galaxies represents a probe of an entirely new cosmological volume.
The SDSS sample consists of more than 100,000
quasar spectra, with a median redshift of $z\sim 1.5$ and absolute
magnitudes M$_i < -22.0$ \citep{schneider10a}.
Just over 19,000 of these quasars lie beyond $z=2.15$;
the new combined sample of \lya\ quasars
is already a factor 3.7 larger than what was available with SDSS.

\begin{figure*}[h]
\begin{center}
\centerline{
  \includegraphics[scale=0.4]{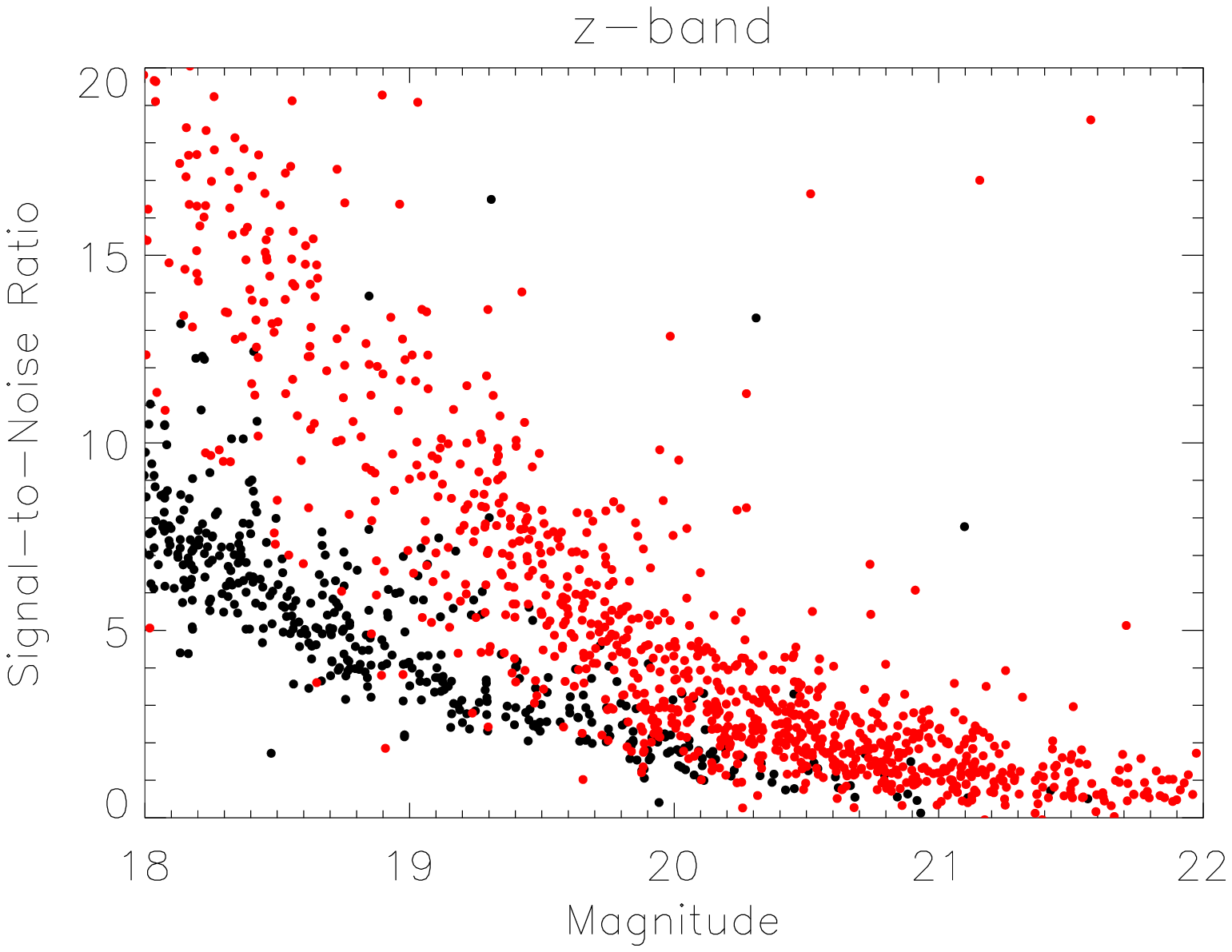}
  \includegraphics[scale=0.4]{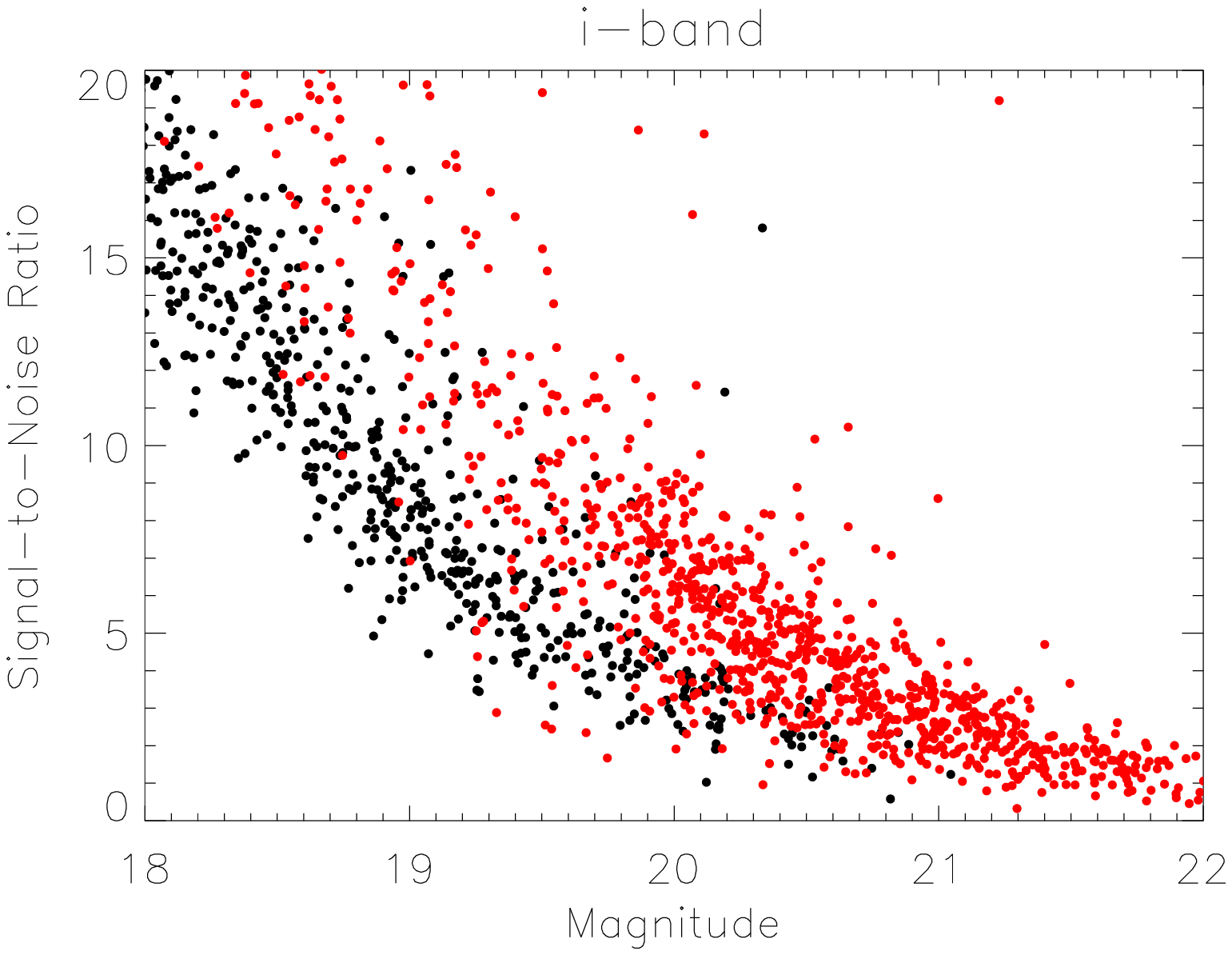}
}
\centerline{
  \includegraphics[scale=0.4]{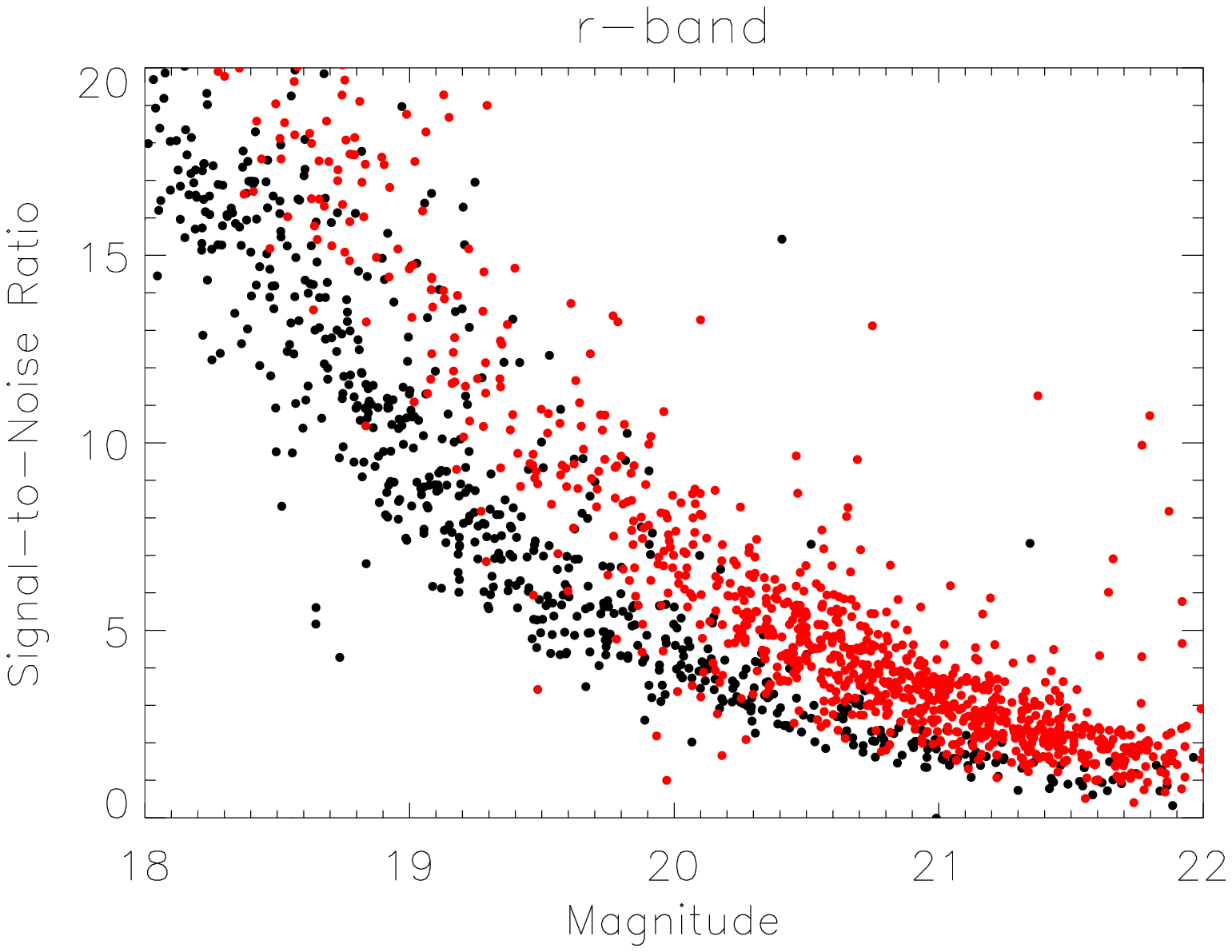}
  \includegraphics[scale=0.4]{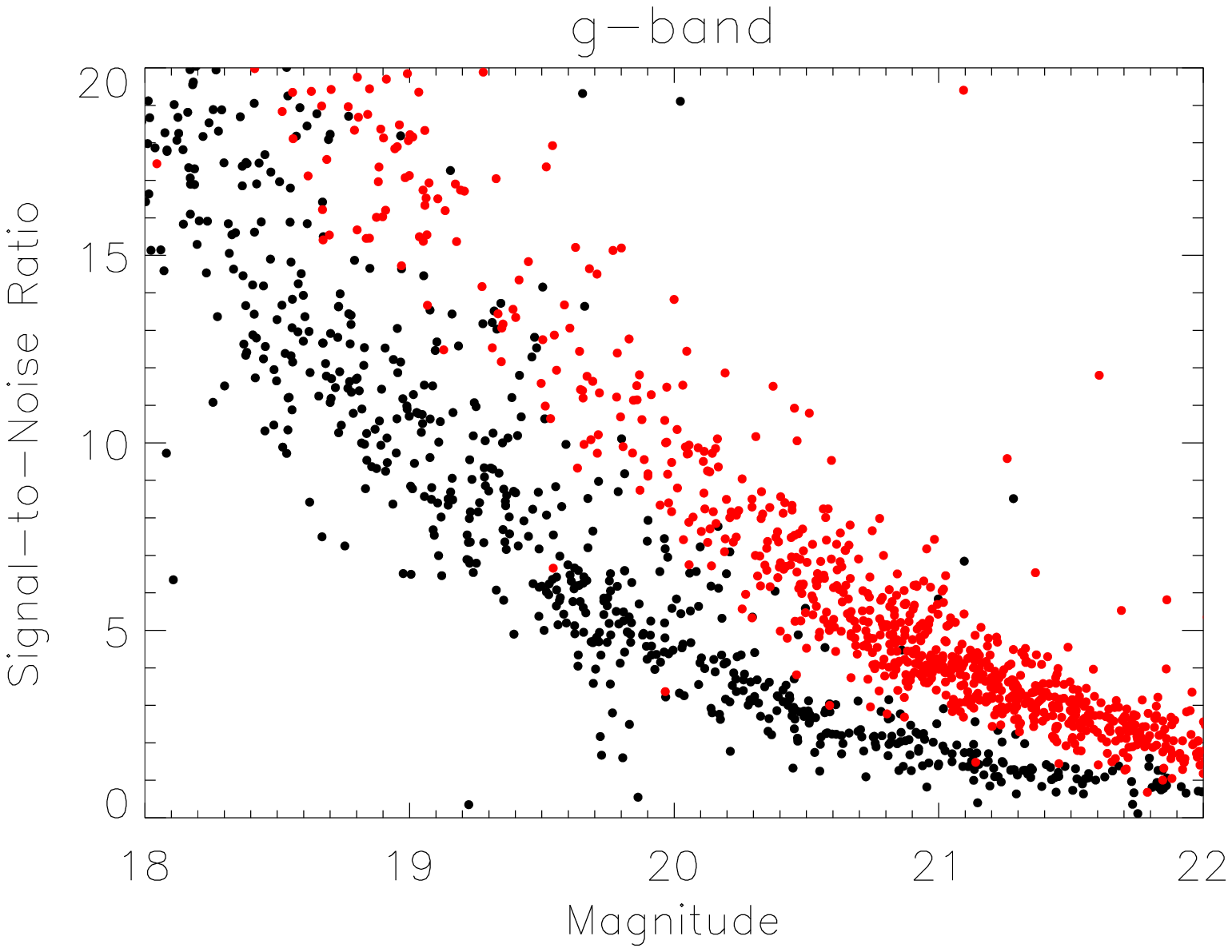}
}
\end{center}
\caption[Compare]{\label{fig:sdssvsboss}
SNR per pixel as a function of extinction-corrected PSF magnitude in each of the four synthetic
bandpass filters ($griz$) matching the photometric passbands of the SDSS imaging survey.  In each case,
the red dots represent BOSS data while the black dots represent SDSS data.  The data were selected from
a random sampling of objects classified as stars over the full SDSS and BOSS samples.
}
\end{figure*}

BOSS produces more objects at high redshift largely due to the
improved throughout of the spectrographs and the decreased sky
background from the smaller fibers.
The improved sensitivity is demonstrated through a comparison
of the typical SNR in BOSS spectra relative to SDSS.
By comparing the quality of data from a random sampling of stellar
point sources from both samples, we find roughly a factor
of two improvement in SNR per pixel in BOSS.
A SNR vs. magnitude scatter plot of individual objects measured in the synthetic bandpass
filters $griz$ is shown in Figure~\ref{fig:sdssvsboss}.
We do not include the synthetic $u$ filter because of the significant increase
in the wavelength coverage of BOSS in that wavelength range.

\begin{figure*}[h]
\begin{center}
\centerline{
  \includegraphics[scale=0.4]{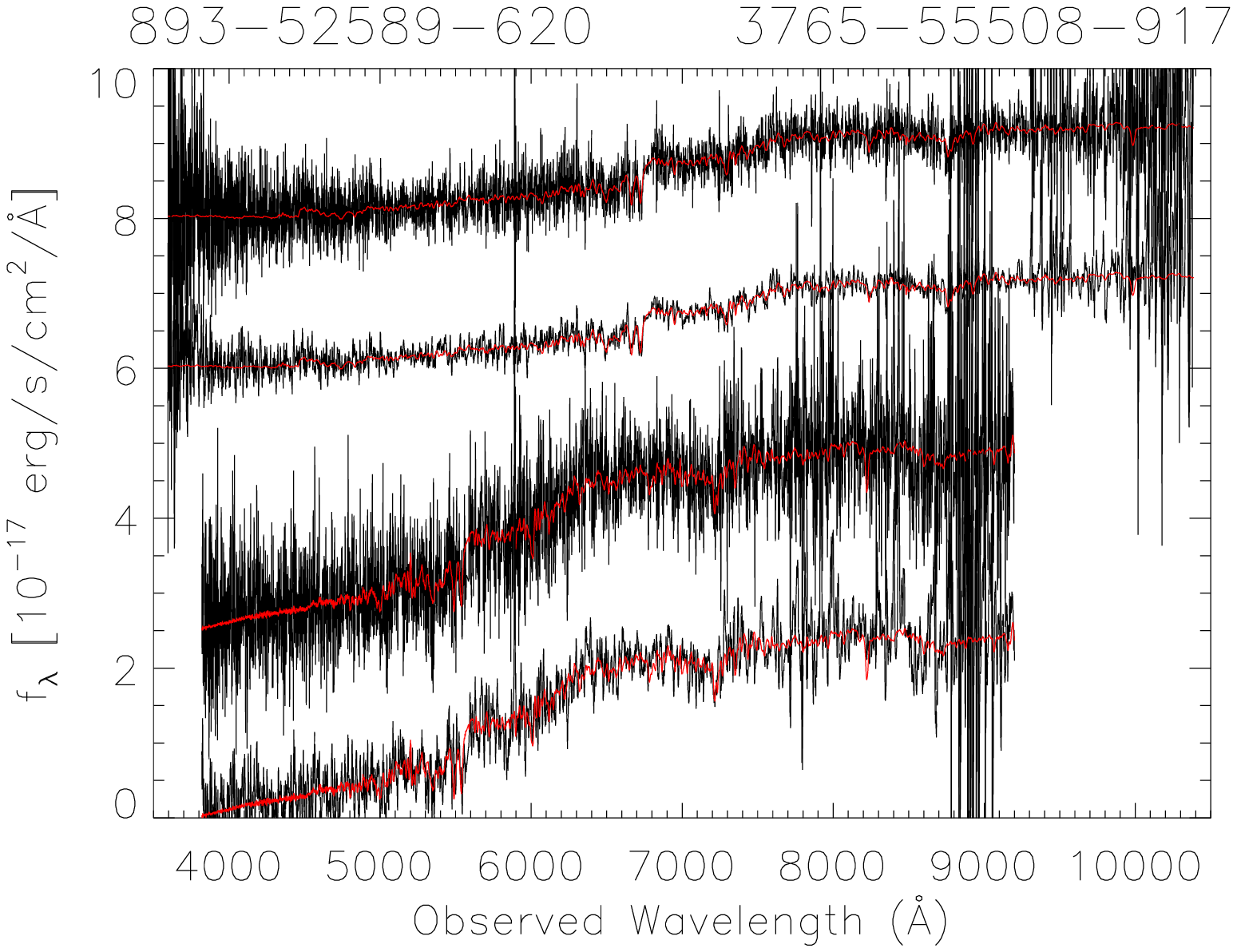}
  \includegraphics[scale=0.4]{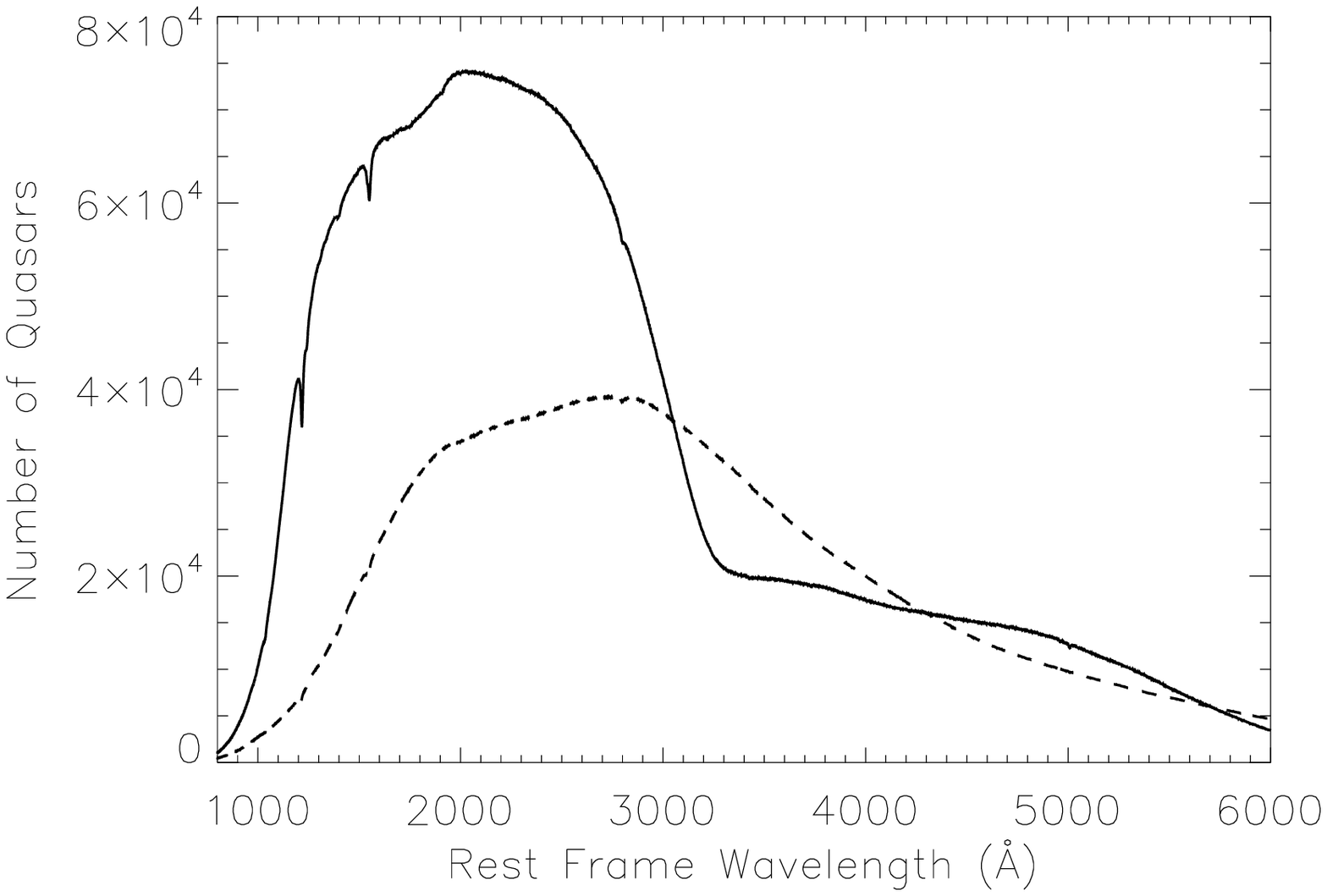}
}
\end{center}
\caption[Histogram]{\label{fig:histograph}
{\bf Left:  }Comparison of the BOSS spectrum of one of the faintest CMASS targets ($z=0.694$,
$i_{\rm fib2}=21.42$, mean(SNR/pixel)$=1.82$) to the SDSS spectrum of one of the faintest LRG targets ($z=0.395$,
$i_{\rm fib}=19.94$, corresponding to $3\2pr$ fiber, mean(SNR/pixel)$=2.38$).  From top to bottom, with offsets of
8, 6, 2, and 0 $\times 10^{-17}$ erg s$^{-1}$ cm$^{-2}$\AA$^{-1}$ respectively,
are the raw BOSS spectrum, the BOSS spectrum smoothed with a five pixel boxcar median filter,
the raw SDSS spectrum, and the SDSS spectrum smoothed with a five pixel boxcar median filter.
In all cases, the red curve represents the best-fit spectral template used to determine the redshift.
The plate, MJD, fiber combination for the two objects are displayed at the top of the panel.
{\bf Right:  }Corresponding spectral coverage of the quasar sample in BOSS (solid curve)
relative to SDSS (dashed curve).
The histogram represents the number of objects sampled at each restframe wavelength.
}
\end{figure*}

The ambitious extension to redshifts beyond $z=0.6$ for the CMASS sample was made possible
not only by improvements to the spectrographs, but also by the proven reliability of the
idlspec2D pipeline to classify objects with spectra as low as SNR$\sim 2$ per coadded pixel in SDSS.
An example of a successful classification of a faint SDSS LRG is shown in Figure~\ref{fig:histograph}.
For comparison, a CMASS galaxy with reliable redshift is shown in the same figure.
The BOSS galaxy has about the same SNR, but has almost 1.5 magnitudes less flux entering the fiber.

As important as increasing the number of \lya\ quasars
is the superior wavelength coverage of BOSS relative to SDSS.
The expanded coverage from $\lambda \greatsim 3780$ \AA\
to $\lambda \greatsim 3615$ \AA\
makes the redshift range $2.15<z<2.3$ more accessible for \lya\ forest studies.
As shown in Figure~\ref{fig:progress},
19\% of the \lya\ quasars for BOSS lie between $2.15 < z < 2.3$.
Figure~\ref{fig:histograph} shows an increase in the number of observations at each wavelength
in the \lya\ forest for BOSS relative to SDSS.

\subsection{Early Science Results from BOSS Data}

In addition to the technical papers summarized in \S\ref{sec:intro},
members of the SDSS-III collaboration have published a number of papers using the BOSS data.
While cosmology was the primary driver, the size and quality of the data sample
enable many other studies of galaxy evolution and quasar physics.

\citet{chen12a} applied principal component analysis
(PCA) to the rest-frame 3700-5500 \AA\ spectral region to estimate the 
properties of the BOSS galaxies. They find the fraction of galaxies 
with recent star formation in the CMASS sample first declines with increasing stellar mass,   
but flattens above a stellar mass of $10^{11.5}M_\odot$ at $z \sim 0.6$.
This is contrast to $z \sim 0.1$, where the fraction of galaxies with recent
star formation declines monotonically with stellar mass 
\citep[right panel of Figure 16 in][]{chen12a}. The PCA fits as well as galaxies properties, 
such as stellar masses, stellar velocity dispersions, are included as part of the DR9 catalog.
\citet{shu12a} present the velocity-dispersion probability function for each galaxy in the LOWZ and CMASS samples.
Correlating velocity dispersion with absolute magnitude and redshift, they find that galaxies
of constant mass exhibit higher intrinsic scatter in velocity dispersion at higher redshifts.
\citet{maraston12a} present the first BOSS galaxy mass functions at redshift $0.4 <z< 0.7$
using the BOSS spectroscopic redshifts and $ugriz$ photometry.
A comparison to galaxy formation models shows that data-derived mass functions compare
well with models at the typical BOSS redshift ($z=0.55$), even at the high mass end.
However, the data do not show the evolution to zero redshift that is predicted in
models of hierarchical mass build-up.
The BOSS spectroscopy and SDSS imaging data has also been used to study the luminosity functions
of galaxies surrounding LRGs \citep{tal12a}.
The data require a multiple component fit to produce
a luminosity function consistent with the distribution of central galaxies and the most luminous satellite galaxies.
The result stands in contrast with models that predict mass growth of central galaxies through major mergers.

Many observations with the Hubble Space Telescope (HST) have been taken inside the
BOSS footprint.
These high spatial resolution data offer a counterpart to the BOSS spectra, allowing, for example,
a study of the morphology and size of the CMASS sample of galaxies \citep{masters11a}.
The study reveals that the majority of CMASS galaxies ($74 \pm 6$\%) exhibit early-type morphology
as intended in the target selection algorithms.
An additional color cut excluding galaxies with $g-i < 2.35$ increases the population of
early-type galaxies to $\ge 90$\%, producing a subsample that is similar to that of the SDSS-I and II LRG galaxies.
In an approach similar to that from the Sloan Lens Advanced Camera for Surveys
\citep[SLACS;][]{bolton06a,bolton08a,auger09a},
\citet{brownstein12a} present a catalog of strong galaxy-galaxy gravitational lens systems 
discovered spectroscopically in the CMASS sample and observed with HST.
Known as the BOSS Emission-Line Lensing Survey (BELLS),
the HST imaging and BOSS spectra enable a reconstruction of the mass profile of the lensing galaxies.
\citet{bolton12b} perform a joint analysis on the SLACS and BELLS samples and determine that galaxies
tend to have steeper mass profiles with decreasing redshift.
Unlike the the studies of \citet{tal12a},
the results indicate that major dry mergers play a major role in the evolution of massive galaxies.

Finally, highlighting the potential for multi-wavelength studies,
the first detection \citep{hand12a} of the kinematic
Sunyaev-Zel'dovich effect \citep{sunyaev72a} was discovered in a correlation of BOSS
galaxies with maps made by the Atacama Cosmology Telescope \citep{hincks10a}.
This is surely the first such result in a stream of studies that will
combine BOSS spectra with data from projects such as WISE, XMM, Chandra and Planck.

Results from the main quasar sample are now also beginning to appear.
The first such study constrains the two-point correlation function over the redshift range $2.2<z<2.8$
\citep{white12a}.
The redshift space correlation function of quasars is described by a power-law that
requires a bias factor of $b \sim3.5$ for the CORE sample,
implying that these quasars reside in halos with a typical mass of $\sim 10^{12} \,  M_{\sun}$.
Members of the BOSS collaboration will soon release results
that constrain the quasar luminosity function,
determine a composite quasar spectrum \citep[as in ][]{vandenberk01a}, report
variable absorption in BAL quasars, and many other studies of quasar physics.
In addition to results from the primary BOSS galaxy and quasar targets,
analysis of data from the ancillary programs described in Appendix~\ref{appendix:ancillary}
continues and will result in an even more diverse array of findings than
would be possible with the main samples alone.

\subsection{BOSS Cosmology}

As explained in \S\ref{subsec:imaging}, we intend to complete spectroscopy with BOSS
over the $\sim $10,000 $\sqdeg$ footprint that falls at high Galactic latitude and was
imaged by SDSS-I and II.
The primary goal of the galaxy redshift survey is to
achieve BAO distance scale constraints that
are close to the limit set by cosmic variance out to $z= 0.6$;
the only substantial (factor-of-two) improvement possible at these redshifts would
be to cover the remaining $3\pi$ steradians of the sky.
The BOSS \lya\ forest survey will pioneer a new method of measuring three-dimensional
structure in the high-redshift universe and provide the first BAO measurements
of distance and expansion rate at z $> 2$.

The first small scale clustering statistics of the LOWZ sample from the
BOSS data are reported in \citet{parejko12a}.
Measurements of clustering in the CMASS sample from the
first year of BOSS data are found in \citet{white10a}.
That analysis includes
constraints on the halo occupation distribution.
With the full DR9 sample, the BOSS collaboration carried out a series
of more ambitious clustering analysis.
For example, \citet{reid12a} measure the growth of structure and expansion rate at $z=0.57$
from anisotropic clustering, and \citet{sanchez12a}
explore the cosmological implications of the large-scale two-point correlation function.
The analysis for these cosmology studies shares much of the same framework as the galaxy BAO study.
Toward that goal, \citet{manera12a} generate a large sample of mock galaxy catalogs to
characterize the statistics of large scale structure,
\citet{ross12b} present an analysis of potential systematics in the galaxy sample,
and \citet{tojeiro12a} measure structure growth using passive galaxies.

The efforts described in the previous paragraph made possible the first BOSS BAO
measurement \citep{anderson12a}.
Using 264,283 CMASS galaxies over $3275\,\sqdeg$
and a redshift range $0.43<z<0.7$, we determined the angle-averaged galaxy
correlation function and galaxy power spectrum in an effective volume of more
than two Gpc$^3$, the largest sample of the Universe ever surveyed at this density.
In combination with the SDSS LRG sample, the BAO detection
provides a 1.7\% measurement of the distance to $z=0.57$,
the most precise distance constraint ever obtained from a galaxy survey.
The resulting constraints on various cosmological models continue to support
a flat Universe dominated by a cosmological constant.

Measurements of the three-dimensional correlation function using
the first year of BOSS \lya\ spectra are found in \citet{slosar11a}.
This analysis also reports the first detection of flux correlations
across widely separated sightlines, a measurement
of redshift-distortion due to peculiar velocities,
and constraints on the linear bias parameter of optical depth.
The complete five year quasar sample should provide the means
to extend the \citet{slosar11a} analysis and observe the BAO feature
at $z>2$ for the first time.

As described in \citet{eisenstein11a},
we have used a Fisher matrix formalism to predict the BAO constraints
from the five year BOSS galaxy and \lya\ forest surveys
\citep[drawing on][]{seo07a,mcdonald07a}, and to predict the impact
of these constraints on cosmological parameter determinations.
For the galaxy survey we forecast $1\sigma$ precision of 1.0\% on $D_A(z)$
and 1.8\% on $H(z)$ at $z=0.35$ (bin width $0.2 < z < 0.5$)
and corresponding errors of 1.0\% and 1.7\% at $z = 0.6$
(bin width $0.5 < z < 0.7$).
Errors at the two redshifts are essentially uncorrelated, while
the errors on $D_A(z)$ and $H(z)$ at a given redshift are correlated at the 40\% level.
Dividing into smaller bins would yield a finer grained view
of $w(z)$ evolution, with larger errors per bin but the same
aggregate precision.
For the \lya\ forest we update the forecast slightly from \citet{eisenstein11a},
predicting a measurement precision of 1.9\%
at a redshift approximately $z=2.5$ on an overall dilation factor scaling both
$D_A(z)$ and $H^{-1}(z)$.
This result was confirmed with a full simulation of the survey using mock
quasar spectra \citep{legoff11a}.
$D_A(z)$ and $H(z)$ have a higher degree of degeneracy than in the
galaxy sample, with a correlation coefficient of 0.6; we therefore
do not report projections for individual constraints here.
In detail, \lya\ forest projections depend on the level of clustering bias
and redshift-space distortion in the forest, which are only
now being measured precisely in ongoing analyses of the DR9 quasar sample.

The impact of these measurements on cosmological parameter determinations
depends on the assumptions about external constraints from other
data sets and about the underlying space of theoretical models.
We have computed forecasts \citep[Appendix A of][]{eisenstein11a}
within the framework adopted by the DETF \citep{albrecht06a},
which assumes an inflationary cold dark matter cosmology,
allows departures from spatial flatness, and employs a dark energy
equation of state described by $w(a) = w_p + w_a(a_p-a)$, where
$a=(1+z)^{-1}$ and $w_p = w(z_p)$ is the value of $w$ at the
``pivot'' redshift $z_p \approx 0.5$ where it is best constrained by BOSS.
When combining the forecast BOSS BAO constraints with Planck priors
for CMB anisotropy measurements and ``Stage II'' priors that
approximately describe the constraints from existing dark energy
experiments, we project $1\sigma$ errors of 0.030 on $w_p$,
0.320 on $w_a$, $2\times 10^{-3}$ on the curvature parameter
$\Omega_k$, and 0.9\% on the Hubble constant $H_0$.
Combining BOSS data with measurements from other, upcoming
``Stage III'' experiments, such as supernova, weak lensing, and
cluster abundance data from the Dark Energy Survey, will further
tighten these constraints, breaking parameter degeneracies and
allowing cross-checks of results from different methods.
Within BOSS itself, analyses that exploit the full shape and
redshift-space anisotropy of the galaxy correlation function or
power spectrum \citep[e.g.;][]{sanchez12a,reid12a}, or that use
other statistical approaches sensitive to higher order correlations,
can achieve tighter constraints on dark energy and curvature,
can test whether the rate of cosmic structure growth is consistent
with the predictions of General Relativity, and provide sensitivity
to neutrino masses, inflation physics, extra radiation backgrounds,
and other departures from the simplest forms of the $\Lambda$CDM scenario.
These approaches make stronger demands
on theoretical models of non-linear gravitational evolution and
galaxy bias than does BAO, but the BOSS clustering measurements allow detailed,
high-precision tests of these models.

The success of BOSS, as documented in this paper, should be viewed as strong
encouragement to future efforts in constraining BAO over large volumes and large redshifts.
BOSS data will have broad impact on our understanding of large scale
structure, massive galaxy evolution, quasar evolution, and the
intergalactic medium.  The ultimate impact on fundamental cosmology
depends in part on what Nature has in store.
Any clearly detected deviations from a flat universe with a cosmological
constant would have profound implications,
taking us closer to understanding one of the most remarkable
scientific discoveries of our time.

Funding for SDSS-III has been provided by the Alfred P. Sloan Foundation, the Participating Institutions,
the National Science Foundation, and the U.S. Department of Energy Office of Science.
The SDSS-III web site is http://www.sdss3.org/.

SDSS-III is managed by the Astrophysical Research Consortium for the Participating Institutions of the
SDSS-III Collaboration including the University of Arizona, the Brazilian Participation Group,
Brookhaven National Laboratory, University of Cambridge, Carnegie Mellon University, University of Florida,
the French Participation Group, the German Participation Group, Harvard University, the Instituto de Astrofisica de Canarias,
the Michigan State/Notre Dame/JINA Participation Group, Johns Hopkins University,
Lawrence Berkeley National Laboratory, Max Planck Institute for Astrophysics,
Max Planck Institute for Extraterrestrial Physics, New Mexico State University,
New York University, Ohio State University, Pennsylvania State University, University of Portsmouth,
Princeton University, the Spanish Participation Group, University of Tokyo, University of Utah,
Vanderbilt University, University of Virginia, University of Washington, and Yale University.

\bibliographystyle{apj}
\bibliography{archive}

\begin{appendix}

\section{Appendix A:  Flags for Primary Target Classes}\label{appendix:mainflags}

The algorithms for targeting various classes of
galaxy and quasar targets described in \S\ref{sec:TS} and several additional classes
of object for calibration are all recorded in the {\bf BOSS\_TARGET1} flag bit.
A brief description of those objects is found below.
The density and survey area containing each type of object is found in Table~\ref{tab:galflags}. 
The {\bf BOSS\_TARGET2} keyword serves as a placeholder for additional targets,
but has not yet been used in SDSS-III.
\\

{\bf Galaxy Target Selection:}
As described in \S\ref{subsec:galaxyTS}, 
the galaxy sample used to derive new BAO constraints was divided into
two principal samples based on redshift.
The LOWZ sample (represented by the flag {\bf GAL\_LOZ})
targets the redshift interval $0.15 <z< 0.43$
while the CMASS sample targets the redshift interval $0.43<z<0.7$.
The CMASS sample is divided into four subsamples of galaxies,
with significant overlap between several subsets:

\begin{itemize}

\item
{\bf GAL\_CMASS} is the core of the high redshift BOSS galaxy sample, with selection criteria
described in the main text.  The flag was assigned to a
test sample in the BOSS commissioning
data found in boss1 and boss2 and should not be used in those chunks.  It was assigned
to the full sample of CMASS objects starting in boss3.

\item
{\bf GAL\_CMASS\_COMM }represent galaxies that were targeted using a commissioning
selection algorithm to investigate possible targeting strategies.
Objects were chosen so that $d_{\perp}> 0.55$, $i < 20.14+ 1.6(d_{\perp} - 0.8)$,
$17.5 < i < 20.0$ and
$r - i < 2$ where $r$ and $i$ are cmodel magnitudes.  Additionally,
objects were required to have
$i_{fib2} < 22$ and $i_{\rm PSF} - i_{model} > 0.2 + 0.2(20.0 - i_{model})$.
This sample is complete only for the first two BOSS commissioning chunks, boss1 and boss2.
The flag was never phased out and appears in DR9 from objects in later stages in the tiling,
representing a subset of galaxies that would otherwise satisfy the {\bf GAL\_CMASS\_COMM}
criteria.  It should not be used for boss3 or any chunks thereafter.

\item
{\bf GAL\_CMASS\_SPARSE }targets have a color-magnitude cut
that is shifted relative to {\bf GAL\_CMASS} to $i < 20.14 + 1.6 \times (d_{\perp} - 0.8)$
to allow for completeness studies.
Galaxies are either flagged {\bf GAL\_CMASS} or {\bf GAL\_CMASS\_SPARSE}, but not both.
This flag was introduced in boss3, after the commissioning period for galaxy target selection.

\item
{\bf GAL\_CMASS\_ALL }is used internally in the target selection code, representing
the union of the {\bf GAL\_CMASS} and {\bf GAL\_CMASS\_SPARSE} targets for boss3 and thereafter.
Because the sample of galaxies identified by this flag can be reconstructed by the union of 
{\bf GAL\_CMASS} and {\bf GAL\_CMASS\_SPARSE}, it provides no additional information and
should rarely be needed by the user.
\end{itemize}

{\bf Quasar Target Selection:}
The use of quasar target flags evolved significantly throughout the first two years of the
survey as different algorithms were tested to maximize the targeting efficiency.
All programs with the {\bf QSO\_} prefix in Table~\ref{tab:galflags} were targeted as potential
quasars.
A detailed explanation of each program is given in \citet{ross12a};
we simply report the densities and survey area for each program.
A few target classes bear special mention:

\begin{itemize}

\item
The {\bf QSO\_UKIDSS} and {\bf QSO\_KDE\_COADD} target classes
were only used in the boss1 chunk of commissioning data
on Stripe 82, covered by a small number of plates between $0\deg<\alpha<40\deg$.

\item
The {\bf QSO\_CORE\_ED} and {\bf QSO\_CORE\_LIKE} classes were implemented in the second year.

\item
The {\bf QSO\_CORE} and {\bf QSO\_BONUS} classes were deprecated after the boss2 chunk and replaced
by the {\bf QSO\_CORE\_MAIN} and {\bf QSO\_BONUS\_MAIN} classes for the rest of the survey,
beginning with chunk3.
\end{itemize}

{\bf Calibration Programs:}
There are five target classes dedicated to calibration programs within BOSS.
The most prevalent target flag is {\bf STD\_FSTAR}, used to denote
objects chosen as likely F stars as described in \S\ref{subsec:platedesign}.
These objects are used both for flux calibration and correction for telluric absorption. 
The other four target classes were only used in the Stripe 82 commissioning
plates in the boss1 chunk.
These contain objects to aid in
the development of templates for classification and redshift determination \citep{bolton12a}.
In order for the classification to be as robust and informative as possible,
the set of template spectra were selected to cover a wide range in the space of
intrinsic spectral variation.
Several classes contain objects previously observed in SDSS; the extended wavelength coverage
of BOSS requires repeat spectra of these objects to cover 361 nm -- 1014 nm
in the observer frame.
The targets are divided into four categories:

\begin{itemize}

\item
{\bf TEMPLATE\_GAL\_PHOTO }were galaxies selected to uniformly cover the space
of [($g-r$), ($r-i$), $i$].  Objects were first chosen with dereddened cmodelmag $i< 20$,
$g_{fib}$, $r_{fib}$, and $i_{fib} > 16.5$,
$SNR > 5.0$ for $gri$ model magnitudes to limit scatter in color measurements,
and an absolute value of the difference between the
model magnitude and cmodel magnitude in $i$ less than 0.8.
Next, for each galaxy, the total number of galaxies within a radius of 0.1 magnitudes
in [($g-r$), ($r-i$), $i$] space was computed and taken as an estimate of the local density
in that space.  A subsample was then drawn at random, with the probability of retaining
any given draw proportional to the inverse of the local density.
This generates an approximately uniform sampling of the populated
region within the color-color-magnitude space, where colors $g-r$
and $r-i$ were defined using model magnitudes and $i$ was determined
using cmodel magnitudes.

\item
{\bf TEMPLATE\_QSO\_SDSS1 }were selected from the DR7 catalog of 
quasars \citep{schneider07a}.
The sample is restricted to objects with
dereddened PSF magnitudes in $ugriz <20.0$
and $gri$ fiber magnitudes $> 16.5$.
A random subsample was chosen after imposing a uniform distribution in redshift out to
$z<3.2$ and taking all quasars at redshifts $z>3.2$.
The PCA redshift templates in BOSS are based exclusively on this sample.

\item
{\bf TEMPLATE\_STAR\_PHOTO }were selected in similar fashion to the galaxy sample above,
with dereddened PSF magnitudes brighter than 19.25 in any one of $ugriz$,
$SNR > 5.0$ in all filters using PSF magnitudes, and
fiber magnitudes fainter than 16.5 in each of $gri$.
A local density estimator within the PSF magnitude space of [($u-g$), ($g-r$), ($r-i$), ($i-z$)]
was computed from the number of stars within a 4-sphere of radius 0.1 magnitude centered on each star.
A random subsample of stars was drawn with the probability of retaining any given draw proportional
to the local density to the power -0.7 (compared to -1 for the galaxy sample).
This heuristic factor is chosen to retain some preferential weighting in favor of
the core stellar locus, while still achieving significant representation
from the less densely populated regions of color space.

\item
{\bf TEMPLATE\_STAR\_SPECTRO }were selected from the SDSS (non-SEGUE) spectroscopic 
database for re-observation to provide a sample of spectra that cover the major classes of
stellar objects that lie away from the main stellar locus.
The objects were required to have been both targeted as stars photometrically and classified
as stars spectroscopically.
The objects are required to have dereddened PSF magnitudes brighter than 19.25 in
$ugriz$ and fiber magnitudes fainter than 16.5 in each of $gri$.
Stars were then selected at random to represent each of the following SDSS target categories
identified by the DR9
{\bf LEGACY\_TARGET1}\footnote{http://www.sdss3.org/dr9/algorithms/bitmask\_legacy\_target1.php}
and {\bf LEGACY\_TARGET2}\footnote{http://www.sdss3.org/dr9/algorithms/bitmask\_legacy\_target2.php} flags:
{\bf STAR\_BHB},
{\bf STAR\_CARBON},
{\bf STAR\_BROWN\_DWARF},
{\bf STAR\_SUB\_DWARF},
{\bf STAR\_CATY\_VAR},
{\bf STAR\_RED\_DWARF},
{\bf STAR\_WHITE\_DWARF},
{\bf STAR\_PN},
{\bf REDDEN\_STD},
{\bf SPECTROPHOTO\_STD},
and {\bf HOT\_STD}.
\end{itemize}

\begin{deluxetable}{lrrr}
\centering
\tablewidth{0pt}
\tabletypesize{\footnotesize}
\tablecaption{\label{tab:galflags} BOSS Programs with {\bf BOSS\_TARGET1} Flag}
\tablehead{
\colhead{Program} & \colhead{Bit}
                  & \colhead{Density} & \colhead{Survey Area} \\
\colhead{} & \colhead{Number}
                  & \colhead{(deg$^{-2}$)} & \colhead{(deg$^{2}$)}
}
\startdata
GAL\_LOZ		& 0	& 20--30	& 10,000	\\
GAL\_CMASS		& 1	& 110		& 9,750	\\
GAL\_CMASS\_COMM	& 2	& 115		& 500	\\
GAL\_CMASS\_SPARSE	& 3	& 5		& 9,750	\\
GAL\_CMASS\_ALL		& 7	& 115		& 9,750	\\
QSO\_CORE		& 10	& 14	& $\sim 250$	\\
QSO\_BONUS		& 11	& 17	& $\sim 250$	\\
QSO\_KNOWN\_MIDZ	& 12	& 1.5	& 10,000	\\
QSO\_KNOWN\_LOHIZ	& 13	& 0.25	& $\sim 100$	\\
QSO\_NN			& 14	& 20	& 10,000	\\
QSO\_UKIDSS		& 15	& 1.0	& $\sim 50$	\\
QSO\_KDE\_COADD		& 16	& 27	& $\sim 50$	\\
QSO\_LIKE		& 17	& 28	& 10,000	\\
QSO\_FIRST\_BOSS	& 18	& 1.0	& 10,000	\\
QSO\_KDE		& 19	& 28	& 10,000	\\
STD\_FSTAR		& 20	& 4.6  	& 10,000	\\
TEMPLATE\_GAL\_PHOTO	& 32	& 4.2   & $\sim 50$	\\
TEMPLATE\_QSO\_SDSS	& 33	& 2.6	& $\sim 50$	\\
TEMPLATE\_STAR\_PHOTO	& 34	& 3.1	& $\sim 50$	\\
TEMPLATE\_STAR\_SPECTRO	& 35	& 1.2	& $\sim 50$	\\
QSO\_CORE\_MAIN		& 40	& 20	& 10,000	\\
QSO\_BONUS\_MAIN	& 41	& 19	& 10,000	\\
QSO\_CORE\_ED		& 42	& 10	& 9,000	\\
QSO\_CORE\_LIKE		& 43	& 11 	& 9,000	\\
QSO\_KNOWN\_SUPPZ	& 44	& 0.25	& $\sim 100$
\enddata
\end{deluxetable}

\newpage

\section{Appendix B:  Target Selection and Scientific Motivation for Ancillary Science Programs}\label{appendix:ancillary}
The first category of ancillary science programs lies in the region of Stripe 82
(the celestial equator in the SGC)
that has completed observations and will be included in DR9.
The second set includes programs that are dispersed throughout the full BOSS
footprint and will continue to accumulate data throughout the survey.
Additional ancillary programs may be added in the future;
those programs will be described in future data release papers.
All ancillary targets are assigned fibers at a priority that is lower than the
primary galaxy and quasar targets; sample selection
for these programs is therefore not complete.
Here we provide a comprehensive description of the motivation,
target selection, and the target densities for these programs.
For programs that are divided into subprograms,
we also provide a description of the ancillary target flags
and selection for each subprogram.
The target bits and statistics for the ancillary science targets are reported
in Table~\ref{tab:ancillary1} and Table~\ref{tab:ancillary2} at the end of this appendix.
The target bit names appear in bold font in what follows.

\subsection{Stripe 82}

Several of these BOSS ancillary programs are contained completely
in the Stripe 82 region.  We finished all plates in Stripe 82
in 2011, leading to a group of ancillary programs that are
included in its entirety in DR9.
Starting with the nearest objects,
we explain the rationale and target selection behind each of these programs below.
\\

{\bf The Transient Universe Through Stripe 82:}
The repeat imaging of Stripe 82 allows identification of transient
and variable phenomena of all sorts
\citep[for example][]{anderson08a,blake08a,becker08a,kowalski09a,bhatti10a,becker11a,sako11a}.
In this program, several classes of variable point sources and high-proper motion stars discovered
in Stripe 82 photometry were targeted for spectroscopy. These objects include
flaring M stars \citep{kowalski09a},
faint high proper motion stars \citep{scholz09a,schmidt10a}, candidate low-metallicity M dwarfs, 
and diverse samples of variable stars \citep{roelofs07a,anderson08a,blake08a}.
BOSS spectra were obtained to characterize the statistical properties of
these categories of transient objects.
Approximately 600 objects were targeted across Stripe 82.
The classes of transient objects were selected as follows:

\begin{itemize}

\item
{\bf FLARE1, FLARE2}
consists of $\sim 200$ flaring M stars selected from the \citet{bramich08a} and \citet{ivezic07a}
catalogs with $g_{\rm PSF}<21.4$, $i_{\rm PSF}<19$, $0.3<(i_{\rm PSF}-z_{\rm PSF})<1.3$, and exhibiting a flare event in the
Stripe 82 imaging data with amplitude $\Delta u >1$~mag \citep{kowalski09a}.

\item
{\bf HPM} consists of $\sim 100$ high proper motion stars also selected
from the catalogs of \citet{bramich08a} and \citet{ivezic07a} with an emphasis on faint
objects with high proper motions ($\mu>0.1$~mas yr$^{-1}$).
The goal of this target selection was to identify nearby low-mass stars and white dwarfs.
Candidate nearby low-mass stars include faint stars ($19<z_{\rm PSF}<20$) with $(i_{\rm PSF}-z_{\rm PSF})>1.5$,
including objects with photometric detections in $z$ band only.
Candidate nearby white dwarfs include stars with $(g_{\rm PSF}-r_{\rm PSF})\sim0$ and $g_{\rm PSF}>19$.

\item
{\bf VARS} consists of $\sim 200$ variables with $g_{\rm PSF}<21.5$,
$(g_{\rm PSF}-r_{\rm PSF}) > -0.5$, and RMS variability in $g$ band $>0.1$~mag.
These objects were selected to lie outside the stellar and quasar loci in
$(g_{\rm PSF}-r_{\rm PSF}) - (r_{\rm PSF}-i_{\rm PSF})$ color-color space as defined in \citet{fan99a}.

\item
{\bf AMC} consists of ten candidate Am CVn stars selected based on their variability and colors
($(u_{\rm PSF}-g_{\rm PSF})>0.4$ and $(g_{\rm PSF}-r_{\rm PSF})>0$).
\end{itemize}

As part of this ancillary program, a number of spectral templates were observed as well.
These include a random sample of approximately 100 M stars, designated {\bf MTEMP},
selected to span the spectra range M0 to M8 following the color criteria outlined
in \citet{west05a} ($17.5<i_{\rm PSF}<18.5$, $(r_{\rm PSF}-i_{\rm PSF}) > 0.5$, $(i_{\rm PSF}-z_{\rm PSF})>0.3$).
Approximately 70 candidate low-metallicity M stars (flag {\bf LOW\_MET}) were also targeted;
their colors were selected to be slightly outside the low-mass star stellar locus defined in \citet{west05a}.
The purpose of these sub-samples is to enable direct comparisons between the prevalence of line emission
in the photometrically-selected flare stars and field M stars selected only on the basis of their colors.
\\

{\bf Host Galaxies of SDSS-II SNe:}  
While many SNe~Ia identified in the SDSS-II program \citep{frieman08a}
were spectroscopically confirmed and used in the
first year cosmology results \citep{kessler09a},
most SNe and their host galaxies have not been observed spectroscopically.
With redshifts of the host galaxies, SN lightcurves can be fit with fewer degrees of
freedom, leading to identifications of SN type with higher confidence.
The new classifications will allow a much larger number of SNe~Ia to be placed on
the Hubble diagram \citep{campbell12a}.
In this ancillary program, candidate SN hosts were drawn from a database containing 21,787 
potentially variable objects \citep{sako08a} determined from
the repeat imaging of Stripe 82.  The next stage of selection required
coincidence of a signal in at least two passbands after vetoing
regions of bright stars and variability from known
active galactic nuclei (AGN) or variable stars.
In total the list includes 4099 candidates
of different SN types and different confidence levels as determined
from a Bayesian classification \citep{sako11a} of the lightcurve shapes.
Fibers were assigned to 3743 of these host galaxies to obtain a
redshift and thereby improve the SN classification.
Approximately one third of the targets have lightcurves that do not
resemble SNe and are included as a control sample.
The redshifts and new classification will lead to
a nearly complete sample of SNe~Ia to $z<0.4$ and an enhanced cosmological analysis.
The new classification will also enable a large statistical study of the correlated properties
between SNe~Ia and their host galaxies \citep[e.g.,][]{kelly10a, sullivan10a, lampeitl10a, brandt10a}.
Subsamples of SNe from this ancillary program have already demonstrated correlations
of residuals from the Hubble Diagram with spectroscopically-derived host properties \citep{dandrea11a}.
The candidates were selected after visual inspection and divided according to the following ancillary
program flags:
   {\bf SN\_GAL1}: fibers assigned to the position of the core of the
   nearest candidate host galaxy (this accounted for 95\% of the cases);
   {\bf SN\_GAL2}: fibers assigned to the position of the core of the
   second nearest candidate host galaxy (when the first was clearly a
   background object or a misclassified star; 4\% of the targets); 
   {\bf SN\_GAL3}: fibers assigned to the position of the core of the
   third nearest candidate host galaxy (1\% of the targets); 
   {\bf SN\_LOC}: already have SDSS host spectra so the fiber was assigned to the location
   of the original SN candidate;
   {\bf SPEC\_SN}: SNe identified from SDSS spectra \citep[e.g. ][]{krughoff11a} rather than photometric variability.
\\
 
{\bf Brightest Cluster Galaxies (BCGs):}
Over 3000 groups and clusters have been identified photometrically in 
Stripe 82 \citep{geach11a,murphy12a}.  
The clusters were selected from $ugriz$ photometry generated from the
coaddition of Stripe 82 images \citep{annis12a} that is $\sim 2$ mag deeper than 
the rest of the area in the BOSS footprint.
These clusters have photometric redshifts in the range $0<z<0.6$
(median $z=0.32$) and are expected to 
reside in dark matter halos with masses in excess of $2.5 \times 10^{13} M_{\sun}$.
Each cluster is assigned a ``BCG'', which is simply defined as the
brightest member associated with the cluster detection.
To confirm the cluster redshifts,
we obtained spectra of the likely BCGs with magnitudes $17 < i_{\rm fib2} < 21.7$
and colors that vary with redshift according to the cluster detection algorithm \citep{murphy12a}.
1505 galaxies were requested and 1345 were observed that were not included in the
main SDSS-I or -II galaxy, SDSS LRG, BOSS LOWZ, BOSS CMASS, 2SLAQ, or WiggleZ samples.
This new sample of spectroscopically-confirmed clusters will enable a wide
variety of science including the weak lensing of groups and clusters,
the link between star formation and AGN activity in BCGs,
and the LRG population of dark matter halos.
\\

{\bf High Quality Galaxy Spectra:}
Two plates (3615 and 3647) are being observed repeatedly throughout the survey
several times per year to test reproducibility of extracted spectral parameters
and to create exceptionally high SNR spectra of a subset of objects in the main BOSS sample.
These observations will produce accumulated exposure times of more than 20 hours
on all objects on the plates, including 96 LOWZ and 503 CMASS galaxies
at $\alpha = 37\deg$ on Stripe 82.
These plates contain identical spectroscopic targets but are drilled for different airmass.
The targets acquired 16.5 hours of total exposure between December 1, 2009 and July 15, 2011
and will be observed in each of the three remaining years of the survey.
The final observations will enable detailed spectral population studies
of intermediate redshift galaxies.
The data also provide a platform for tests of the consistency in spectroscopic
parameter extraction \citep[e.g.,][Thomas et al., in prep]{shu12a}
and of systematic errors in the models of continuum fitting
in the \lya\ forest region of the quasar sample \citep[e.g.,][]{lee12b}.
\\

{\bf No Quasar Left Behind:}
We observed unresolved sources that had not been previously observed in
SDSS spectroscopy that exhibit photometric variability statistically
similar to that of spectroscopically confirmed quasars.
Approximately 1500 targets were selected from 11,000 variable sources
with $16.2 < i_{\rm PSF} < 20.5$ identified according
to the technique outlined in \citet{butler11a}.
This study complements the sample of ``Variability Selected Quasars'' described below but targets
brighter objects without color cuts, leading to a higher density of lower redshift quasars.
The adopted variability-based selection criteria correctly identify 96\% of
previously known quasars, and more than 80\% of targets are expected to be quasars in the
redshift range $0<z<5$. The brightest of these ($i_{\rm PSF} < 19$) were observed to
test the completeness of the color-selected SDSS sample \citep{ross12a}.
The fainter subset represents a nearly
complete sample that was selected from uniform photometry.
In addition, the program provides an
invaluable training sample for optimizing quasar selection algorithms based on
photometric variability, vital for future synoptic surveys.
\\

{\bf Variability Selected Quasars:}
Variability is used to improve the selection efficiency for quasars around $z = 2.7$
and $\sim 3.5$, where they lie in the region of color-color space that is occupied by the stellar locus. 
The selection method is quite similar to that
used for the variability target selection used to select the main
sample of quasars in Stripe 82 as described at the end of \S\ref{subsec:QSOTS}. 
The sample also complements the sample of ``No Quasar Left Behind''
but includes fainter quasars and imposes a loose color selection to select primarily objects at $z>2.15$.
In this method, objects were included with 
$i_{\rm fib2}>18$, $(g_{\rm PSF}-i_{\rm PSF})<2.2$, $(u_{\rm PSF}-g_{\rm PSF})>0.4$, and $c_1<1.5$ or $c_3<0$.
Colors $c_1$ and $c_3$ are defined in \citet{fan99a} as 
\begin{eqnarray}
c_1 &=& 0.95(u-g)+0.31(g-r)+0.11(r-i) \, ,\nonumber \\
c_3 &=& -0.39(u-g)+0.79(g-r)+0.47(r-i)\, .\nonumber 
\end{eqnarray} 
A variability neural network is used to quantify the likelihood that each object is a quasar.
The neural network takes as an input (1) the $\chi^2$s between the light curve
in each band and a model assuming no variability and (2) a structure function derived from the
lightcurve as in \citet{schmidt10b}.
Targets require a probability of being a quasar from the variability neural network greater than 0.95.
Candidate quasars are chosen from $\sim$60 imaging epochs from the last ten years,
resulting in about 15 deg$^{-2}$ quasar candidates.
After removal of previously known quasars and removal of candidates that were already
included in the main quasar selection, 3.4 deg$^{-2}$ were selected for this sample.
These data address the completeness of color selections and
identifies obscured objects that would not be selected otherwise
\citep{palanque-delabrouille11a}.
These data are also being used to demonstrate the efficiency of variability
target selection for possible implementation
in future surveys such as BigBOSS \citep{schlegel11a}, a ground-based dark energy experiment
to study BAO and the growth of structure with deeper observations than BOSS.
\\

{\bf Reddened Quasars:}
Quasar candidates likely to be intrinsically reddened with E(B-V) $>0.5$
were selected from SDSS photometry and either FIRST or the Two Micron
All Sky Survey \citep[2MASS;][]{skrutskie06a}.
The goal was a better determination of the distribution of
reddenings in intervening absorbers and at the quasar redshift
\citep[e.g.,][]{hopkins04a,menard08a},
and investigating reddening as a function of quasar properties.
Targets were selected based on color selection and stellar morphology.
All targets were required to have
$17 < z_{\rm PSF} < 21$ with Galactic extinction colors satisfying
$(g_{\rm PSF} - i_{\rm PSF})-3\sigma_{(g_{\rm PSF}-i_{\rm PSF})} >
0.9$, where $\sigma$ is the measurement error in the color.
Targets selected from SDSS plus FIRST were denoted
{\bf RQSS\_SF} or {\bf RQSS\_SFC}
\citep[the trailing C indicates a CHILD object in SDSS, as in][]{stoughton02a}
and were additionally required to match within $2\2pr$ of a FIRST source.
Targets selected from SDSS plus 2MASS
({\bf RQSS\_TM} or {\bf RQSS\_TMC})
were additionally required to match within $3\2pr$ of a 2MASS
source having $11 < K < 15.1$, $(J - K) > 1.25$, and a 7-dimensional
color distance \citep{covey07a} greater than 50
(this is the distance in units of $\chi^2$ to the point on
the stellar locus with the same $(g-i)$ color as the object).
The program was discontinued after the first year due to low yield,
and targets are included on only 13 completed plates.
It is superceded by the program ``K-band Limited Sample of Quasars''.
\\

{\bf K-band Limited Sample of Quasars:}
Potential quasars are identified via photometering UKIDSS \citep{lawrence07a} at the position of
SDSS sources to a limiting magnitude $K_{\rm AB}<19.0$ ($K_{\rm vega}<17.1$).
The intention was to identify highly red and reddened quasars not included in the
main selection methods.  The sample allows an investigation, for instance, of the incidence of
BAL quasar samples, or of highly reddened gravitationally lensed quasars.
In the original selection, approximately 2000 targets were identified that have
$(g_{\rm PSF} - i_{\rm PSF}) < 1.153 \times (i_{\rm PSF} - K_{\rm AB}) - 1.401$ to avoid the color locus of the
stellar main sequence, and $(u_{\rm PSF} - g_{\rm PSF}) > 0.4$ to exclude UV excess quasars
(i.e., unreddened quasars at redshifts of about $z< 2.15$). Here, the
$K$-band magnitudes are on the Vega system, while the SDSS photometry
is AB.  The
targets not included in Stripe 82 were effectively incorporated into the
BONUS sample through the XDQSO selection \citep{bovy11a, bovy12a} and this
ancillary program was not continued for the remainder of the BOSS footprint.
This ancillary program superseded the ancillary program
``A Large Sample of Reddened Quasars'', as the superior UKIDSS photometry
could better target red, reddened and high redshift quasars than 2MASS. In addition, the
cut of $(u_{\rm PSF} - g_{\rm PSF}) > 0.4$ better complemented the BOSS mission to
target mostly higher redshift quasars.

\subsection{Full BOSS Survey} 

The remaining BOSS ancillary science programs are distributed
over some or all of the full 10,000 $\sqdeg$ footprint and are selected
from single-epoch imaging data.
Observations will continue through the end of the survey, and
additional programs will be added in the future.
As with the Stripe 82 programs, we list these programs in order
of increasing distance.
\\

{\bf Very Low Mass Stars and Brown Dwarfs:}
Very low mass stars and brown dwarfs (spectral types M8, M9 and L)
are ideal tracers of the kinematic properties of the Milky Way thin disk.
While the program ``The Transient Universe Through Stripe 82'' uses variability information
over Stripe 82 to identify rare classes of stars, this program uses
photometric selection over
the much larger BOSS survey area to identify new low mass stars and brown dwarfs.

SDSS I and II yielded a wealth of spectroscopic data of these ultracool dwarfs
\citep{schmidt10a,west11a},
but were limited by the small number of spectra observed for
stars with spectral types later than M7.
Additional observations of these objects are essential to understand the
properties of magnetic activity in these ultracool dwarfs
\citep[extending the results of][]{west08a}.
The data will also enable us to use kinematics to understand the distribution of ages,
especially at the stellar/sub-stellar boundary.
Finally, the sample will contain a class of L dwarfs
which are peculiarly blue in the near-infrared
but have typical L dwarf colors in SDSS $i-z$ \citep{schmidt10c}.  

The program was divided into different densities in Stripe82
(5 deg$^{-2}$) and the rest of the BOSS footprint (1 deg$^{-2}$).
In order to select the cleanest possible sample, we included 2MASS
magnitudes as 
part of our selection criteria. Our Stripe 82 criteria were:
\begin{itemize}
\item $i_{\rm PSF}-z_{\rm PSF}>1.14$
\item $i_{\rm PSF}<21$
\item $i_{\rm PSF}-J > 3.7$
\item $1.9<z_{\rm PSF}-J<4$
\end{itemize}
The criteria for the rest of the BOSS footprint are:
\begin{itemize}
\item $i_{\rm PSF}-z_{\rm PSF}>1.44$
\item $i_{\rm PSF}<20.5$
\item $i_{\rm PSF}-J > 3.7$
\item $1.9<z_{\rm PSF}-J<4$
\end{itemize}
Here, the $J$ band photometry, unlike the SDSS photometry, is Vega-based.  

The targets are divided into subsamples with different priorities.
L dwarfs are both less common and fainter, so our priority was to detect L dwarfs
over late-M dwarfs.
Targets likely to be bright L dwarfs ({\bf BRIGHTERL}; $i_{\rm PSF}<19.5$, $i_{\rm PSF}-z_{\rm PSF}>1.14$)
are assigned first priority and fainter L dwarfs
({\bf FAINTERL}; $i_{\rm PSF}>19.5$, $i_{\rm PSF}-z_{\rm PSF}>1.14$)
are assigned second priority.
The bright M dwarfs ({\bf BRIGHTERM}; $i_{\rm PSF}<19.5$, $i_{\rm PSF}-z_{\rm PSF}<1.14$) are assigned
third priority and {\bf FAINTERM} dwarfs are assigned lowest priority.
\\

{\bf Low-Mass Binary Stars:}
Ultra-wide, low-mass binaries probe how dynamical interactions affect
and shape the star formation process and the environment in which the processes occur.
The binaries are also ideal coeval laboratories to constrain and calibrate properties of
low-mass stars, as they were presumably born at the same time of the same material.
The Sloan Low-mass Wide Pairs of Kinematically Equivalent Stars \citep[SLoWPoKES;][]{dhital10a}
catalog is the largest sample of wide, low-mass binaries.
The objects in the catalog were identified from common proper motions and photometric
distances but lacked information regarding radial velocities.
The BOSS spectra will be used to measure radial velocities for the components and
confirm their physical association.
Targets for this program were selected in two ways: (i) SLoWPoKES systems of spectral
type M0 or later which had robust SDSS/USNO-B proper motions \citep{munn04a} and
(ii) systems of spectral type M4 or later, without proper motions.
Both sets of targets were selected to have angular separations of 65--180$\2pr$
to avoid fiber collision and to be brighter than $i_{\rm fib2}=20$ to achieve the critical SNR.
The pairs from the second sub-sample are valuable for probing the
lower-mass mid- and late-type M stars, which were underrepresented in SLoWPoKES.
Although the lack of proper motion matching for these pairs makes them more susceptible
to chance alignments, radial velocities from BOSS spectroscopy will confirm their common motions.
From the combined sample, 500 pairs were randomly selected for targeting with BOSS,
and both components will be observed.
The spectra will be used to measure and calibrate the metallicity and mass--age--activity
relationship for low-mass stars \citep[e.g.,][]{dhital12a}, especially at
the mid--late M spectral types.
There was an error in correcting the positions of the target for their proper motions
in the first year, affecting targets in plates numbered less than 3879 and between 3965--3987;
few of these targets resulted in usable spectra.
\\
{\bf White Dwarfs and Hot Subdwarf Stars:}
The SDSS multi-color imaging efficiently distinguishes hot white
dwarf and subdwarf stars from the bulk of the stellar and quasar
loci \citep{harris03a}.  Special target classes in SDSS produced the
largest spectroscopic samples of white dwarfs
\citep{kleinman04a,eisenstein06a}.  However, much of SDSS white dwarf
targeting required that the objects be unblended, which caused many
brighter white dwarfs to be skipped \citep[for a detailed discussion,
see \S~5.6 of][]{eisenstein06a}.  This BOSS ancillary program relaxes
this requirement and imposes color cuts to focus on warm and hot
white dwarfs.  Importantly, the BOSS spectral range extends further
into the UV, allowing full coverage of the Balmer lines.  We require
targets to be point sources with good $u$, $g$, and $r$ photometry
(following the clean point source selection from the DR7 documentation)
and USNO counterparts. We restrict to regions inside the DR7 footprint
with Galactic extinction of $A_r<0.5$ mag.  Targets must satisfy
$g<19.2$, $(u-r)<0.4$, $-1<(u-g)<0.3$, and $-1 < (g-r) <0.5$, using
extinction-corrected model magnitudes.  Additionally, targets that
do not have $u-r<-0.1$ and $g-r<-0.1$ must have USNO proper motions
of more than $2\2pr$/century.  Objects satisfying the selection criteria
and not observed in previously in SDSS are denoted by the {\bf
WHITEDWARF\_NEW} target flag, while those with SDSS spectra are
assigned the {\bf WHITEDWARF\_SDSS} flag.  Some of the latter objects are
re-observed with BOSS in order to obtain the extended wavelength
coverage.  This color selection includes DA stars with temperatures
above $\sim$14,000~K and helium atmosphere white dwarfs above
$\sim$8000~K, as well as many rarer classes of white dwarfs.  Hot
subdwarfs (sdB and sdO) will be included as well.  Many of these
stars are excellent spectrophotometric standards and can be tested
in comparison to the BOSS F-star calibration.
\\

{\bf Distant Halo Giant Stars:}
Rare giants in the outer halo of the Milky Way are selected using a
targeting strategy tested in SEGUE2
\citep[the SDSS-III counterpart to SEGUE;][Rockosi et al., in prep]{yanny09a}.
Observations of the new BOSS targets are expected to
increase the number of known halo stars beyond 60 kpc by a factor of ten.
The full sample {\bf RED\_KG} will enable measurements of substructure of the Milky
Way in position-velocity phase-space, a signature of the hierarchical assembly of the stellar halo.
The data will thus provide tests of $\Lambda$CDM predictions for galaxy assembly in a
part of the Galaxy expected to be almost entirely composed of debris from recent accretions \citep{xue11a}.
The sample will also enable dynamical mass estimates for the Milky Way out to 150 kpc.
Objects with $17<g_{\rm fib2}<19.5$ are selected from a color box, with
($u_{\rm PSF}-g_{\rm PSF}$, $g_{\rm PSF}-r_{\rm PSF}$) corners of
(0.8,2.35), (0.8,2.65), (1.4,3.0), and (1.4,3.9).
This selection identifies stars near the tip of the red giant branch in the
region of the $u-g/g-r$ diagram where giants separate from the locus
of foreground dwarfs \citep{yanny09a}.
In addition, total proper motion must be $< 11$ mas/year, zero within
three standard deviations of the proper motion errors, to reject
nearby stars.  The $3\2pr$ fiber magnitude must be $i > 16.5$ as an
additional mechanism to reject bright targets.
A smaller sample of targets is found on the commissioning plates.
Denoted {\bf RVTEST}, this sample targets stars previously observed spectroscopically
as a test of the reproducibility of velocity measurements.
\\

{\bf Bright Galaxies:}
Bright galaxies were commonly missed in the original SDSS spectroscopic survey 
due to fiber collisions, bright limits
(objects with model magnitudes $r>15$, $g>15$, or $i>14.5$ were excluded),
and errors in the deblending of overlapping images \citep{strauss02a}.
Approximately 10\% of the brightest galaxies were not spectroscopically observed \citep{fukugita07a}.
To improve the completeness of this spectroscopic sample,
objects were chosen with Petrosian radius $> 1\2pr$ in $r$ to reject stars,
with no saturated pixels, and with extinction-corrected, Petrosian \citep{petrosian76a, strauss02a}
$r$ magnitude between 10 and 16.
Targets were also required to have an extinction-corrected Petrosian magnitude $i<20$ and $z<20$
to exclude misidentified satellite tracks which would not show up in the other bands.
Galaxies without spectra ($\sim$24,000 from the original list of $\sim$93,000) where then
visually vetted to remove foreground stars that remained in the sample,
detector artifacts (e.g. internal reflections) that were misidentified, and other sources of confusion.
In cases where a foreground star was misidentified as the galaxy center,
the target position was moved to the correct position.
In cases of merging galaxies, we visually identified multiple targets corresponding to the
centers of each galaxy.
The list was cross-correlated with the Third Reference Catalog of Bright Galaxies
\citep[RC3][]{devaucouleurs91a, corwin94a}, and any targets that did not
appear in the original SDSS spectroscopic survey were added to the target list
(0.05\% of the final list).
Finally, targets within $2\2pr$ of a star that appears in the Tycho-2
Catalog \citep{hog00a} were removed.  The final sample includes 8637
galaxies over the BOSS footprint. 
\\

{\bf High Energy Blazars and Optical Counterparts of Gamma-ray Sources:}
We targeted candidate optical counterparts of sources detected
(or likely to be detected) by NASA's Fermi Gamma-ray Space Telescope \citep{atwood09a},
with the goal to spectroscopically confirm and provide redshifts for
candidate gamma-ray blazars, with model magnitude $m<21$ in any of the three bandpasses
$g$, $r$, or $i$. 
We also require targets to have $3\2pr$ fiber magnitudes $m > 16.5$
to minimize impact of fiber cross-talk.
Ranked in approximate order of priority,
fibers are assigned to targets from the following subprograms:

\begin{itemize}

\item
{\bf BLAZGXR}: about 300 blazar candidates are assigned at highest priority to DR7
optical sources within Fermi gamma-ray error ellipses.
Targets must also lie within the $< 1\pr$ radius error circle for X-ray sources
in the ROSAT All-Sky Survey (RASS) \citep{voges99a,voges00a} and within $2\2pr$ of
a FIRST \citep{becker95a} radio source.

\item
{\bf BLAZGRFLAT}: about 175 blazar candidates detected with Fermi and the Combined
Radio All-Sky Targeted Eight GHz Survey \citep[CRATES;][]{healey07a}.
Objects from the DR7 catalog within $2\2pr$ of a CRATES radio source and within a
Fermi error ellipse were targeted.

\item
{\bf BLAZGXQSO}: 95 further candidate X-ray and gamma-ray emitting quasars/blazars,
including photometric quasar/blazar candidates \citep{richards09a},
as well as confirmed DR7 quasars/blazars \citep{schneider10a} revisited to
assess optical spectral variability.
Targets are selected that lie within $< 1\pr$ of a RASS X-ray source and within Fermi error ellipses.

\item
{\bf BLAZGRQSO}: 185 candidate radio and gamma-ray emitting quasars/blazars,
including both photometric candidates \citep{richards09a}, and DR7 confirmations
\citep{schneider10a} revisited to assess optical spectral variability.
Targets are selected that lie within $2\2pr$ of a FIRST radio source
and within Fermi error ellipses.

\item
{\bf BLAZGX}: 75 targets that are candidate high-energy counterparts but which
lack typical (e.g., radio emission, unusual optical color, etc.)
blazar properties were targeted to probe unknown classes of gamma ray sources.
The optically brightest objects from DR7 within the Fermi error ellipses and
within $1\pr$ of a RASS X-ray source were preferentially targeted.

\item
{\bf BLAZXR}: 1100 targets are selected that may plausibly emerge as Fermi sources,
but are still below the detection limits in the early Fermi source catalogs.
The approach is similar to the ``ROSAT\_A'' target selection scheme described in
\citet{anderson03a} and the ``pre-selection'' approach of \citet{healey08a}
that provided many of the gamma-ray counterpart associations reported in the first Fermi
catalogs \citep{abdo10a,abdo10b}.
Targets are chosen from the DR7 photometry catalog with radio coincidence
(within 2$\2pr$ of a FIRST source) and X-ray coincidence ($< 1\pr$ of a RASS source).
This sample overlaps heavily with the BONUS quasar sample, but includes quasars at lower redshift.

\end{itemize}

In addition, there were ten miscellaneous candidate blazar spectra taken in an early trial
of this program.  These targets were assigned subcategory names using
the following flags: {\bf BLAZGVAR}, {\bf BLAZR}, and {\bf BLAZXRSAM}. 
\\

{\bf An X-ray View of Star-Formation and Accretion in Normal Galaxies:}
The extended wavelength coverage and improved throughput of BOSS relative to SDSS enable
studies of the relationship between star formation and black hole accretion in galaxies
using the key diagnostic H$\alpha$/N[II] to $z \sim 0.49$.
For this study, a target list was derived from the matched Chandra Source Catalog
\citep[CSC; version 1;][]{evans10a} and SDSS DR7 photometric catalogs.
High quality matches underwent visual inspection in X-ray and optical imaging and were
required to have positional matches between the SDSS DR7 catalog and the CSC as defined by the
\citet{rots09a} matching method that incorporates positions, positional errors, and sky coverage.
Matches with Bayesian probability $< 0.5$ suffer from a larger number of multiple matches and are discarded.
In addition, targets were required to have model magnitude $16.5<r<20.75$ and Chandra off-axis angles $\theta<10\pr$.
Objects with existing SDSS spectroscopy, proper motions from \citet{munn04a, munn08a} exceeding
11 mas/year \citep[selection criteria are described further in][]{haggard10a},
or poor-quality X-ray measurements are removed from the sample.
Because only sources with Chandra coverage were included, the
target have a non-uniform distribution over the DR7 footprint.
To avoid overlap between the targets selected for this program and the program
``Remarkable X-ray Source Populations'' described below, the target lists were cross-correlated
(using a $2\2pr$ match radius) and 754 duplicates were removed from this program designation.
\\

{\bf Remarkable X-ray Source Populations:}
This program targets remarkable serendipitous X-ray sources from the 
Second XMM-Newton Serendipitous Source Catalog \citep[2XMMi;][]{watson09a}
and the Chandra Source Catalog \citep[CSC;][]{evans10a} that do not 
already have available identification spectra. Source types of primary
interest include AGN (often obscured), X-ray binaries, magnetic 
cataclysmic variables, and strongly flaring stars. 

XMM-Newton sources were required to lie within $14\pr$ of the pointing center
of an XMM-Newton observation, to have a 2XMMi detection likelihood greater 
than 12 as defined in \citet{watson09a},
to have a statistically significant match to an SDSS photometric counterpart,
and not to have any 2XMMi problem flags set. Similarly, Chandra 
sources were required to have an SDSS 
counterpart and not to have any indication of Chandra source
confusion.  All XMM-Newton and Chandra sources having Galactic latitudes
of $|b|<20\deg$ were furthermore removed to emphasize extragalactic sources.

The four XMM-Newton source types for this program were all selected
using SDSS model magnitudes and are determined as follows: 

\begin{itemize}

\item
{\bf XMMHR:} 1030 XMM-Newton sources were selected that have unusual 
2XMMi hardness ratios in the HR2-HR3 plane. These sources were
also required to have $i<20.5$ (to ensure reasonable BOSS spectral
quality) and to have 2--12 keV X-ray to $i$-band flux ratios greater
than 0.03 (to minimize stellar contamination).  Optical
fluxes are defined as $f_\nu \Delta \nu$, where $\Delta \nu$ is
the width of the bandpass.  

\item
{\bf XMMBRIGHT:} 826 XMM-Newton sources were selected that have bright 
\hbox{2--12~keV} fluxes (brighter than 
$5\times 10^{-14}$~erg~cm$^{-2}$~s$^{-1}$). 
These sources were also required to have $i<20.0$.

\item
{\bf XMMRED:} 627 optically red XMM-Newton sources were selected that have 
SDSS colors of $g-i>1.0$. These sources were also required to have 
$i<19.3$ and to have 2--12 keV X-ray to $i$-band flux ratios greater
than 0.03. 

\item
{\bf XMMGRIZ:} 149 XMM-Newton sources were selected that have ``outlier'' 
SDSS colors. Specifically, we selected SDSS point-source counterparts
that have $g-r>1.2$ or $r-i>1.0$ or $i-z>1.4$. These sources were also 
required to have \hbox{$i=18$--21.3} and to have 0.5--2 keV X-ray to
$i$-band flux ratios greater than 0.1. 
\end{itemize}

\noindent
There is some overlap among these XMM-Newton 
source-selection approaches (e.g., a bright XMM-Newton source might 
also have optically red colors). At each selection step, we removed 
sources already selected in previous XMM-Newton selection steps 
(following the ordering above). There are 2632 selected XMM-Newton 
sources in total. 

As with the XMM-Newton sources, Chandra source types were all selected
using SDSS model magnitudes and are determined as follows:

\begin{itemize}
\item
{\bf CXOBRIGHT:} 387 Chandra sources were selected that have bright
\hbox{2--8~keV} fluxes (brighter than 
$5\times 10^{-14}$~erg~cm$^{-2}$~s$^{-1}$). 
These sources were also required to have $i<20.0$.

\item
{\bf CXORED:} 635 optically red Chandra sources were selected that have 
SDSS colors of $g-i>1.0$. These sources were also required to have 
$i<19.3$ and to have 2--8 keV X-ray to $i$-band flux ratios greater
than 0.03.

\item
{\bf CXOGRIZ:} 66 Chandra sources were selected that have ``outlier'' 
SDSS colors. Specifically, we selected SDSS point-source counterparts
that have $g-r>1.2$ or $r-i>1.0$ or $i-z>1.4$. These sources were also 
required to have \hbox{$i=18$--21.3} and to have 0.5--2 keV X-ray to
$i$-band flux ratios greater 
than 0.1. 
\end{itemize}

\noindent
Again, there is some overlap among these Chandra source-selection
approaches, and again at each step we removed sources already selected 
in previous Chandra selection steps (following the ordering above). 
There are 1088 selected Chandra sources in total. We furthermore removed 
selected Chandra sources that were also selected XMM-Newton sources;
this reduced the number of selected Chandra sources to 952. 
\\

{\bf Star-Forming Radio Galaxies:}
Joint analysis of SDSS,
FIRST, and the NRAO VLA Sky Survey \citep[NVSS;][]{condon98a}
has shown that low redshift radio
AGNs play an essential role in regulating the growth of massive
galaxies \citep[e.g.][]{best05b,best07a}.
However, much less is known about the detailed interplay of gas
cooling and radio feedback in more luminous radio galaxies at
higher redshifts.
Current samples are incomplete, in particular for radio galaxies with
significant on-going star formation.
This ancillary program selects radio galaxies with blue colors at $z > 0.3$
that would otherwise be missed from the LOWZ and CMASS samples.
Galaxy targets were selected from DR7 according to the following criteria:
extended morphology in SDSS photometry; (2) clean $ugriz$ model photometry
and $17 < i < 19.9$; (3) $i_{fib2} < 21.7$;
(4) [$(g-r) >  1.45$] OR [$(u-g) < 1.14*(g-r)$] where photometry is determined from model magnitudes.
The later criterion is designed to color-select objects at $z > 0.3$.
We cross-matched this sample with the FIRST catalog (July 2008 version) and
selected all objects within $3\2pr$ of FIRST sources with fluxes $> 3.5 $~mJy.
Most targets were within $1.5\2pr$.
Finally, we rejected objects spectroscopically observed by SDSS-I/II,
and objects meeting the target selection criterion for the galaxy samples.
In total there were 4610 targets;
we randomly sampled these to produce a final list of 4170 ancillary targets.
\\

{\bf Galaxies Near SDSS Quasar Sight-Lines:}
Obtaining accurate redshifts of galaxies projected near the lines of sight to quasars with
existing SDSS spectra allows a study of the properties of galaxies that are
associated with intervening quasar absorption systems.
BOSS enables a study at small galaxy-quasar separations
that could not be done with SDSS due to fiber collisions.
Galaxy targets are selected by a $ugr$ color cut to lie in a redshift range ($z>0.35$) where
MgII at 2800 \AA\ is detectable in SDSS spectra.
The sample of spectroscopically confirmed quasars was selected with model magnitude $g<19.2$
and redshift $0.7<z<2.1$ from the SDSS DR7 quasar catalog \citep{schneider10a}.
Galaxies were chosen that lie between $0.006 \pr$
and $1 \pr$ of a quasar with spectroscopy, with model magnitudes 
$17.5 < i < 19.9$,
$A_g < 0.3$, and $(g-r) > 1.65$ OR $(u-g) <  1.14 \times (g-r) - 0.4$ to select
objects with $z>0.35$.  The sample is weighted to have similar
numbers of galaxies at separations $b<0.5 \pr$ and $0.5 \pr<b<1 \pr$.
A sample of approximately 3000 galaxies will help determine
whether absorption is correlated with
galaxy type, the covering fraction of absorption as a function of radius,
and the size of velocity offsets.
\\

{\bf Luminous Blue Galaxies at $0.7<z<1.7$:}
Studies from the second Deep Extragalactic Evolutionary Probe \citep[DEEP2;][]{davis03a} reveal that the most luminous,
star-forming blue galaxies at $z\sim 1$ appear to be a population that
evolves into massive red galaxies at lower redshifts \citep{cooper08a}.
Sampling 2000 color-selected galaxies in Stripe 82
and from the CFHT-LS Wide fields (W1, W3, and W4) allows a measure of the clustering of the rarest,
most luminous of these blue galaxies on large scales.
Such a measurement has not previously been conducted, as prior galaxy evolution-motivated surveys
have a limited field of view and mostly target fainter galaxies.

The galaxy targets were color-selected based on the CFHT-LS photometric-redshift catalog \citep{coupon09a}.
Different color selections were explored using either the
($u_{\rm PSF}-g_{\rm PSF}$, $g_{\rm PSF}-r_{\rm PSF}$) color-color diagram down to $g_{\rm PSF}<22.5$ or the
($g_{\rm PSF}-r_{\rm PSF}$, $r_{\rm PSF}-i_{\rm PSF}$) color-color diagram down to $i_{\rm PSF}<21.3$.
Detailed description of the color selection and redshift measurement is in \citet{comparat12a}.
Using this dataset, photometric redshifts can be re-calibrated in the CFHT-LS W3 field,
thereby reducing biases in redshift estimates at $z>1$.
Measurement of the galaxy bias of these luminous blue galaxies will
be presented in Comparat et al (2012b, in preparation).
This dataset has been important in motivating the ``extended-BOSS'' project
(PI, J.-P. Kneib), a survey that is proposed to begin in Fall
2014 as part of the successor to SDSS-III.
The data will also be used to improve the targeting strategy of future projects such as
BigBOSS \citep{schlegel11a}.
\\

{\bf Broad Absorption Line (BAL) Quasar Variability Survey:}
Thousands of BAL quasars were discovered in the SDSS-I and -II
\citep[e.g.][]{gibson09a}. In some cases, repeat spectroscopy showed 
variable absorption, providing clues to the nature of the BAL 
phenomenon \citep[e.g.,][]{lundgren07a, gibson08a, gibson10a}. 
Returning with BOSS to obtain repeat spectra on a much larger 
sample of these quasars allows a large-scale study of BAL 
variability on multi-year timescales in the rest frame. The 
resulting data provide insight into the dynamics, structure, 
and energetics of quasar winds. First results from this ancillary 
project are presented in \citet{filizak12a}. 

The targets for this ancillary project were selected before the
decision was made to re-target known quasars at $z>2.15$ (see \S\ref{subsec:QSOTS}), 
and thus there is some overlap between these two samples. However, 
this ancillary project also provides many unique targets at 
$z<2.15$. The main sample of BAL quasars chosen for study contains
2005 objects assigned the ancillary target 
flag {\bf VARBAL}; this sample is about two orders of 
magnitude larger than those used previously to investigate 
BAL variability on multi-year timescales. These 2005 
objects were selected to be optically bright ($i_{\rm PSF}<19.28$ 
with no correction for extinction) and to have at least 
moderately strong absorption in one of their BAL troughs 
\citep[with a ``balnicity index'' of BI$_0>100$~km~s$^{-1}$ as measured 
by][]{gibson09a}. In addition, only quasars that are in 
redshift ranges such that strong BAL transitions are fully 
covered by the SDSS-I, SDSS-II and BOSS spectra (from outflow 
velocities of \hbox{0--25000~km~s$^{-1}$}) were included; see \S4 
of \citet{gibson09a} for further explanation. The corresponding 
redshift ranges are 
$1.96<z<5.55$ for Si~IV BALs,
$1.68<z<4.93$ for C~IV BALs,
$1.23<z<3.93$ for Al~III BALs, and
$0.48<z<2.28$ for Mg~II BALs.
Finally, for those objects in the \citet{gibson09a} catalog 
that have measurements of the SNR at 
rest-frame 1700~\AA\ (SN$_{1700}$), we require that SN$_{1700}>6$; 
this criterion ensures that high-quality SDSS/SDSS-II spectra 
are available for these targets. 

In addition to the primary {\bf VARBAL} sample objects described 
above, the BAL quasar variability survey also targets 
102 additional BAL quasars selected with other approaches. 
These targets may violate one or more of the selection criteria 
utilized for the {\bf VARBAL} targets, but they have been 
identified as worthy of new observations nonetheless. The relevant 
source types for these additional BAL quasars are the following: 

\begin{itemize}

\item
 {\bf LBQSBAL} and {\bf FBQSBAL} are BAL quasars identified in
 the Large Bright Quasar Survey \citep[LBQS; e.g.,][]{hewett95a} and the 
 FIRST Bright Quasar Survey \citep[FBQS; e.g.,][]{white00b}, respectively. 
 They thus have LBQS or FBQS spectra predating the SDSS-I and -II
 spectra by up to a decade or more.

\item
 {\bf OTBAL} (Overlapping-Trough BAL quasars) are BAL quasars with 
 nearly complete absorption at wavelengths shortward of Mg~II in 
 one epoch and which in one case have already shown extreme 
 variability \citep[e.g.,][]{hall02a}. 

\item
 {\bf PREVBAL} are BAL quasars observed more than once by 
 SDSS-I and -II. They thus already possess more than one observation 
 epoch for comparison to BOSS spectra.  

\item
 {\bf ODDBAL} are BAL quasars selected to have various unusual 
 properties \citep[e.g.,][]{hall02a}. For these objects, variability 
 (or the lack thereof) between SDSS-I and -II and BOSS may help to 
 unravel the processes responsible for their unusual spectra.
\\
\end{itemize}

{\bf Variable Quasar Narrow-line Absorption:}
Quasar absorption lines are plentiful in SDSS I and II and have been documented in a
catalog of all lines and systems \citep[QSOALS;][]{york05a}.
This catalog (York et al. 2012, in prep.), now updated through DR7, contains 60,000
uniformly detected quasar absorption line systems in which two or more transitions from common
metal absorption lines (e.g., Mg II, Fe II, C IV) are identified at the same redshift.
This dataset has been used to study the statistics of quasar absorption lines \citep{york06a}
and to confirm correlations with quasars \citep{wild07a} as well as foreground galaxies projected
along the line of sight \citep{lundgren09a}.

It has been shown that smaller equivalent width BALs are more prone
to variation on short timescales \citep[e.g.,][]{barlow94a,lundgren07a}.
A large survey of variability in narrow absorption lines (NALs) is therefore required in order
to examine if this trend applies across a larger range in equivalent widths.
Complementary to the ancillary BAL quasar variability study described above, this program seeks to
compile the largest dataset of multi-epoch observations of quasar sight lines with known {\it narrow}
absorption along the line of sight.
Detections of variability in NAL systems hold great promise for identifying high-velocity
intrinsic quasar absorption and mini-BAL emergence and for providing limits on the sizes of
cold gas clouds in the extended haloes of luminous galaxies in the foreground. 

The targets of this program include quasars with $16.5<i_{\rm fib2} < 17.9$ and redshift $0.7<z<2.2$
from the DR7 quasar catalog \citep{schneider10a}, which would otherwise be ignored by the primary
BOSS target selection.
Sight lines with known BALs \citep{gibson08a} are ignored, as this parameter space is being covered by
the separate BOSS BAL variability program.
Each of the sight lines targeted in this program contains a NAL system detected at $>4 \sigma$
(including multiple unambiguous transitions of Mg II, C IV, or both), which have been
identified in \citet{york05a}.
However, this program is not limited only to cases in which previously identified NALs disappear in
later epochs, since NALs should have the same probability of emerging along these lines of sight with
and without identified NALs.
In total, this program targets about 3000 quasars with a target density of $\sim 0.35$ deg$^{-2}$.

As one of the science goals of this program is to determine the extent of variable NALs in velocity
space relative to the quasar, the target list includes sight lines with NALs over a wide range in velocity.
The following sub-groups allow for the identification of quasars and absorbers with particular characteristics.

\begin{itemize}
\item
{\bf QSO\_RADIO\_AAL}: radio-loud with 1 associated absorption system
(AAL; v$\leq$5000 km s$^{-1}$ in the quasar rest frame).

\item
{\bf QSO\_RADIO\_IAL}: radio-loud with 1 intervening absorption system
(IAL: v$>$5000 km s$^{-1}$ in the quasar rest frame).

\item
{\bf QSO\_AAL}: radio-quiet source with 1 AAL

\item
{\bf QSO\_IAL}: radio-quiet source with 1 IAL

\item
{\bf QSO\_RADIO}: radio-loud source with multiple AALs and/or IALs

\item
{\bf QSO\_AALs}: radio-quiet source with multiple AAL and/or IALs

\item
{\bf QSO\_noAALs}: radio-quiet source with no AALs and multiple IALs
\\
\end{itemize}

{\bf Double-lobed Radio Quasars:}
Objects identified as optical point sources near
the midpoint of pairs of FIRST radio sources are observed as 
potential double-lobed radio quasars.
Such quasars are important for studying quasar evolution and interactions
of radio jets with their local environment.
Candidates are selected by identifying FIRST pairs with a separation less than
$60\2pr$ and no SDSS optical counterpart within $2\2pr$ of either source.
SDSS point sources located within a search radius that ranges between $2\2pr$ and $5.3\2pr$
(depending on the separation distance of the FIRST pair) from the midpoint are targeted.
FIRST pairs with a flux ratio $> 10$ are rejected because true double-lobed sources
are unlikely to have a high ratio of lobe-lobe flux density. 
The final catalog includes objects not spectroscopically observed with SDSS,
not targeted in the main BOSS sample,
and with $17.8 < i_{\rm PSF} < 21.6$ (Galactic extinction-corrected).
\\

{\bf High Redshift Quasars:}
High-redshift quasars trace the evolution of early generations of supermassive black holes,
provide tests for models of quasar formation and AGN evolution, and probe evolution
in the intergalactic medium (IGM).
However, the BOSS quasar survey \citep{ross12a} selects objects only to $z \sim 3.5$.
Light emitted by high redshift quasars at wavelengths shorter than \lya\ is absorbed
by the IGM, meaning that for redshifts $z>5.7$, quasars are detected in only the
$z$ band, the reddest filter in the SDSS imaging survey. 
We use areas with overlap imaging, thereby reducing contamination from cosmic
rays and improving the photometry, to select high redshift quasar
candidates in three redshift ranges to fainter magnitudes than in the SDSS survey.

For the main survey, PSF magnitudes for objects with multiple detections are extracted from the
Neighbors table in the DR7 Catalog Archive Server, and the detections are coadded in each band.
Target selection is performed on the coadded photometry.
For objects within Stripe 82, PSF magnitudes were extracted from the coadded image
catalogs described in \citet{annis12a}.
These catalogs combine roughly 20 epochs instead of just two, and permit selection of
objects at fainter magnitudes.
The first part of the program targets objects with similar color cuts
imposed in the SDSS quasar survey \citep{richards02a}.
The SDSS quasar target selection defined two inclusion regions in
$gri$ and $riz$ color space for targeting $z>3.6$ and $z>4.5$ quasars, respectively
\citep{richards02a}. The first two high-$z$ quasar ancillary programs
are straightforward extensions of these color
selection criteria to fainter magnitudes, with limits of $i_{\rm PSF}<21.3$
in the main survey regions and $i_{\rm PSF}<21.5$ (compared to $i_{\rm PSF}<20.2$ for SDSS).
The ancillary target flag {\bf QSO\_GRI} is assigned to objects meeting the $gri$
color criteria, and the flag {\bf QSO\_RIZ} to those meeting the $riz$ criteria.
In both cases, the primary color cut follows a diagonal line in the respective color
plane (e.g., $r-i < A(g-r) + B$ for the $riz$ cut), and the intercept of this line
is shifted slightly upwards for brighter objects. This approach allows objects with similar brightness
to SDSS quasars to be selected with more relaxed color criteria.
This sample will provide a probe of the quasar luminosity function at
high redshift and improve small-scale clustering measurements.
A final class of targets ({\bf QSO\_HIZ}) in this program are candidates
for quasars with redshifts between 5.6 and 6.5. At these redshifts, quasars have 
extremely red $i-z$ colors; this program targets all objects with $(i-z)>1.6$
and $z<20.8$ and that have no detections in the other SDSS bands.
\\

{\bf High-redshift Quasars from SDSS and UKIDSS:}
The final ancillary program described here targets high redshift quasar
candidates through a combination of color cuts combining SDSS $ugriz$ PSF photometry
and UKIDSS $YJHK$ aperture photometry ($1\2pr$ radius apertures). The addition of IR photometry from
UKIDSS provides leverage to separate quasars at $z\sim5.5$ from the
red end of the stellar locus -- indeed, the SDSS quasar survey was limited
to $z<5.4$ by the strong overlap in $ugriz$ colors of higher redshift quasars
with red stars. Quasar candidates were selected by matching stellar objects
from the SDSS DR7 UKIDSS DR3 databases. The initial sample is drawn
from SDSS with the cuts $r-i>1.4$, $i-z>0.5$, and $z<20.2$. Likely stars
are rejected with the criteria $(H-K)_{\rm Vega}<0.53$ or
$(J-K)_{\rm Vega} < 1.3(Y-J)_{\rm Vega} + 0.32$, taking advantage of the fact
that quasars are redder than M stars at longer wavelengths and bluer at shorter wavelengths.
The remaining candidates are then prioritized based on $izYJK$ colors.
For Stripe 82, the color criteria used for prioritizing targets were
slightly relaxed,
owing to the coadded $ugriz$ photometry available from \citet{annis12a}, which
greatly reduced the initial stellar contamination from the $riz$
selection. Objects
selected with the relaxed criteria on Stripe 82 are given the {\bf HIZQSO82}
target bit.

\begin{deluxetable}{lcrrr}
\centering
\tablewidth{0pt}
\tabletypesize{\footnotesize}
\tablecaption{\label{tab:ancillary1} BOSS Ancillary Programs with {\bf ANCILLARY\_TARGET1} Flag}
\tablehead{
\colhead{Primary Program} & \colhead{SubProgram}  & \colhead{Bit}
                  & \colhead{Density} & \colhead{Survey Area} \\
\colhead{} & \colhead{}  & \colhead{Number}
                  & \colhead{(deg$^{-2}$)} & \colhead{(deg$^{2}$)}
}
\startdata
Transient Universe    & AMC                    &  0  & 0.05  &  220\tablenotemark{a}  \\
Transient Universe    & FLARE1                 &  1  & 0.2  &  220\tablenotemark{a}  \\
Transient Universe    & FLARE2                 &  2  & 0.7  &  220\tablenotemark{a}  \\
Transient Universe    & HPM                    &  3  & 0.5  &  220\tablenotemark{a}  \\
Transient Universe    & LOW\_MET               &  4  & 0.3  &  220\tablenotemark{a}  \\
Transient Universe    & VARS                   &  5  & 0.9  &  220\tablenotemark{a}  \\
High Energy Blazars   & BLAZGVAR               &  6  & $\ll 1$  &  7650\tablenotemark{b}  \\
High Energy Blazars   & BLAZR                  &  7  & $\ll 1$  &  7650\tablenotemark{b}  \\
High Energy Blazars  & BLAZGXR                 &  8  & $\ll 1$  &  7650\tablenotemark{b}  \\
High Energy Blazars   & BLAZXRSAM              &  9  & $\ll 1$  &  7650\tablenotemark{b}  \\
Remarkable X-ray Sources & XMMBRIGHT           & 11  & 0.1  & 7650\tablenotemark{b}   \\
Remarkable X-ray Sources & XMMGRIZ             & 12  & 0.02  & 7650\tablenotemark{b}   \\
Remarkable X-ray Sources & XMMHR               & 13  & 0.1  & 7650\tablenotemark{b}   \\
Remarkable X-ray Sources & XMMRED              & 14  & 0.08  & 7650\tablenotemark{b}   \\
BAL Variability  & FBQSBAL             & 15  & 0.003  &  5740  \\
BAL Variability  & LBQSBAL             & 16  & 0.002  &  5740  \\
BAL Variability  & ODDBAL              & 17  & 0.007  &  5740  \\
BAL Variability  & OTBAL               & 18  & 0.003  &  5740  \\
BAL Variability  & PREVBAL             & 19  & 0.004  &  5740  \\
BAL Variability  & VARBAL              & 20  & 0.4   &  5740  \\
Bright Galaxies  & BRIGHTGAL           & 21  & 1.1  &  7650\tablenotemark{b}  \\
Variable Quasar Absorption  & QSO\_AAL               & 22  & 0.08   &  7650\tablenotemark{b}  \\
Variable Quasar Absorption  & QSO\_AALS              & 23  & 0.2   &  7650\tablenotemark{b}  \\
Variable Quasar Absorption  & QSO\_IAL               & 24  & 0.05   &  7650\tablenotemark{b}  \\
Variable Quasar Absorption  & QSO\_RADIO             & 25  & 0.04   &  7650\tablenotemark{b}  \\
Variable Quasar Absorption  & QSO\_RADIO\_AAL        & 26  & 0.02   &  7650\tablenotemark{b}  \\
Variable Quasar Absorption  & QSO\_RADIO\_IAL        & 27  & 0.01   &  7650\tablenotemark{b}  \\
Variable Quasar Absorption  & QSO\_NOAALS            & 28  & 0.01   &  7650\tablenotemark{b}  \\
High Redshift Quasars       & QSO\_GRI               & 29  & 2.7   &  220\tablenotemark{a}  \\
High Redshift Quasars       & QSO\_GRI               & 29  & 0.8   &  2500  \\
High Redshift Quasars       & QSO\_HIZ               & 30  & 0.7   &  220\tablenotemark{a}  \\
High Redshift Quasars       & QSO\_HIZ               & 30  & 0.2   &  2500  \\
High Redshift Quasars       & QSO\_RIZ               & 31  & 1.2   &  220\tablenotemark{a}  \\
High Redshift Quasars       & QSO\_RIZ               & 31  & 0.6   &  2500  \\
Reddened Quasars & RQSS\_SF               & 32  & 1.0   &  $\sim50$  \\
Reddened Quasars & RQSS\_SFC              & 33  & 0.3   &  $\sim50$  \\
Reddened Quasars & RQSS\_STM              & 34  & 0.2   &  $\sim50$  \\
Reddened Quasars & RQSS\_STMC             & 35  & 0.1   &  $\sim50$  \\
SN Host Galaxies & SN\_GAL1             & 36  & 13.8   &  220\tablenotemark{a} \\
SN Host Galaxies & SN\_GAL2             & 37  & 0.4   &  220\tablenotemark{a}  \\
SN Host Galaxies & SN\_GAL3             & 38  & 0.1   &  220\tablenotemark{a}  \\
SN Host Galaxies & SN\_LOC              & 39  & 1.6   &  220\tablenotemark{a}  \\
SN Host Galaxies & SPEC\_SN             & 40  & 0.009   &  220\tablenotemark{a}  \\
Low Mass Binary Stars & SPOKE             & 41  & 0.1   &  7650\tablenotemark{b}  \\
White Dwarf Stars & WHITEDWARF\_NEW    & 42  & 0.5  &  7430\tablenotemark{b}  \\
White Dwarf Stars & WHITEDWARF\_SDSS   & 43  & 0.5  &  7430\tablenotemark{b}  \\
Very Low Mass Stars  & BRIGHTERL             & 44  &  0.07  &  220\tablenotemark{a}  \\
Very Low Mass Stars  & BRIGHTERL             & 44  &  0.08  &  7430\tablenotemark{b}  \\
Very Low Mass Stars  & BRIGHTERM             & 45  &  2.9  &  220\tablenotemark{a}  \\
Very Low Mass Stars  & BRIGHTERM             & 45  &  0.3  &  7430\tablenotemark{b}  \\
Very Low Mass Stars  & FAINTERL              & 46  &  0.3  &  220\tablenotemark{a}  \\
Very Low Mass Stars  & FAINTERL              & 46  &  0.2  &  7430\tablenotemark{b}  \\
Very Low Mass Stars  & FAINTERM              & 47  &  2.2  &  220\tablenotemark{a}  \\
Very Low Mass Stars  & FAINTERM              & 47  &  0.6  &  7430\tablenotemark{b}  \\
Distant Halo Giants  & RED\_KG               & 48  &  0.8  &  10,000  \\
Distant Halo Giants  & RVTEST                & 49  &  $\sim 0.8$  &  $\sim50$ \\
High Energy Blazars  & BLAZGRFLAT                 & 50  & 0.02  &  7650\tablenotemark{b}  \\
High Energy Blazars  & BLAZGRQSO                  & 51  & 0.02  &  7650\tablenotemark{b}  \\
High Energy Blazars  & BLAZGX                     & 52  & 0.01  &  7650\tablenotemark{b}  \\
High Energy Blazars  & BLAZGXQSO                  & 53  & 0.01  &  7650\tablenotemark{b}  \\
High Energy Blazars  & BLAZGXR                    & 54  & 0.03  &  7650\tablenotemark{b}  \\
High Energy Blazars  & BLAZXR                     & 55  & 0.1  &  7650\tablenotemark{b}  \\
Star Forming Radio Galaxies  & BLUE\_RADIO         & 56  & 0.4  & 10,000  \\
X-ray View of Star-Formation and Accretion & CHANDRAV1  & 57  & 0.2  &  7650\tablenotemark{b}  \\
Remarkable X-ray Sources & CXOBRIGHT             & 58  & 0.05  &  7650\tablenotemark{b}  \\
Remarkable X-ray Sources & CXOGRIZ               & 59  & 0.009  &  7650\tablenotemark{b}  \\
Remarkable X-ray Sources & CXORED                & 60  & 0.08  &  7650\tablenotemark{b}  \\
Luminous Blue Galaxies  & ELG          & 61  & 22  &  143  \\
Galaxies near Quasar Sight-Lines  & GAL\_NEAR\_QSO      & 62  & 0.3 & 7650\tablenotemark{b}   \\
Transient Universe & MTEMP             & 63  & 0.5  &  220\tablenotemark{a}  
\enddata
\tablenotetext{a}{Sample is taken from the Stripe 82 region.}
\tablenotetext{b}{Sample is taken from BOSS footprint that overlaps with DR7 imaging data.}
\end{deluxetable}

\begin{deluxetable}{lcrrr}
\centering
\tablewidth{0pt}
\tabletypesize{\footnotesize}
\tablecaption{\label{tab:ancillary2} BOSS Ancillary Programs with {\bf ANCILLARY\_TARGET2} Flag}
\tablehead{
\colhead{Primary Program} & \colhead{SubProgram} & \colhead{Bit}
                  & \colhead{Density} & \colhead{Survey Area} \\
\colhead{} & \colhead{} & \colhead{Number}
                  & \colhead{(deg$^{-2}$)} & \colhead{(deg$^{2}$)}
}
\startdata
High Redshift Quasars with UKIDSS  & HIZQSO82         & 0   & 0.5  &  220\tablenotemark{a}  \\
High Redshift Quasars with UKIDSS  & HIZQSOIR         & 1   & 0.3  & 700   \\
K-Band Selected Quasars & KQSO\_BOSS            & 2   & 1.0  &  220\tablenotemark{a}  \\
No Quasar Left Behind & QSO\_VAR               & 3   & 6.5  &  220\tablenotemark{a}  \\
Variability Selected Quasars  & QSO\_VAR\_FPG       & 4   &  3.4 &   220\tablenotemark{a}  \\
Double-lobed Radio Quasars & RADIO\_2LOBE\_QSO   & 5   & 0.3  &  7650  \\
Brightest Cluster Galaxies   & STRIPE82BCG       & 6   & 6.0  &  220\tablenotemark{a} 
\enddata
\end{deluxetable}

\end{appendix}

\end{document}